\documentclass[11pt,a4paper
amsmath,amssymb,
reprint,%
]{article}

\usepackage{graphicx}% Include figure files
\usepackage{dcolumn}% Align table columns on decimal point
\usepackage{bm}% bold math
\usepackage{hyperref}
\usepackage[utf8]{inputenc}
\usepackage[T1]{fontenc}
\usepackage{mathptmx}
\usepackage{etoolbox}

\usepackage{textcomp} % only to get rid of a warning
\usepackage{gensymb} % \degree
\usepackage{tabularx} % \begin table
\usepackage{caption}
\usepackage{subcaption}
\usepackage{dirtytalk} % \say
\usepackage{array} % see below
\newcolumntype{P}[1]{>{\centering\arraybackslash}p{#1}} % for centred elements table
\newcolumntype{M}[1]{>{\centering\arraybackslash}m{#1}}
\usepackage{tablefootnote} %\tablefootnote 
\usepackage{cleveref} % Section~\ref

\begin{document}
	
	\title{Reducing the Drag of a Bluff Body\\
		 by Deep Reinforcement Learning}
	\author{E. Ballini$^\#$, A. S. Chiappa$^\dagger$, S. Micheletti$^\#$}
	\maketitle 
	\begin{center}
		{\small 
			$^\#$ MOX -- Modellistica e Calcolo Scientifico,
			Dipartimento di Matematica\\ 
			Politecnico di Milano, Piazza L. da Vinci 32, I-20133 Milano, Italy
			{\tt \{enrico.ballini,stefano.micheletti\}@polimi.it}\\[2mm]
			$^\dagger$ EPFL, Brain Mind Institute, Route Cantonale, CH-1015, Lausanne, Switzerland
			{\tt alberto.chiappa@epfl.ch}	
		}
	\end{center}
	
	\date{}
	
	\maketitle
	
	\begin{abstract}
		We present a deep reinforcement learning approach to a classical problem in fluid dynamics, i.e., the reduction of the drag of a bluff body. We cast the problem as a discrete-time control with continuous action space: at each time step, an autonomous agent can set the flow rate of two jets of fluid, positioned at the back of the body. The agent, trained with Proximal Policy Optimization, learns an effective strategy to make the jets interact with the vortexes of the wake, thus reducing the drag. To tackle the computational complexity of the fluid dynamics simulations, which would make the training procedure prohibitively expensive, we train the agent on a coarse discretization of the domain. We provide numerical evidence that a policy trained in this approximate environment still retains good performance when carried over to a denser mesh. Our simulations show a considerable drag reduction with a consequent saving of total power, defined as the sum of the power spent by the control system and of the power of the drag force, amounting to 40\%  when compared to simulations with the reference bluff body without any jet. Finally, we qualitatively investigate the control policy learnt by the neural network. We can observe that it achieves the drag reduction by learning the frequency of formation of the vortexes and activating the jets accordingly, thus blowing them away off the rear body surface.
	\end{abstract}

%% main text
% =====================================================================
% INTRODUCTION 
% =====================================================================
\section{\label{sec:introduction}Introduction} % ----------------------------------------
\begin{figure}[htb]
	\includegraphics[width=\linewidth]{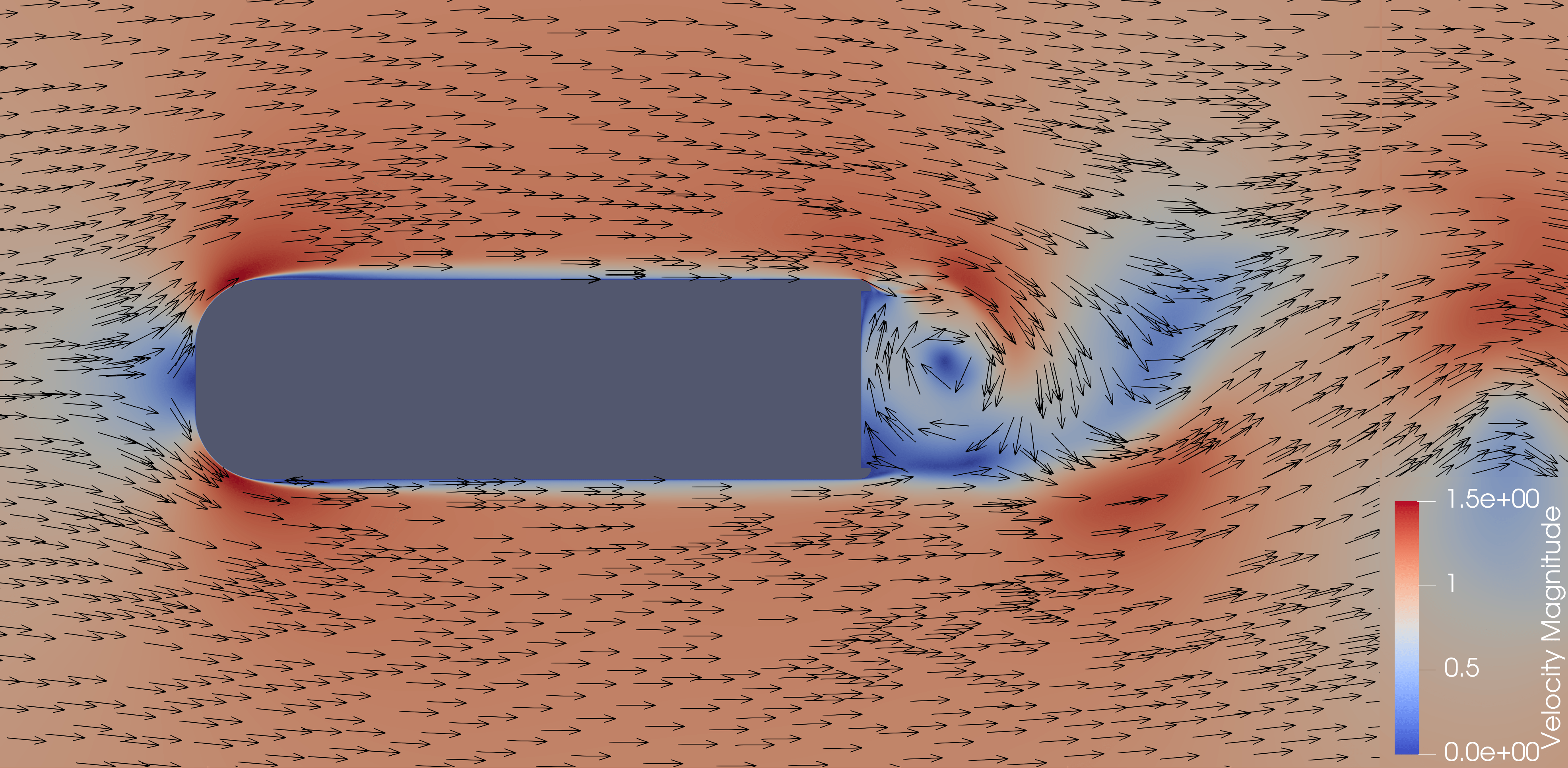} % {img/intro_big.png}
	\caption{Snapshot of the flow around the body. The arrows represent the velocity field and the color map the associated magnitude.}
	\label{fig:bb_intro}
\end{figure}

\begin{figure}[ht]
	\begin{subfigure}[t]{0.5\textwidth}
		\begin{subfigure}[t]{1\textwidth}
			\textbf{(a)}	
		\end{subfigure}
		\includegraphics[width=\linewidth]{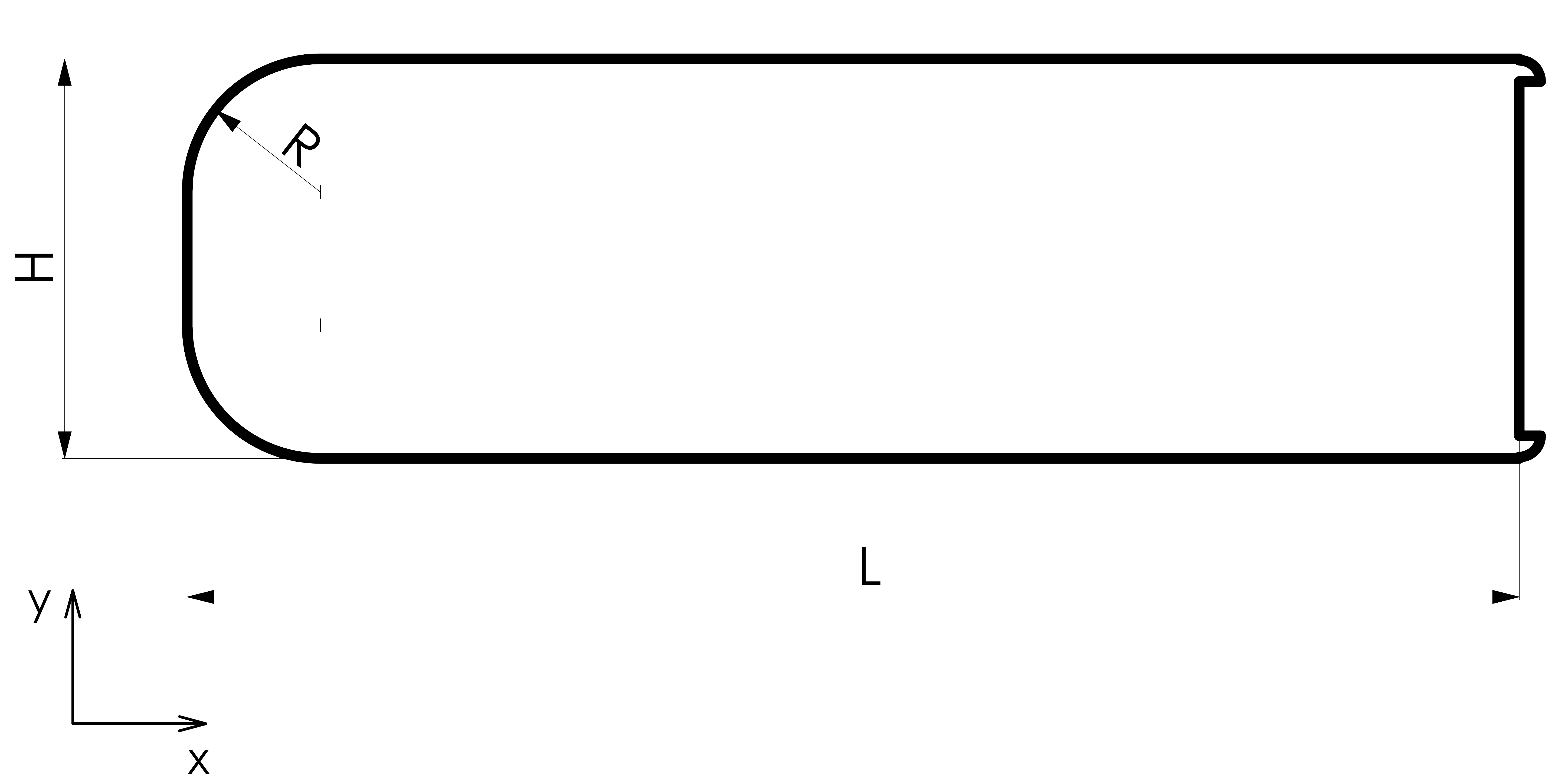}
	\end{subfigure}
	\begin{subfigure}[t]{0.25\textwidth}
		\begin{subfigure}[t]{1\textwidth}
			\textbf{(b)}
		\end{subfigure}
		\includegraphics[width=\linewidth]{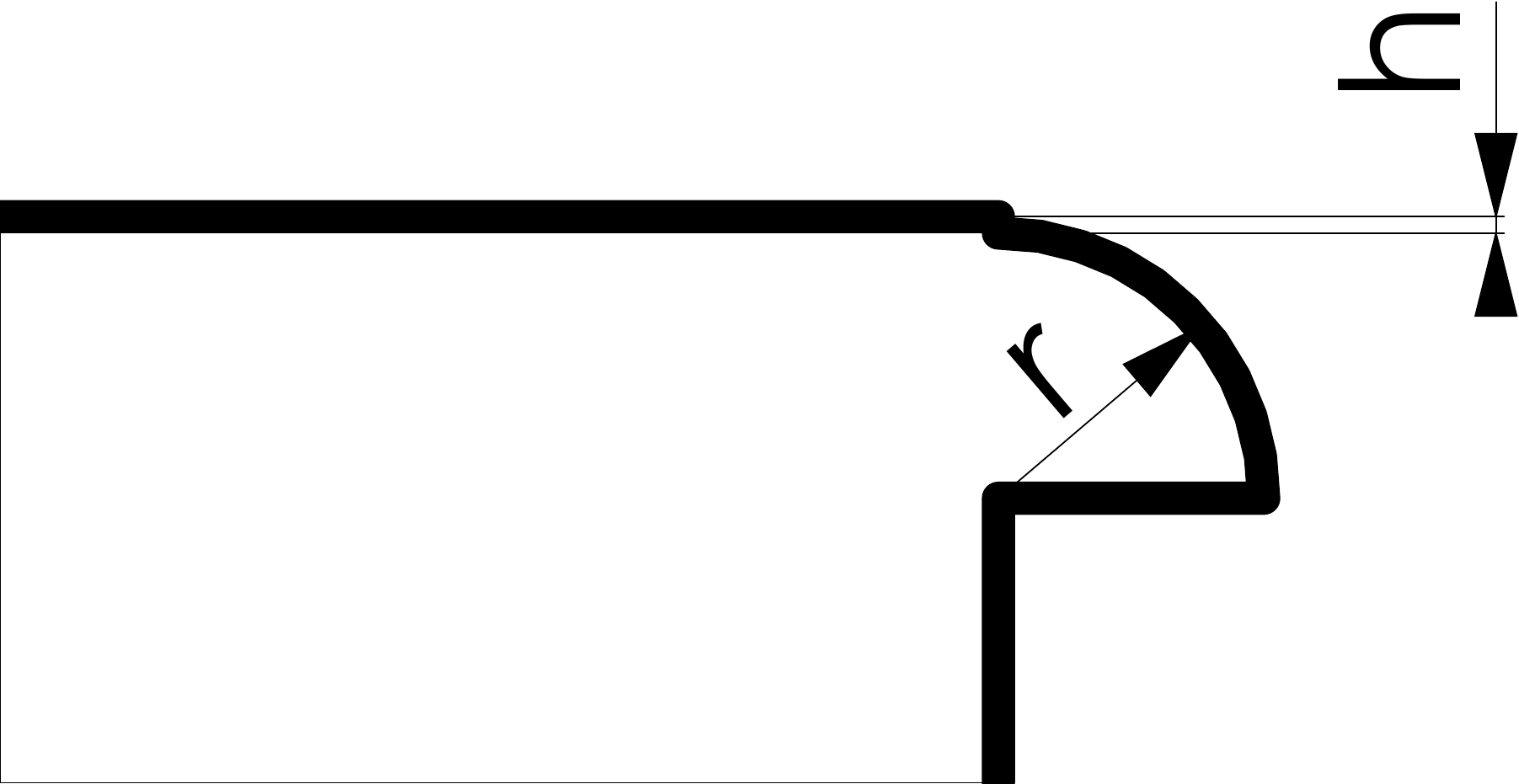}
	\end{subfigure}
	\begin{subfigure}[t]{0.5\textwidth}
		\begin{subfigure}[t]{1\textwidth}
			\textbf{(c)}
		\end{subfigure}
		\includegraphics[width=\linewidth]{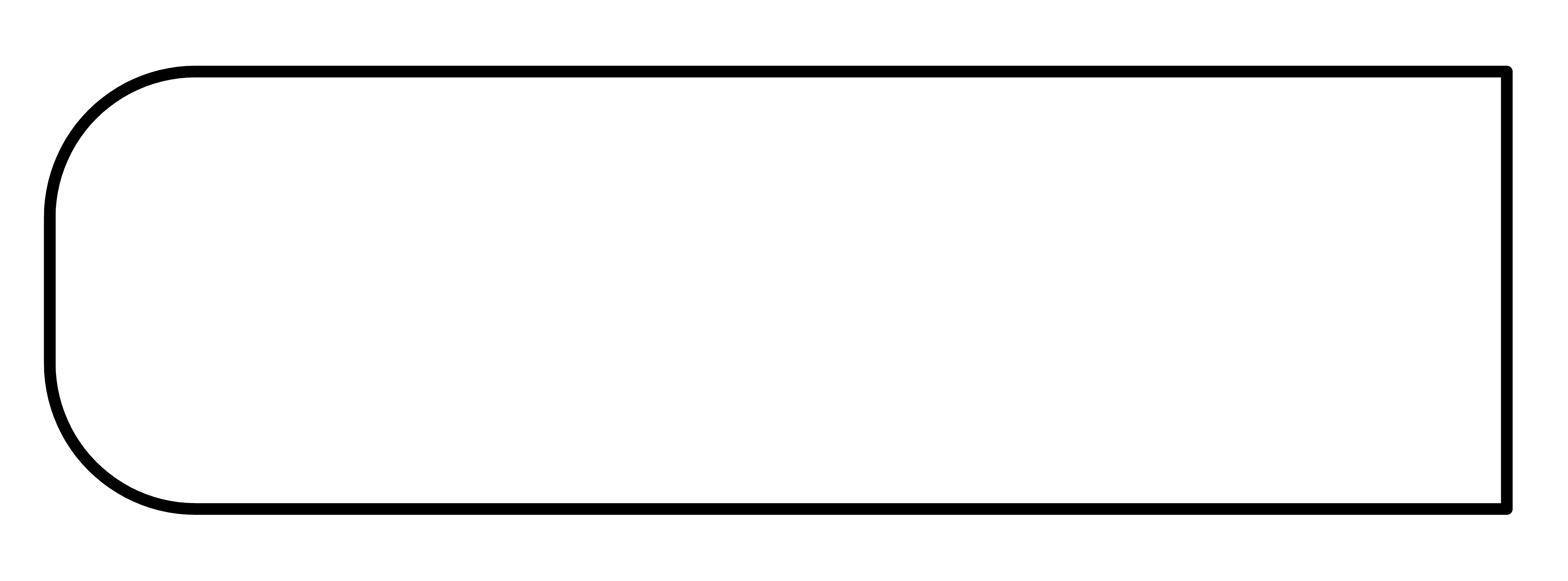}
	\end{subfigure}
	\caption{Schematic representation of the bluff bodies used for the fluid dynamics simulations. \textbf{(a)} Geometry of the bluff body with controlled jets. \textbf{(b)} Curved edge and slit detail. \textbf{(c)} Geometry of the bluff body clean used as reference for the drag reduction evaluation.}
	\label{fig:bluff_bodies}
\end{figure}

% Contextualization:
The reduction of the drag force is a problem of paramount importance in fluid dynamics, because of its ubiquity in aeronautical, naval, and land transport applications. The drag force, acting in the opposite direction to the body's motion, is often the principal source of power consumption. The problem has been deeply studied both for aerodynamic bodies, such as aircraft surfaces, and bluff bodies, i.e., those bodies with a compact geometry that show conspicuous recirculation areas.
In this work, we propose a method for drag reduction with deep reinforcement learning. We consider a 2D bluff body at high Reynolds number, simulating the flow field with unsteady Computational Fluid Dynamics (CFD) (Fig.~\ref{fig:bb_intro}).
The geometry of the body is rectangular, with rounded front edges and with the addition of two small curved edges at the back that alter the initial geometry in a negligible way (Fig.~\ref{fig:bluff_bodies}).
The drag of a bluff body is mainly caused by the vast wake behind it because it strongly influences the pressure distribution around the body. Therefore, our goal is to reduce the drag force by interacting with the wake. This is done by introducing two small nozzles, positioned on the rear surface, that can interact with the wake by emitting fluid at a desired flow rate. The two jets of fluid, coming out of the nozzles, blow on the two small curved edges positioned at the rear corners. We cast the control of the fluid injection rate as a continuous control problem, which can be solved through reinforcement learning. An autonomous agent must learn a policy to optimally control the two nozzles. We remark that the learning is purely simulation-driven, as it does not require any a-priori knowledge of the equations governing the system. The agent learns through experience the function linking its current observation of the environment to the action that maximally reduces the drag force.
The observation consists in the real-time measurement of the pressure on the base surface of the bluff body, at 12 fixed points. In spite of the limited amount of information made available to the agent, the trained control policy successfully reduces the drag force on the bluff body. Compared to a simulation without active jet control, the net drag power saving amounts to 40\%.

An analysis of the control policy learned by the neural network reveals how the jets interact with the wake vortexes to highly reduce the drag force. We found that the agent ignites the jet on the opposite side where the vortex is being generated, pushing its formation downstream. This mechanism increases the base pressure, which is the main cause of the drag reduction.

\section{\label{sec:related_works}Related works}
A long list of approaches to reduce the drag of bluff bodies have been proposed. We can find an overview in \cite{haecheon_2008,eun_2018}. 
% NO JETS:
Examples of some particular applications are:
surface modifications to control the separation point \cite{ hasegawa_2021, % microfiber
	latif_2019, % dimple on top
	kwangmin_2011}, % trip wire on sphere
porous surfaces  \cite{gatto_2018, % holes % NE HAI ALTRI TODO
	klausmann_2017}, % back porousity
addition of other bodies or appendages that interact with the main body \cite{garcia_2021, % flexible plates
	cicolin_2021, % second small cilinder
	siddiqui_2020, % big flap
	karthik_2018, % back flat
	pinelli_2017, % pelskin %
	hirst_2015, % profili posteriori
	hyungmin_2006}, % small rectangles on the edge
plasma actuators \cite{zongnan_2021, % D-shape
	kazemi_2021}, % plasma to ahmed
and closed-loop control of back oscillating flaps \cite{brackston_2016}. % closed-loop, palette piu altro
% JETS open-loop (non controllati):
Many studies concern the use of suitably positioned jets of fluid. The flow can be ejected at constant velocity \cite{zhang_2018},  % stedy blowing
at variable velocity with non-zero mean flow rate \cite{haffner_2020, % unsteady coanda effect and drag reduction for turbulent wake
	wang_2019, % different jets 
	tounsi_2016, % exp on ahmed
	oxlade_2015}, % axis sym high freq 
or at variable velocity with null mean flow rate \cite{qiao_2021}, % loud speaker
% JETS closed-loop (controllati):
The literature regarding controlled jets % closed-loop
is less abundant. Some important studies concern the control of the wake through variable jets \cite{henning_2005, % sythetic, linear black-box and extremunm seeking
	li_2017, % genetic algorithn both open and closed
	longa_2017}. % quello con i risultati simili ai miei
A relevant practical application of the drag reduction of a bluff body is found in heavy vehicles because lower drag means lower fuel consumption. Indeed, solutions focused on both the front and rear have been found for these vehicles \cite{salati_2018, % phd
	choi_haecheon_2014}. % review
Also Machine Learning (ML) has found applications in different branches of fluid dynamics. For an overview, we refer to \cite{sadrehaghighi_2021, brunton_2020, kutz_2017}.
In particular, Reinforcement Learning (RL) \cite{sutton} has also been applied in the fluid dynamics field \cite{garnier_2021, ren_2020}.
Some particular applications are:
PDE control \cite{bucci_2019, farahmand_2017},
flow control \cite{ren_2021, beintema_2020, fan_2020},
control of the movement of objects immersed in a fluid, such as
fish-like swimmers \cite{novati_2017},
microswimmers \cite{muinos_2018},
Zermelo problem \cite{biferale_2019},
drones e gliders \cite{bohn_2019, reddy_2016},
as well as more general applications as 
shape optimization \cite{viquerat_2020},
and the creation of numerical methods to approximate the solution to differential equations \cite{wang_2020}. 
To the best of the authors' knowledge, the application of RL presented in \cite{zheng_2021} is the closest to ours. Indeed, the authors of \cite{zheng_2021}, by providing the RL agent with an accurate measurement of the fluid flow, managed to reduce the oscillation of a cylinder immersed in a low-Reynolds number flow.

% =====================================================================
% FLUID DYNAMICS 
% =====================================================================
\section{\label{sec:fluid_dynamics_model}Fluid dynamics model} 
The numerical experiments simulate the flow around a 2D bluff body of height $H=1$ and length $L=3.3$, as shown in Fig. \ref{fig:bluff_bodies}. All quantities are dimensionless. The front is rounded with a constant radius of $R=0.3$ in order to avoid large separations after the leading edge. The back jets are blown by two slits of height $h=0.03$. 
The jets blow on two $90\degree$ circular edges of radius $r=0.05$. The Reynolds number, based on the height of the bluff body, is $Re=\frac{HU_{\infty}}{\nu}=20000$, where $U_{\infty}$ is the velocity at infinity, and $\nu$ is the kinematic viscosity.
We evaluate the drag reduction resulting from the simulations against a reference clean bluff body depicted in Fig.~\ref{fig:bluff_bodies}. It is nearly the same as the presented bluff body, with the only difference that it has neither jets nor curved edge on the back, so the back is totally plain. 
The fluid motion is modeled by the URANS equations \cite{pope_2000, blazek_2015} and the $k-\omega\ SST$ model for the closure \cite{menter_sst, menter_sst_2}. The numerical schemes configurations and the set of constants used for $k-\omega\ SST$ are reported in Appendix~\ref{app:cfd_settings}.

\subsection{\label{subsec:bc}Boundary conditions}
\begin{figure}[htp]
	\centering
	\includegraphics[width=\linewidth]{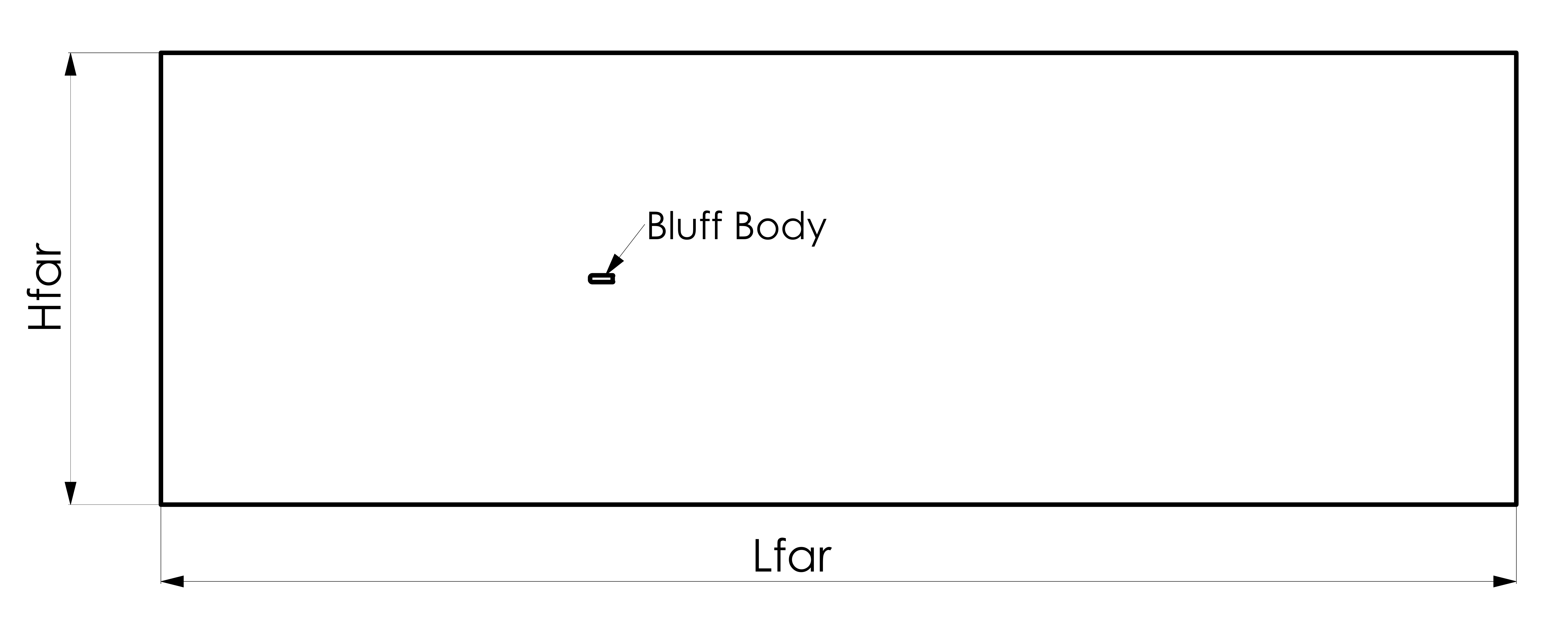}
	\caption{Geometry of the computational domain.}
	\label{fig:domain}
\end{figure}
The external domain, shown in Fig.~\ref{fig:domain}, is set large enough to minimize the effects of the artificial boundaries.
It is 200 units long and 66 units high. The bluff body is positioned at half of the height and at $\frac{1}{3}$ of the length.
While in the numerical experiments the velocity magnitude of the jets will vary according to the output of the policy network (see Section~\ref{sec:rl}), for the assessment of grid and time convergence (see Section~\ref{subsec:grid_time_conv}) it is fixed to a representative value of $2$.
A far-field boundary condition is prescribed on the outer box. The velocity $U_\infty$ is set equal to $1$ and the gauge pressure $P_\infty$ equal to $0$.
The turbulence boundary conditions on the outer box are prescribed by the viscous ratio $r = \frac{\mu_t}{\mu}$ and turbulence intensity $I = \sqrt{\frac{3}{2} k}$.
We set $r = 10$ and $I = 0.05$, which are typical values for an average level of turbulence.
A no-slip boundary condition is applied on the body surface.

\subsection{\label{subsec:mesh}Mesh} 
\begin{figure*}[htp]
	\centering
	\includegraphics[width=\linewidth]{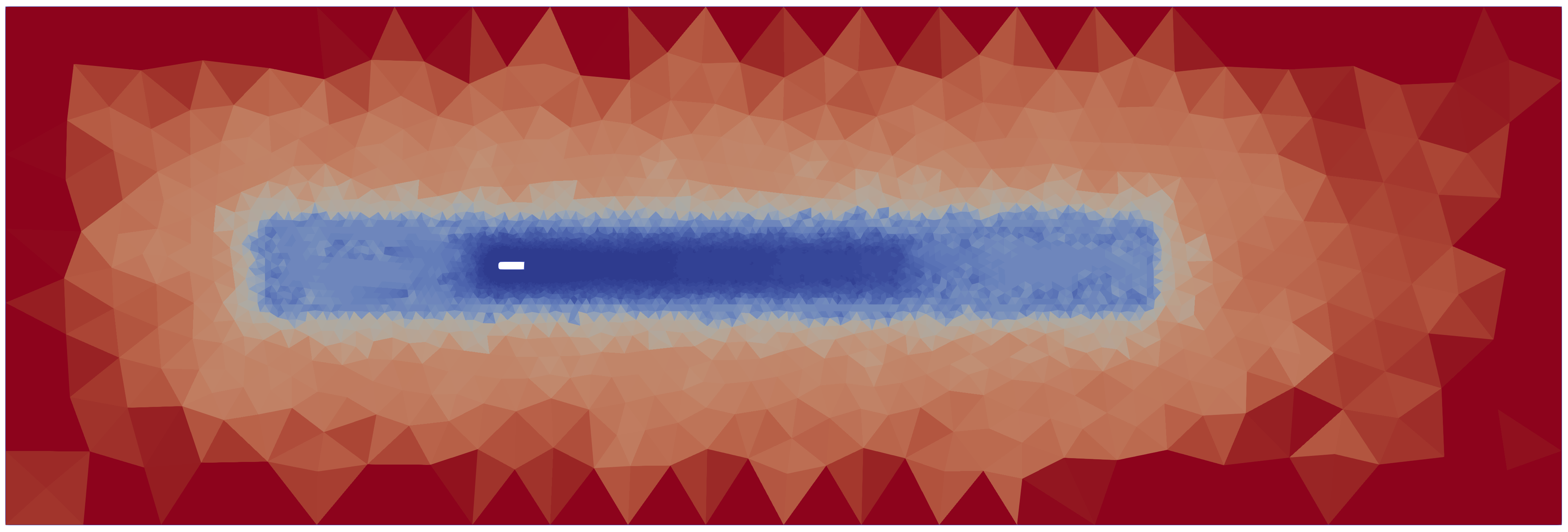}
	\caption{Qualitative representation of the grid cells size. The colors are proportional to the cell area.}
	\label{fig:mesh_colormap}
\end{figure*}

\begin{figure*}[htp]
	\centering
	\includegraphics[width=\linewidth]{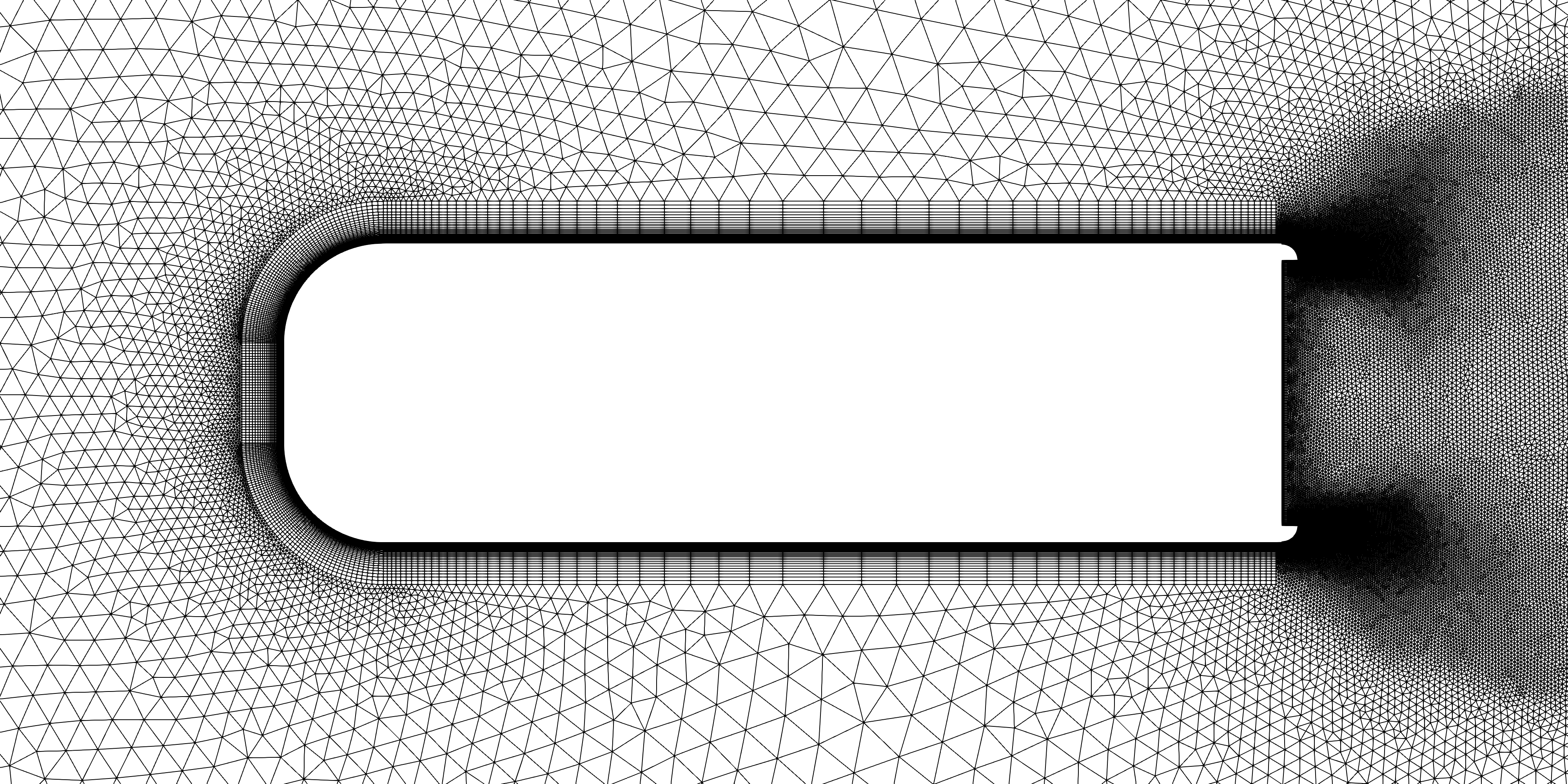}
	\caption{Mesh with 1.48e5 elements: Bluff body detail.}
	\label{fig:mesh_detail_bb}
\end{figure*}

\begin{figure*}[htp]
	\centering
	\includegraphics[width=\linewidth]{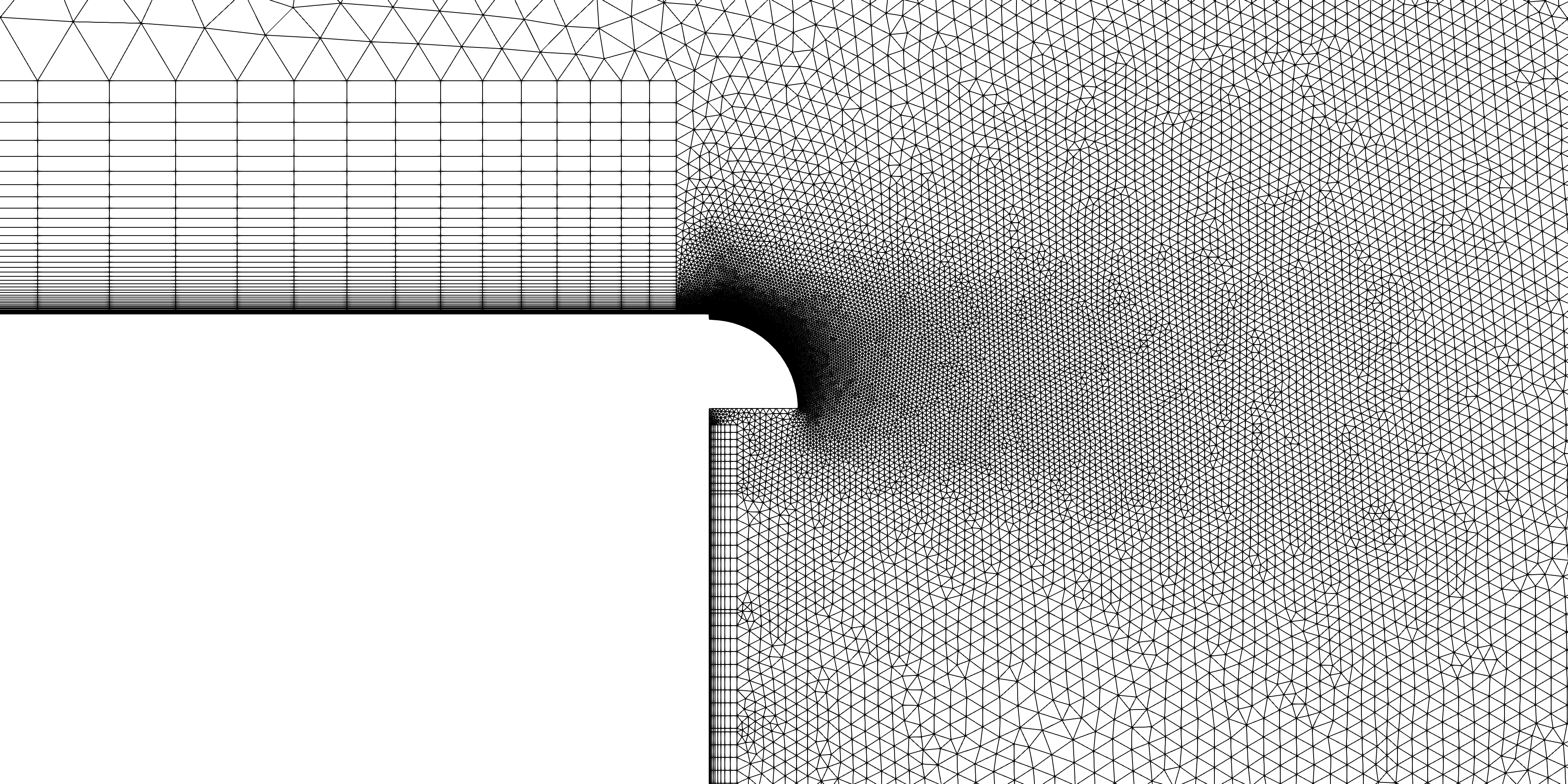}
	\caption{Mesh with 1.48e5 elements: Curved edge detail.}
	\label{fig:mesh_detail_curved}
\end{figure*}

The mesh is hybrid: it is made of a structured region adjacent to the body, and an unstructured region elsewhere (Fig.~\ref{fig:mesh_colormap}-\ref{fig:mesh_detail_bb}). The structured region contains $40$ layers with a growth ratio of $1.1$. This ensures to well resolve the boundary layer and the eddy viscosity creation in the front part.
At the back, the number of layers is reduced to $20$, because the triangular mesh is already fine enough and there is no advantage in increasing the number of layers.
Along the curved edge (Fig.~\ref{fig:mesh_detail_curved}), there are no structured layers since the mesh is sufficiently fine to capture the interaction between the jet and the flow field. This implies that $y^+$ reaches a small enough value by using triangular elements. Moreover, if rectangular elements had been used, the high diagonal flow crossing the elements would have compromised the convergence speed of the resolution. 

\subsection{\label{subsec:grid_time_conv}Grid and time convergence}
The grid and time convergence (or independence) analysis is an important assessment to evaluate the characteristic size of a grid element and the size of the time step. It is carried out as follows: a coarse mesh and a large time step are initially considered and they are subsequently refined. This process continues until both grid and time convergence is achieved. Table~\ref{tab:time_grid_conv_bb} shows the drag coefficient resulting from simulations with different grids and time steps. The table should be read from the top-left corner towards the bottom-right one where the grid element characteristic size and time step decrease and, conversely, the total number of the mesh elements increases. We can observe that the convergence for both parameters is reached with a time step $\Delta t = $ 2.5e-4 and a mesh of $9.44e5$ elements. However, the computational complexity of the CFD simulation with the parameters obtained by the convergence analysis is incompatible with the reinforcement learning training. For this reason, we consider a mesh with 1.48e5 elements and a time step of 4e-3, adequate for the training. 

\begin{table}[th!]
	\caption{Time and grid convergence for the bluff body. }
	\label{tab:time_grid_conv_bb}
	\centering
	%\begin{ruledtabular}
		\begin{tabular}{ccccc}
			%\hline
			\# elements & 3.9e4 & 1.48e5 & 5.81e5 & 9.44e5 \\		
			\hline
			% \multicolumn{5}{l}{$\Delta t$ = 8e-3}\\ [1ex]
			%\hline
			%$\bm{C_D}$
			$\Delta t$ = 8e-3 & 0.9145 & - & - & -\\
			%\hline
			
			%\multicolumn{5}{l}{$\Delta t$ = 4e-3}\\ [1ex]
			%\hline
			% $\bm{C_D}$ 
			$\Delta t$ = 4e-3 & 0.9219  & 0.9749 & - & - \\
			%\hline
			
			%\multicolumn{5}{l}{$\Delta t$ = 2e-3}\\ [1ex]
			%\hline
			$\Delta t$ = 2e-3 & 0.9242 & 0.9785  & - & - \\
			%\hline
			
			%\multicolumn{5}{l}{$\Delta t$ = 1e-3}\\ [1ex]
			%\hline
			$\Delta t$ = 1e-3 & - & 0.9802 & 1.035 & - \\
			%\hline
			
			%\multicolumn{5}{l}{$\Delta t$ = 5e-4}\\ [1ex]
			%\hline
			$\Delta t$ = 5e-4 & - & 0.9811 & 1.038  & 1.045 \\
			%\hline
			
			%\multicolumn{5}{l}{$\Delta t$ = 2.5e-4}\\ [1ex]
			%\hline
			$\Delta t$ = 2.5e-4 & - & - & - &1.045 \\
			%\hline
			
		\end{tabular}
	%\end{ruledtabular}
	
\end{table}
%

% =====================================================================
% REINFORCEMENT LEARNING 
% =====================================================================
\section{\label{sec:rl}Reinforcement Learning} 
\begin{figure}[th]
	\centering
	\includegraphics[width=\linewidth]{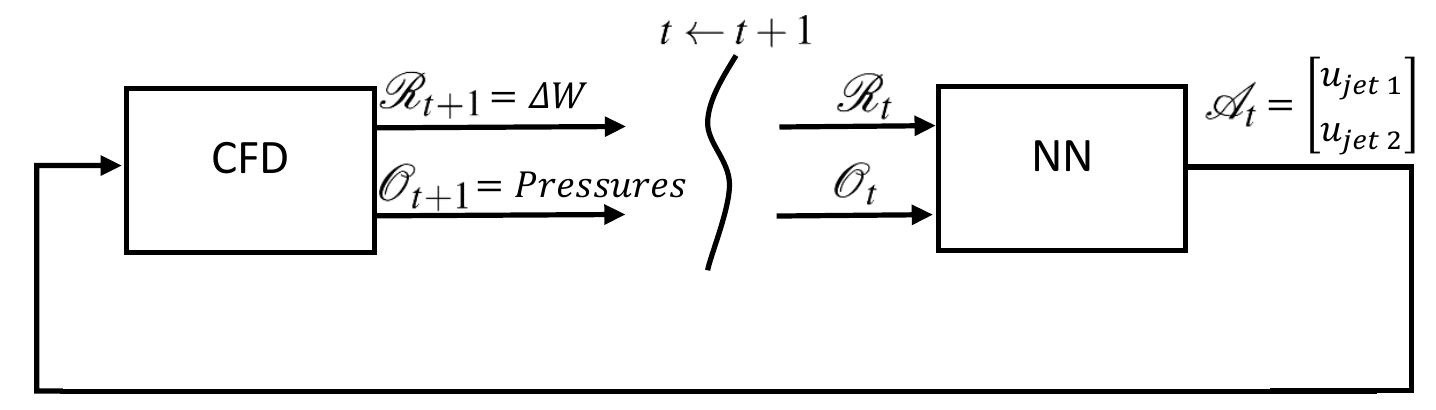}
	\caption{Agent-environment interaction. The left block represents the environment, the right one the agent. The environment is perceived through its observations, $\mathcal{O}_{t}$, that are measurements of the pressure on the rear surface at each time step. The reward is set equal to the power saving, $\Delta W$. The action, $\mathcal{A}_t$, is a vector whose components are the instantaneous velocities of the jets.}
	\label{fig:markov_2}
\end{figure}
\begin{figure}[htp]
	\centering
	\includegraphics[width=0.5\linewidth]{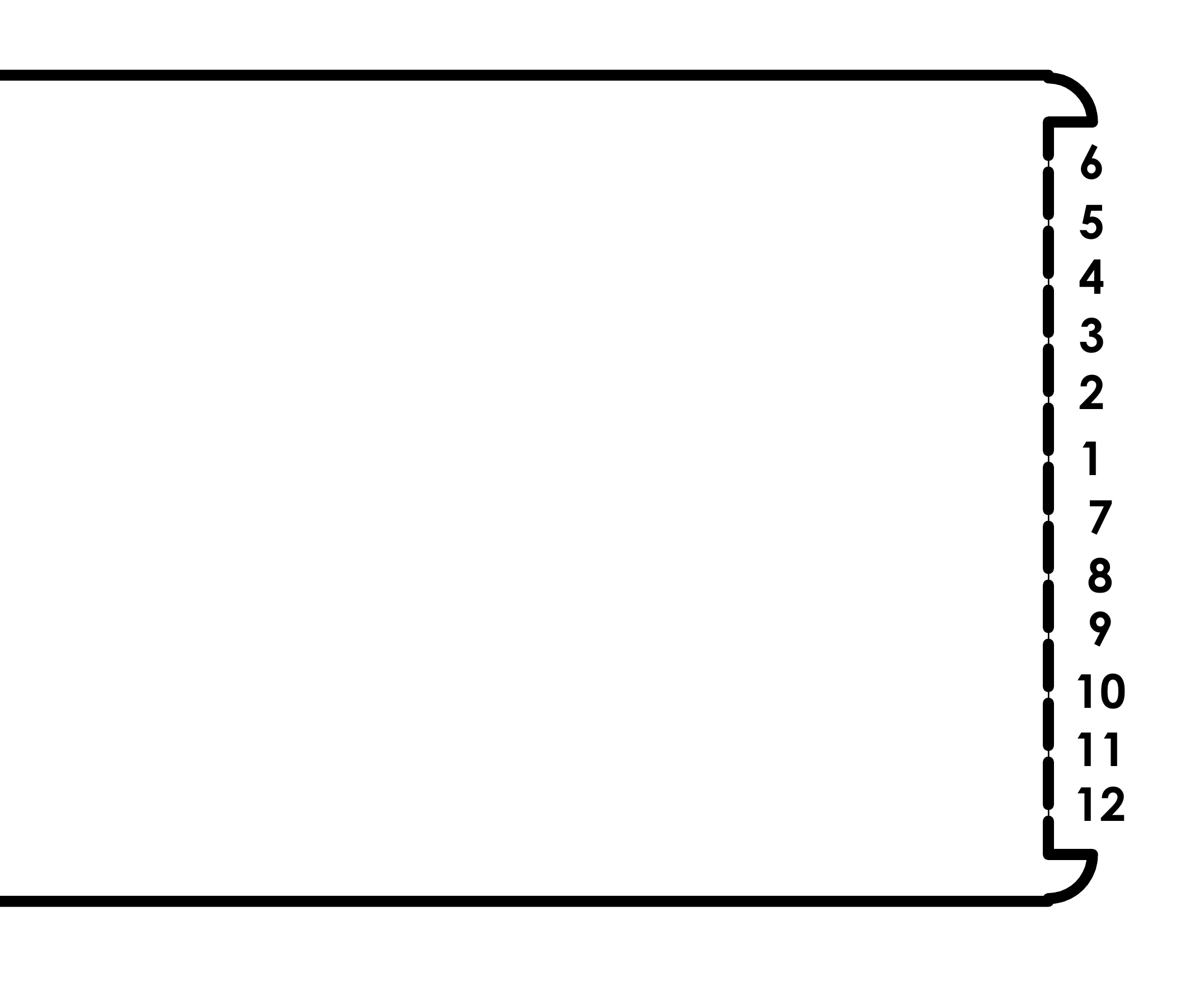}
	\caption{Position and numbering of pressure taps.}
	\label{fig:ptaps}
\end{figure}

We model the control of the flow rate of the nozzles as a partially observable Markov decision process $\mathcal{M} = \langle \mathcal{S}, \mathcal{A}, \mathcal{O}, \mathcal{T}, \mathcal{R}, \gamma \rangle$. Fig.~\ref{fig:markov_2} schematizes the elements of the Markov decision process specified to our problem, which are now described. Of the full state $\mathcal{S}$, corresponding to the whole fluid dynamics simulation, the agent has access to the observation $\mathcal{O}$, which consists of $12$ pressure measurements at the back of the body (Fig.~\ref{fig:ptaps}). Each pressure measurement is averaged over the time between two iterations, to filter out the high-frequency fluctuations. The pressure measurements persist in the observed state for $2$ consecutive iterations, so that the agent has access to a the current and the previous observation (for a total of $24$ values). This gives the agent access to an approximation of the first derivative of the pressure, necessary to understand whether a vortex is in the formation or removal phase. The transition dynamics $\mathcal{T} = P(s_{t+1}\vert s_t, a_t)$ is given by the evolution of the fluid dynamics simulation for $n = 50$ simulation time steps. % CFD-time step
%As a result, a single RL-time step contains many CFD-time steps. 
The value $n$ corresponds to approximately $1/5$ unit of time. The fluid dynamics phenomena show an oscillating behaviour. Therefore, we can consider the lift coefficient, $CL$, and defining its period, called $C_L$-period. The $CL$-period is about $4$ units of time, so there are roughly $20$ % actions (and rewards)
transitions per $C_L$-period, thus there are 20 actions and rewards per $C_L$-period.
The value $n$ is chosen as a compromise between temporal resolution and environment adaptation time. A higher value would imply weak time resolution, thus a coarser control. On the other hand, lower values would not be adequate, because the global phenomena, and thus the reward, are quite insensitive to high-frequency actions, making the credit assignment problem more difficult. The action $\mathcal{A} = (u_{jet_1}, u_{jet_2}) \in [0,4]\times[0,4]$ is the velocity of the fluid emitted by the nozzle for the following transition. The parameter $\gamma$ is the discount factor.
The reward $\mathcal{R}$ is the net power saving, namely the difference between the power of the drag of clean bluff body (see Section~\ref{sec:fluid_dynamics_model}), $W^{clean}$, and the power of the drag of bluff body with controlled jets, $W^{C_D+jets}$:
\begin{equation}\label{eq:reward}
	\mathcal{R}_{t+1} = W^{clean} - W_{t}^{C_D+jets} = W^{clean} - W_{t}^{C_D} - W_{t}^{jets}, 
\end{equation}
with:
\begin{equation}\label{eq:W_CD}
	W_{t}^{C_D} = drag \cdot U_{\infty} = \frac{1}{2}\rho U_{\infty}^2 \left( \frac{1}{\Delta T}\int_{t}^{t+\Delta T}C_D \right) S U_{\infty},
\end{equation}
\begin{equation}\label{eq:W_clean}
	W^{clean} = drag_{clean} \cdot U_{\infty} = \frac{1}{2}\rho U_{\infty}^2 C_{D}^{clean} S U_{\infty},
\end{equation}
\begin{equation}\label{eq:W_jet}
	W_{t}^{jets} = \sum_{i = 1,2} \frac{1}{\Delta T}\int_{t}^{t+\Delta T} \int_{slit} \left( \frac{1}{2} \rho |\bm{u}_{jet_i}|^2 + P_i \right) \bm{u}_{jet_i} \cdot \hat{n}, 
\end{equation}
where $S$ is the surface of the section of the bluff body. $C_D$ consists of the pressure and viscous contributions, and the small thrust due to the mass flow exiting the body. For simplicity, we neglect all possible efficiency factors in the power computation.
The pressure, $P_i$, in Eq.~(\ref{eq:W_jet}) is a gauge pressure referred to the pressure at infinity.
The time average is made over one transition, so $\Delta T$ corresponds to the previously mentioned $n$ time steps.

\subsection{\label{chap:experiment_solution_and_algorithm_settings}Experiments solution and algorithm settings} 

We use the Proximal Policy Optimization (PPO) algorithm \cite{schulman_2017}. The main hyperparameters are shown in Appendix~\ref{app:rl_settings}.\\
The actor policy and the critic are represented by two separate fully-connected neural networks, with $2$ hidden layers of $30$ units each. The policy network outputs the $2$ action values, corresponding to the flow rate of the two jets, while the critic network outputs a single scalar, corresponding to the expected cumulative reward associated with the current state. The network size is selected in accordance to the theorem in \cite{lu_2017, park_2020}. 
The total number of parameters is $1804$ (actor) + $1711$ (critic) for a total equal to $3515$. The activation function, $\sigma$, is the Rectified Linear Unit (ReLU) function.
Due to the computational complexity of the CFD simulation and the considerable number of time steps required to train a policy with reinforcement learning, the tuning of the hyperparameters of PPO and of the network architecture is beyond the scope of this work.

To challenge the computational complexity of the combined CFD simulation and reinforcement learning, the training phase is carried out in two steps. In the first step, the CFD simulation is run on a coarse mesh. This approximate simulation retains most of the properties of the complete one, while remaining computationally inexpensive. Indeed, by using a mesh with 2e4 elements, we reduce the computation time by 92\%, at the cost of an error less than 10\% on the computation of the drag. As the control policy of the nozzles is initialized with random weights, it needs many interactions with the environment before finding an adequate control strategy. %This approximate simulation retains most of the properties of the complete one, such as the periodic formation of vortexes (error on strouhal?) and (other features). 
Once the performance of the agent saturates in this approximate environment, the second step consists in fine-tuning it in the CFD simulation with the full mesh. 
%As the two environments are similar as far the quantities of interest are concerned (drag, strouhal, ..?), we expect few fine-tuning episodes to be necessary

% ================================================
% RESULTS ========================================
% ================================================
\section{\label{sec:results}Results} 
We now show the main results of the whole simulation process.
\begin{figure}[htb]
	\begin{subfigure}{0.5\linewidth}
		\begin{subfigure}[t]{1\linewidth}
			\textbf{(a)}
		\end{subfigure}
		%\centering
		\includegraphics[width=\linewidth]{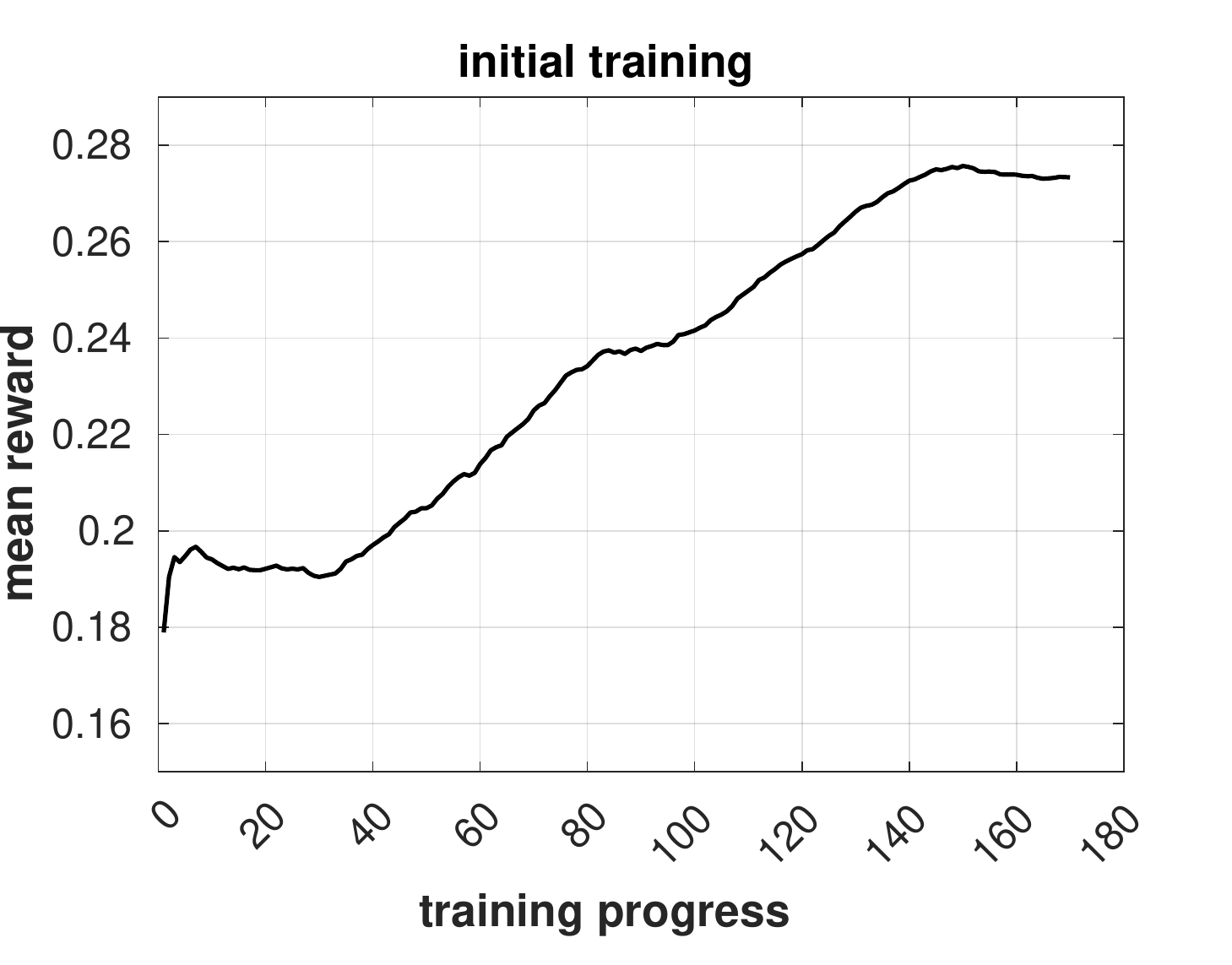}
	\end{subfigure}
	\begin{subfigure}{0.5\linewidth}
		\begin{subfigure}[t]{1\linewidth}
			\textbf{(b)}
		\end{subfigure}
		%\centering
		\includegraphics[width=\linewidth]{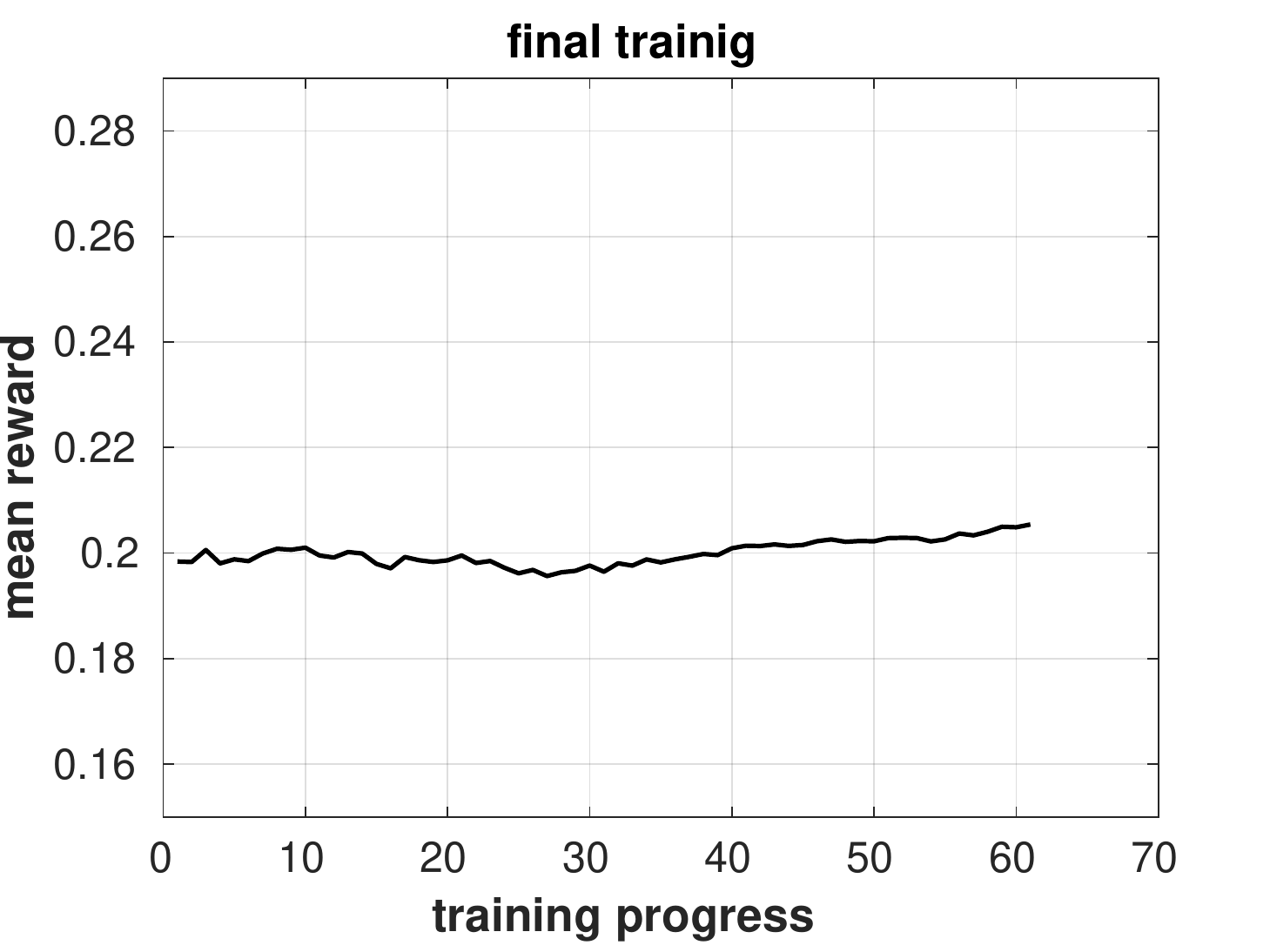}
	\end{subfigure}
	\caption{Training progress. \textbf{(a)} Initial training with a coarse mesh used to quickly adapt the neural network to the environment. The graph represents the mean rewards averaged over the $N$ workers and over all the previous episodes versus the training progress.
		\textbf{(b)} Final training with a finer mesh. The gap between the mean rewards of the two phases is due to an improvement of the CFD accuracy, so a better reward calculation.}
	\label{fig:rewards}
\end{figure}
% 1 training progress corresponds to 1 episode = 4*160/4 time_iter_rl. but the episodes are dummies since the environment is not reinitialized 

%
\subsection{Policy learning}
The graphs in Fig.~\ref{fig:rewards} show the learning curve of the reinforcement learning agent in the two training steps described in Section~\ref{chap:experiment_solution_and_algorithm_settings}. We can observe that, in the first training step, the agent starts the power consumption of the baseline, that is the clean bluff body represented in Fig.~\ref{fig:bluff_bodies}, and by accumulating experience it largely outperforms it. After approximately $170$ training iterations, corresponding to $24$ hours of CFD simulation, the reward reaches a plateau.

%Fig.~\ref{fig:rewards} shows the training progress. On the vertical axis, there are the mean rewards averaged over the $N$ workers and over all the previous episodes. The left panel regards the initial training phase. It shows an almost constant trend except for an initial uncertainty and a final stabilization to the maximum reward.  
% The panel \textbf{(b)} regards the final training where the neural network of the initial training is applied to the environment having the 1.48e5 elements mesh. It has a flatter trend: the difference between the maximum and the minimum is less than $5\%$ of the mean value. Despite the weakly and constantly increasing trend, which may indicate an unfinished training, the process was interrupted because of an excessive time request. 

The plot of the learning curve in the second step of the training, performed on the finer mesh, shows that the impact of the fine tuning on the drag reduction is rather limited. While after approximately $60$ training iteration, that corresponds approximately to $300$ hours of simulations, the drag is further reduced by only 5\%, so most of the power saving was already achieved during the first training step.
%have not to be compared with the maximum mean reward of the initial training because the mesh is so coarse that the CFD simulation is not accurate enough to correctly estimate $C_D$ and, consequently, the power and, in turn, the rewards are roughly computed.
We highlight the fact that the mean rewards of the two training steps are different because of the different simulation environments. In fact, as per our convergence study on the mesh, the drag, and consequently the power of the drag force, computed in the coarse mesh is lower than the drag of the finer mesh. 
We recall that the initial training has the only aim to quickly adapt the neural network to the environment, saving a large amount of computation time. %The gap is due to a better estimation of the $C_D$ that the coarse mesh underestimates. 

The marginal drag reduction due to the final training phase indicates that the neural network obtained after the initial training can be successfully used without any adaptation, boasting similar performance to the adapted network. Already after the first training phase the two jets can interact with the vortexes and obtain a consistent drag reduction even though the pressures on the rear surface for which it was trained are coarsely simulated, suggesting a good transfer learning property.
\begin{figure}[hbt]
	\centering
	\includegraphics[width=\linewidth]{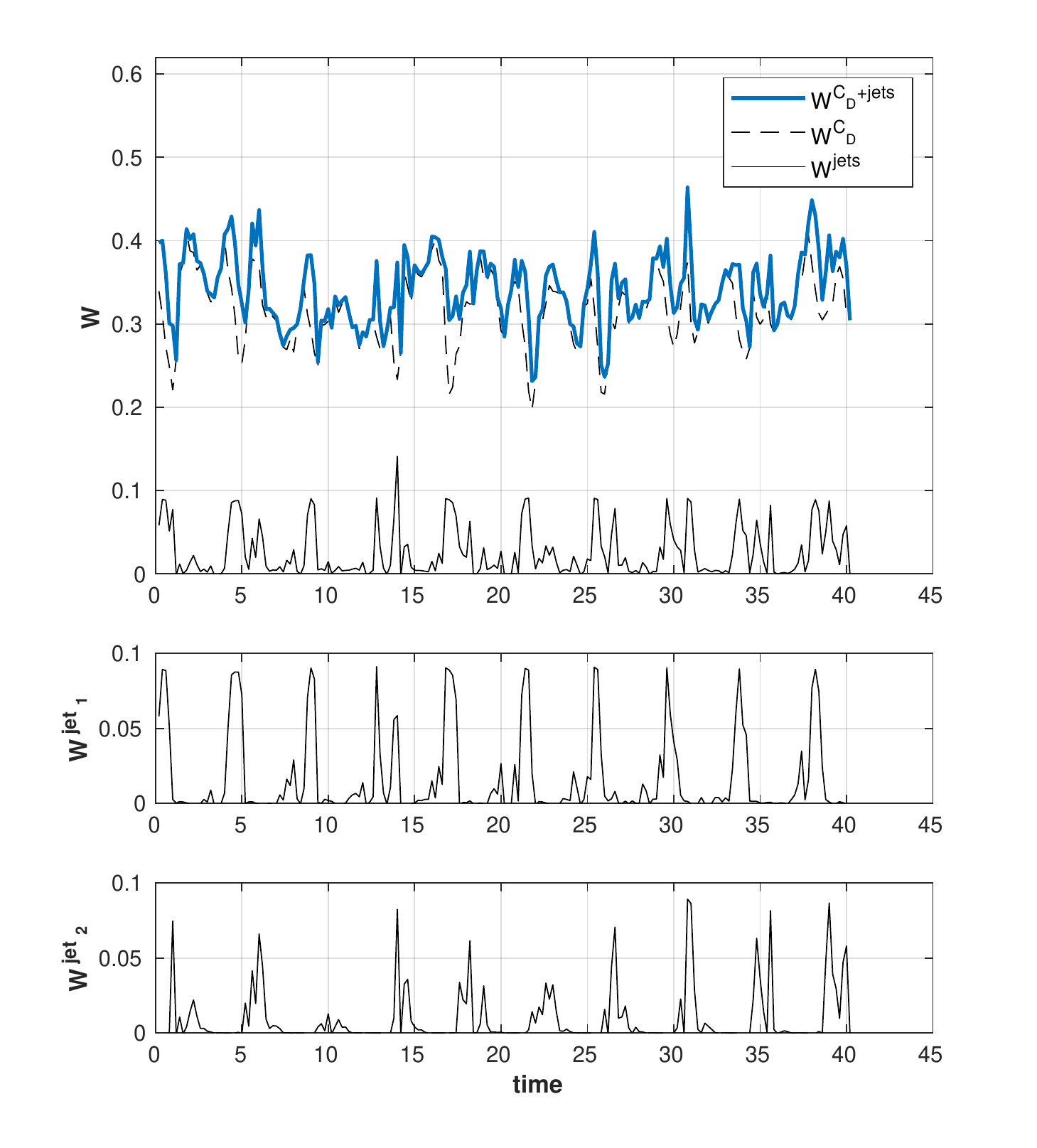}
	\caption{Upper panel: powers $W^{C_D+jets} = W^{C_D} + W^{jets}$, $W^{C_D}$, of the clean bluff body, and $W^{jets} = W^{jet_1}+W^{jet_2}$. Lower panels: powers $W^{jet_1}, W^{jet_2}$, of the upper and lower jet, respectively. The time laps is about $10$ $C_L$-periods.}
	\label{fig:powers}
\end{figure}
The trained neural network is then applied to a simulation to study the learned control for the jets and to evaluate the power saving. The instantaneous powers $W^{C_D}$, $W^{jet_1}$, $W^{jet_2}$, and $W^{C_D+jets} $ are reported in Fig.~\ref{fig:powers}. With $jet_1$ and $jet_2$, we denote the upper and the lower jet, respectively, and with the word $jets$ we indicate the sum of the two jets, as in Eq.~(\ref{eq:W_jet}). Table~\ref{tab:power_savings_comparison} contains the  averaged powers, evaluated from the aforementioned simulation, compared to the powers coming from a reference control with constant jets flow rate and velocities. The total power of the bluff body with controlled jets is about $40 \%$ less than the clean bluff body power and almost $20 \%$ less than the constant jet case.
\begin{table}[h!]
	\caption{Power saving and comparison with a constant-velocity jet case. The first row having $u_{jets} = 0$ corresponds to the clean bluff body. The power consumption of the controlled jet is comparable to the one of the constant-velocity jet equal to $2$, but the $C_D$ reduction is much higher, causing a net power saving of 40\%.}
	\label{tab:power_savings_comparison}
	\centering
	%\begin{ruledtabular}
		\begin{tabular}{cccccc}
			$u_{jets}$ & $C_D$ & $W^{C_D}$ & $W^{jets}$ & $W^{C_D+jets}$ & net\ saving \% \\
			\hline
			0 & 1.18  & 0.589 & 0 & 0.589 & 0 \\
			1 & 1.15 & 0.577 & 0.001 & 0.578 & 2 \\
			2 & 1.01 & 0.506 & 0.020 & 0.526 & 11 \\
			3 & 0.83 & 0.414 & 0.076 & 0.490 & 17 \\
			4 & 0.57 & 0.283 & 0.183 & 0.466 & 21 \\
			5 & 0.42 & 0.209 & 0.364 & 0.573 & 3 \\
			controlled & 0.67 & 0.323 & 0.028 & 0.356 & $40^{+11}_{-12}$ \\ 
		\end{tabular}
	%\end{ruledtabular}
\end{table}
%

% \clearpage %

\begin{figure}[hp!] % press vortex 
	\begin{subfigure}{0.5\textwidth}
		\begin{subfigure}[t]{1\textwidth}
			\textbf{(a)}
			\vspace{0.1cm}
		\end{subfigure}
		%\centering
		\includegraphics[width=\linewidth]{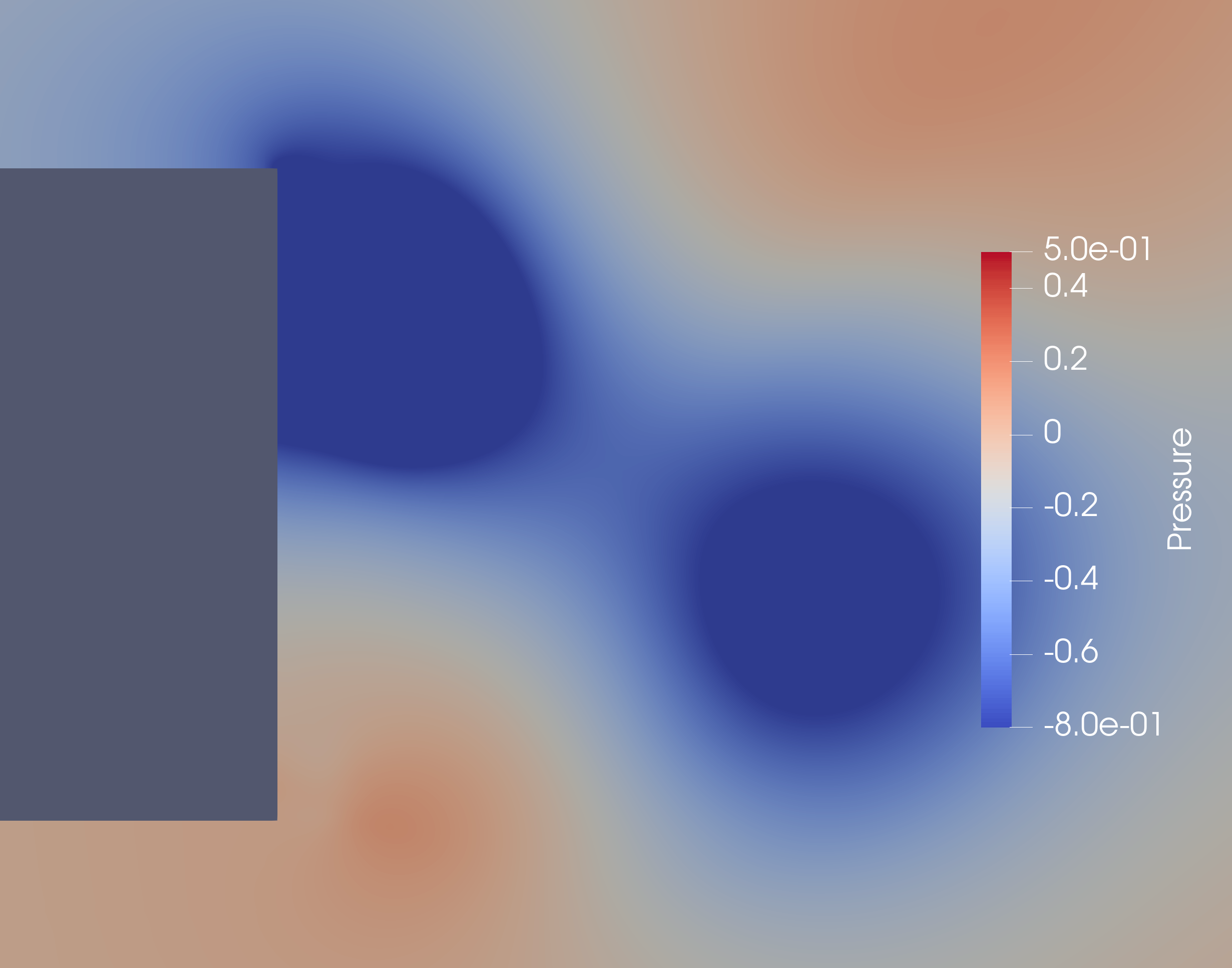}
	\end{subfigure}
	\begin{subfigure}{0.5\textwidth}
		\begin{subfigure}[t]{1\textwidth}
			\textbf{(b)}
			\vspace{0.1cm}
		\end{subfigure}
		%\centering
		\includegraphics[width=\linewidth]{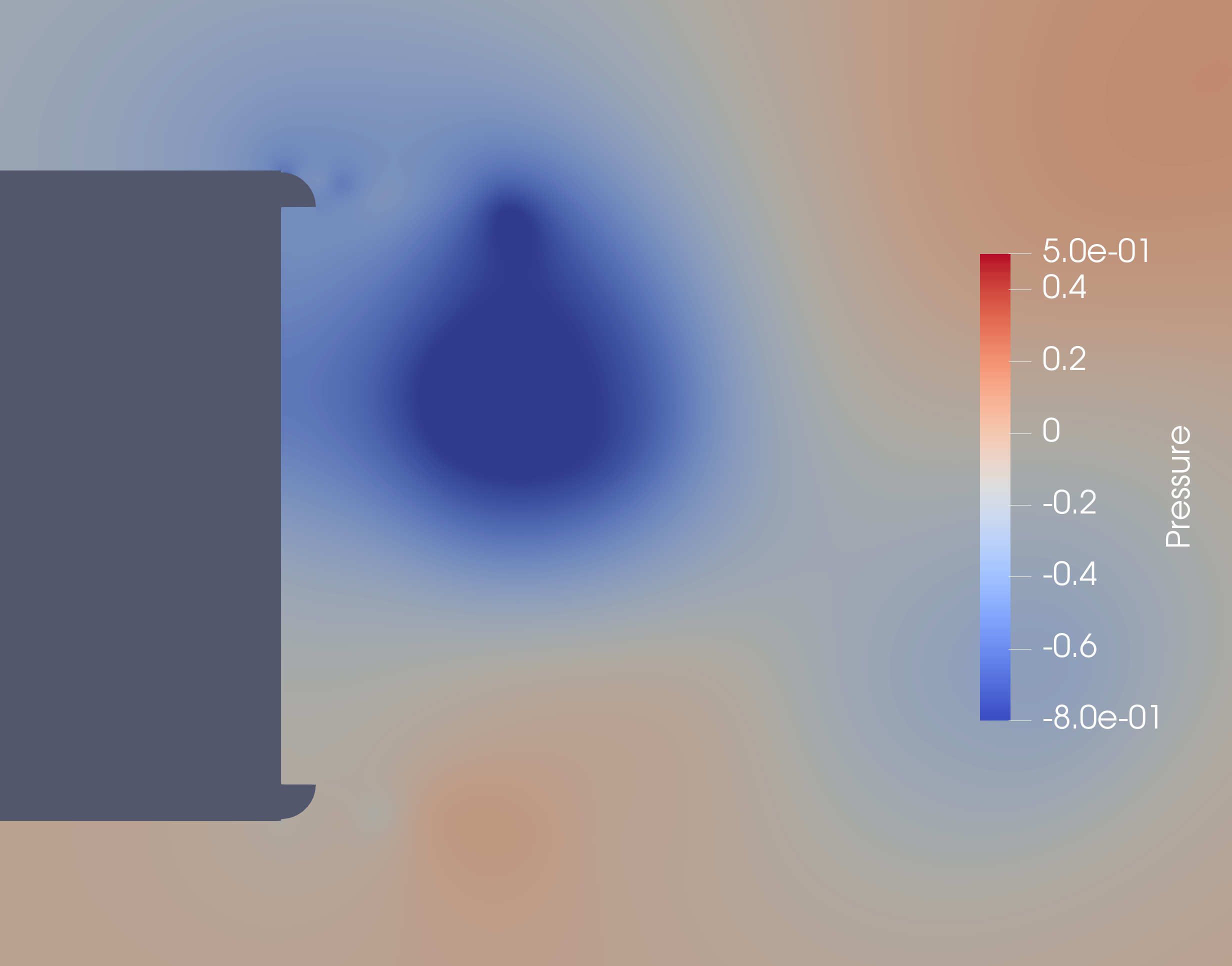}
		%\caption{unit}
		%\label{fig:unit}
	\end{subfigure}
	\caption{Snapshots of pressure distribution behind the bluff bodies. \textbf{(a)} Bluff body clean. The vortex is created close to the wall generating a notable low-pressure region.  \textbf{(b)} Bluff body with controlled jets. The vortex is created far from the wall.}
	\label{fig:p_vortex}
\end{figure}

\begin{figure*}[h] 
	\includegraphics[width=.32\textwidth]{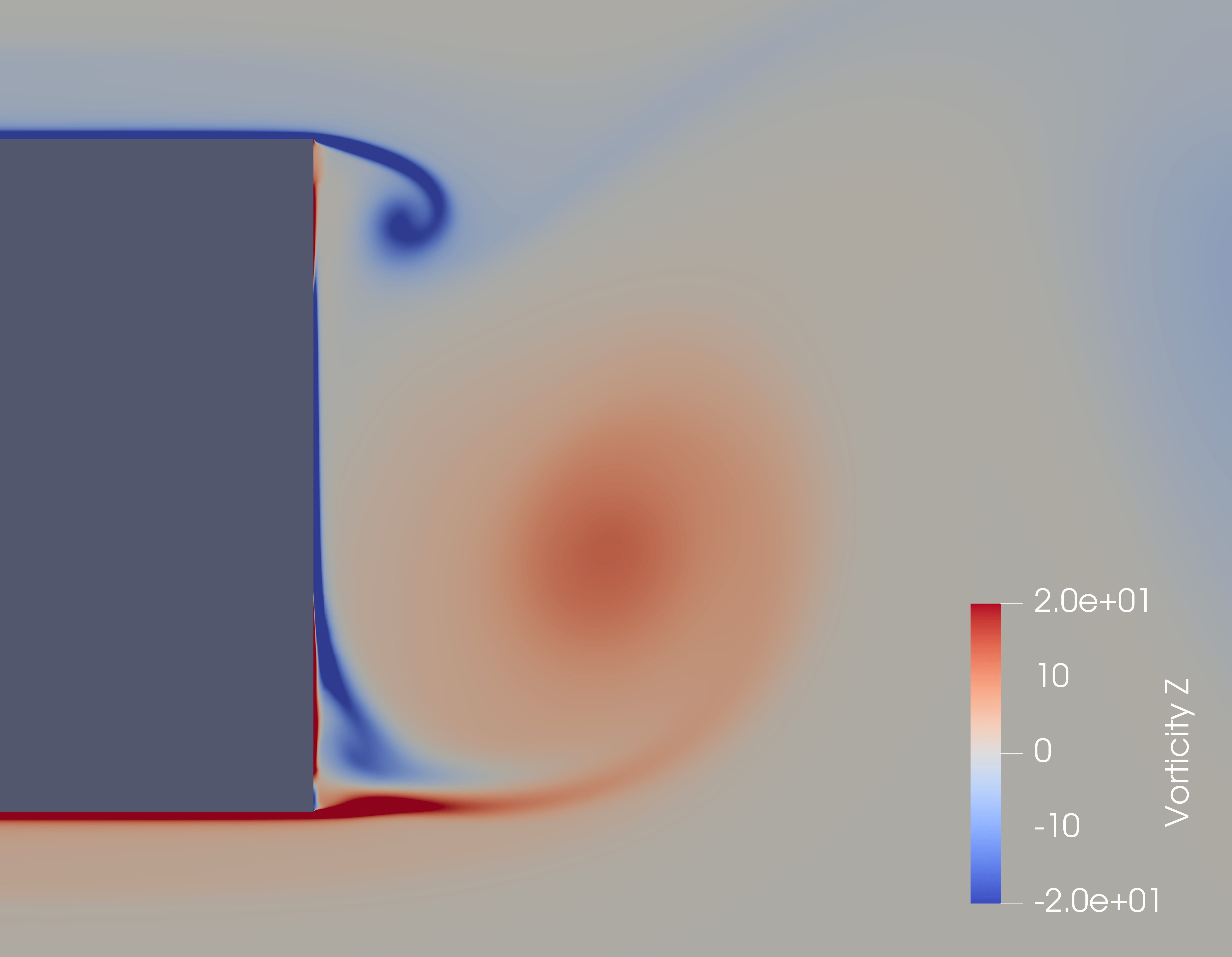}\hfill
	\includegraphics[width=.32\textwidth]{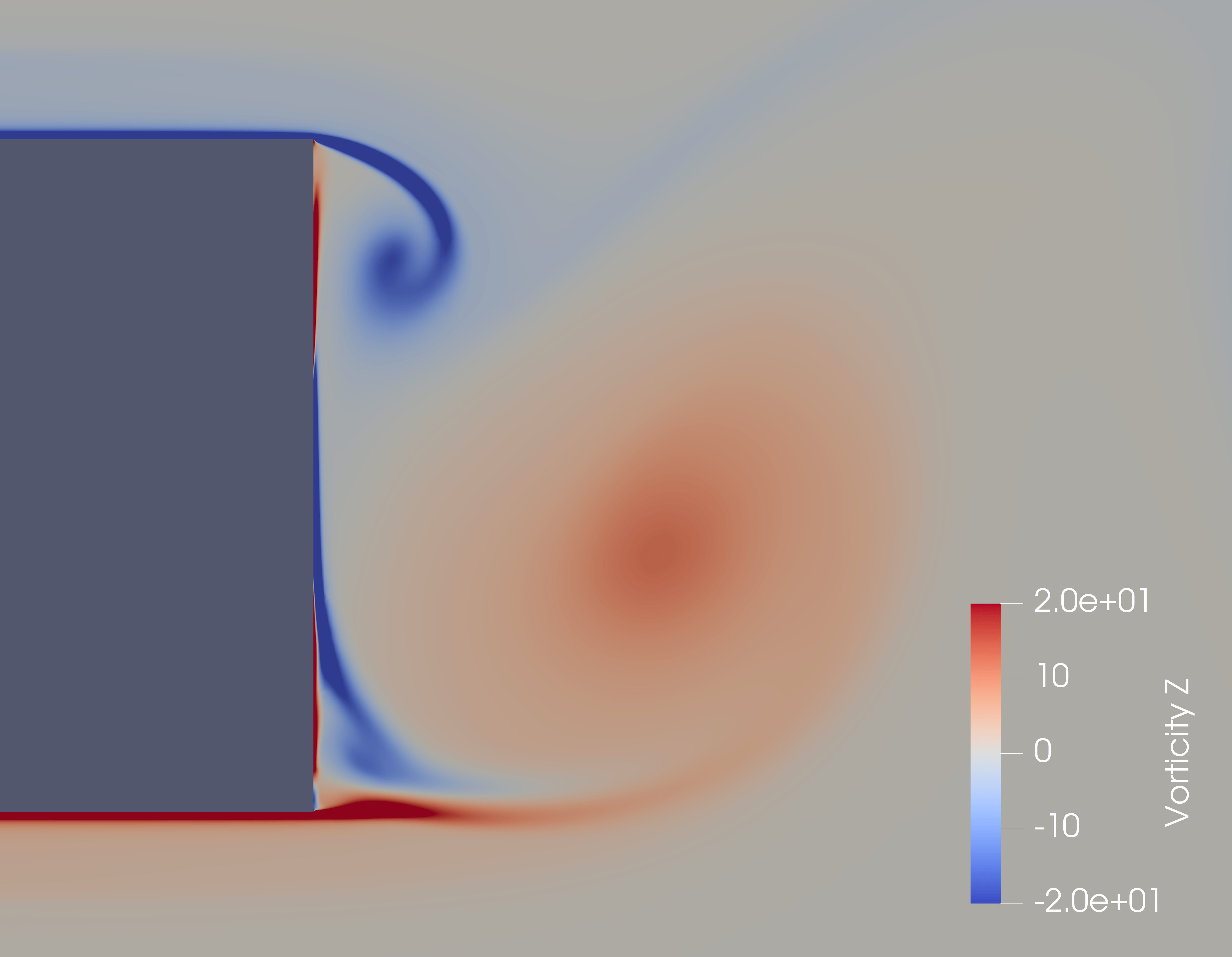}\hfill
	\includegraphics[width=.32\textwidth]{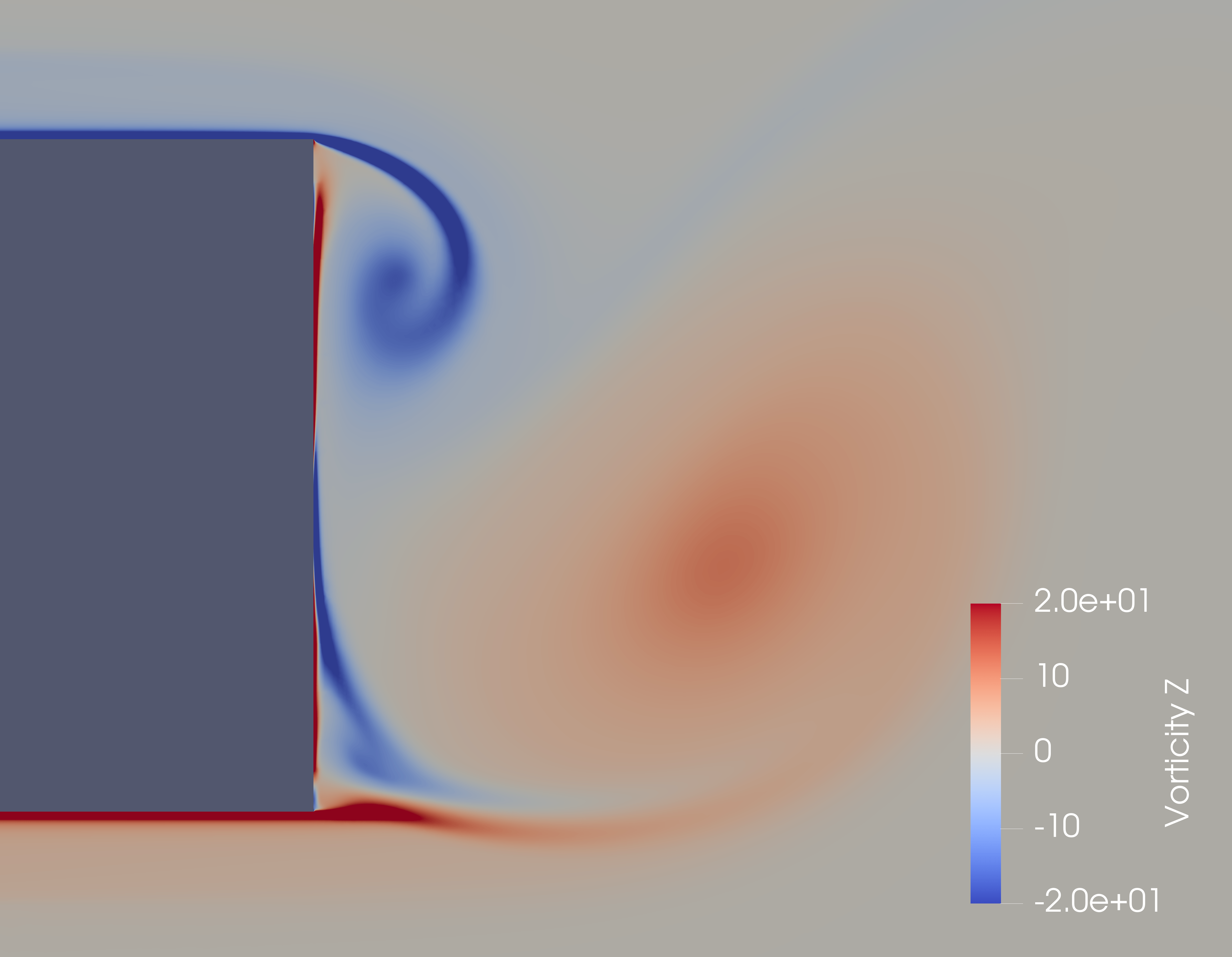}\hfill
	\includegraphics[width=.32\textwidth]{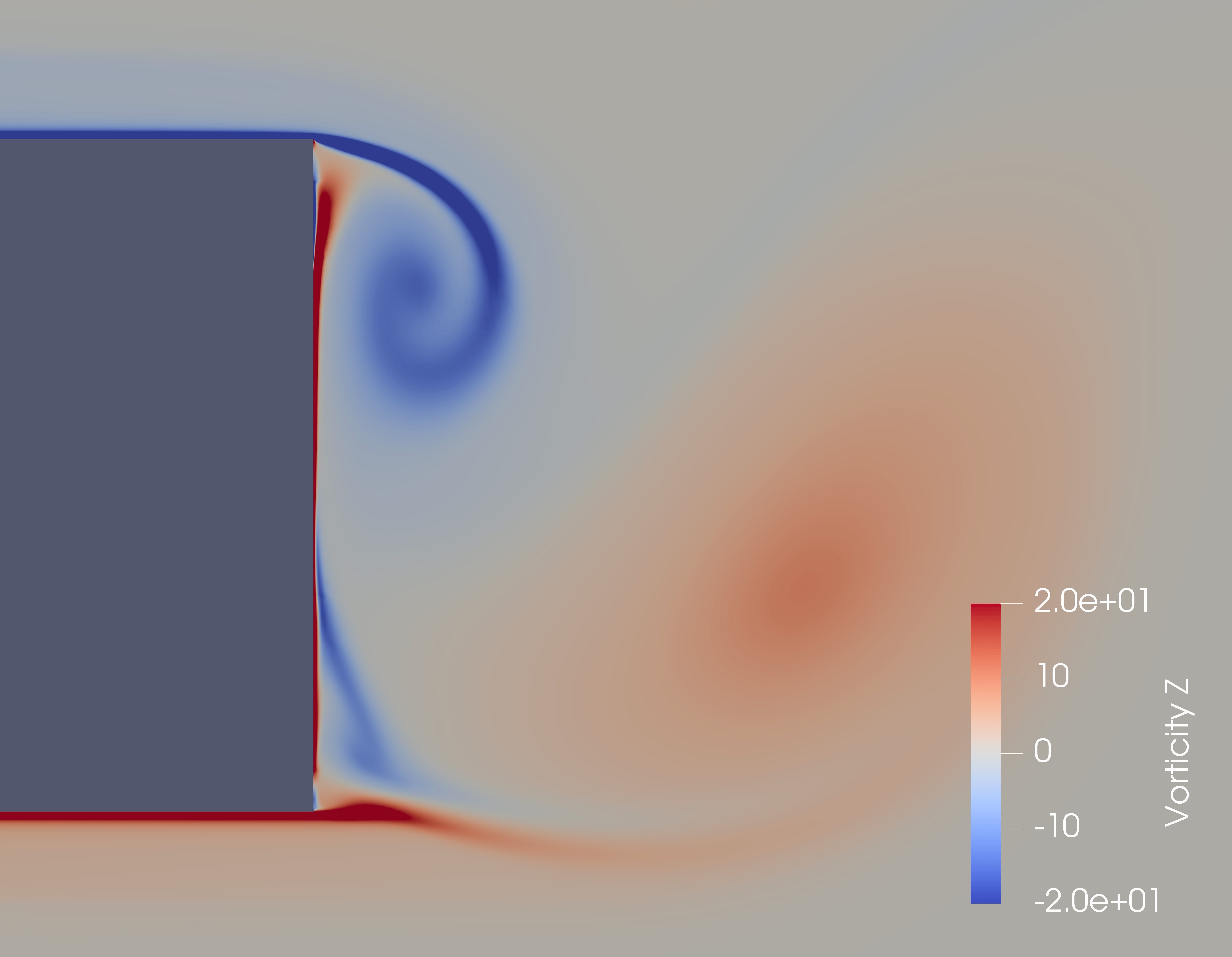}\hfill
	\includegraphics[width=.32\textwidth]{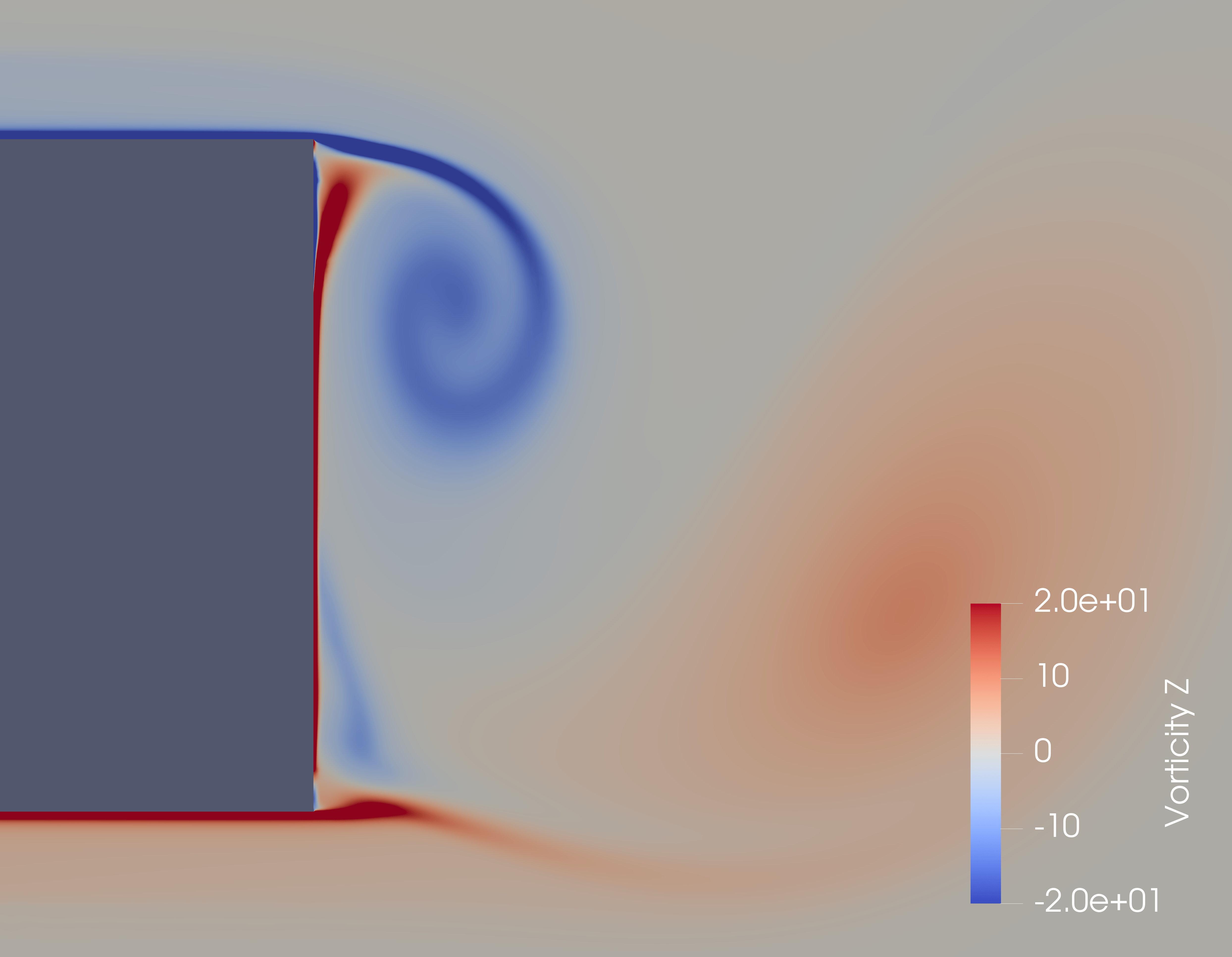}\hfill
	\includegraphics[width=.32\textwidth]{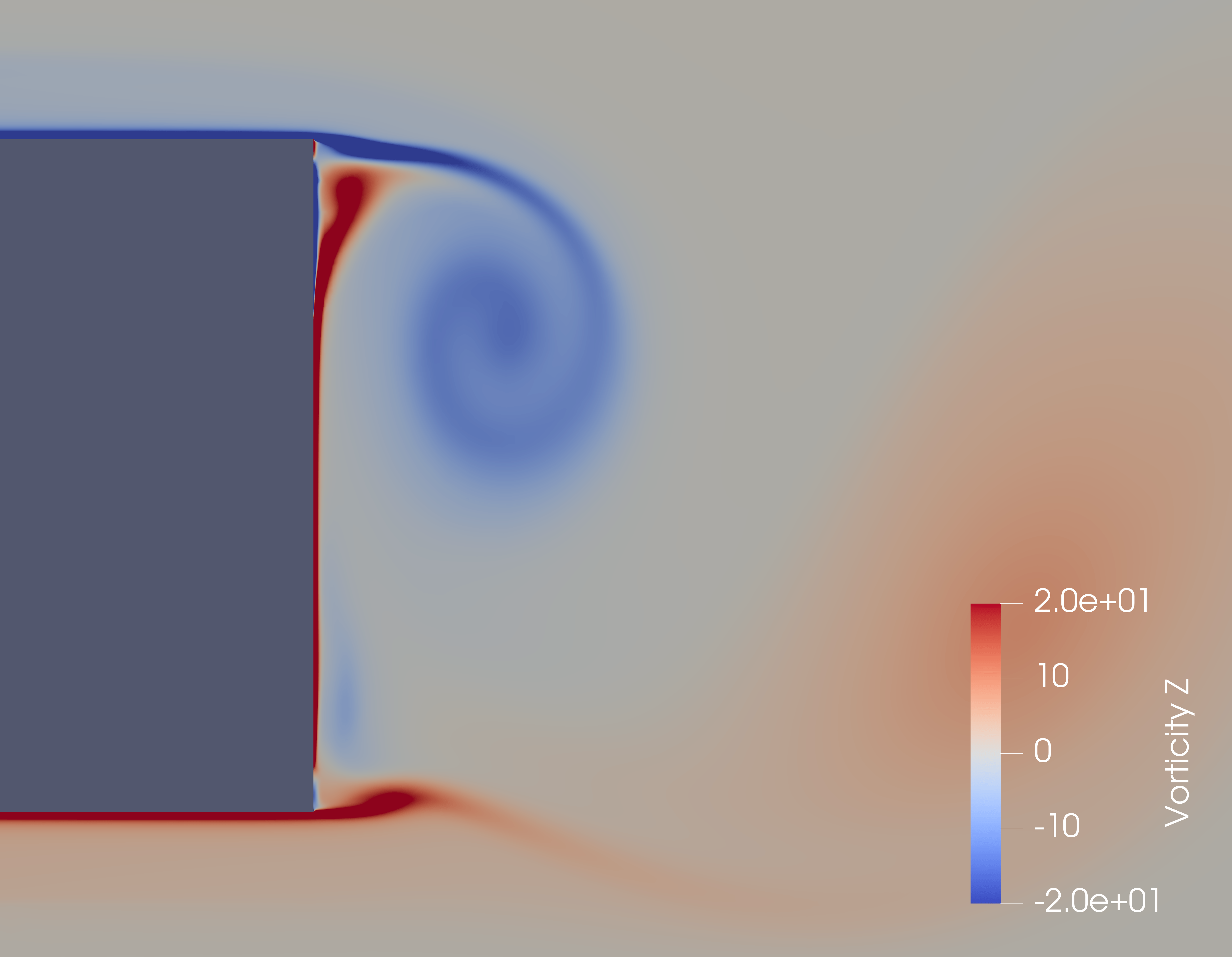}\hfill
	\includegraphics[width=.32\textwidth]{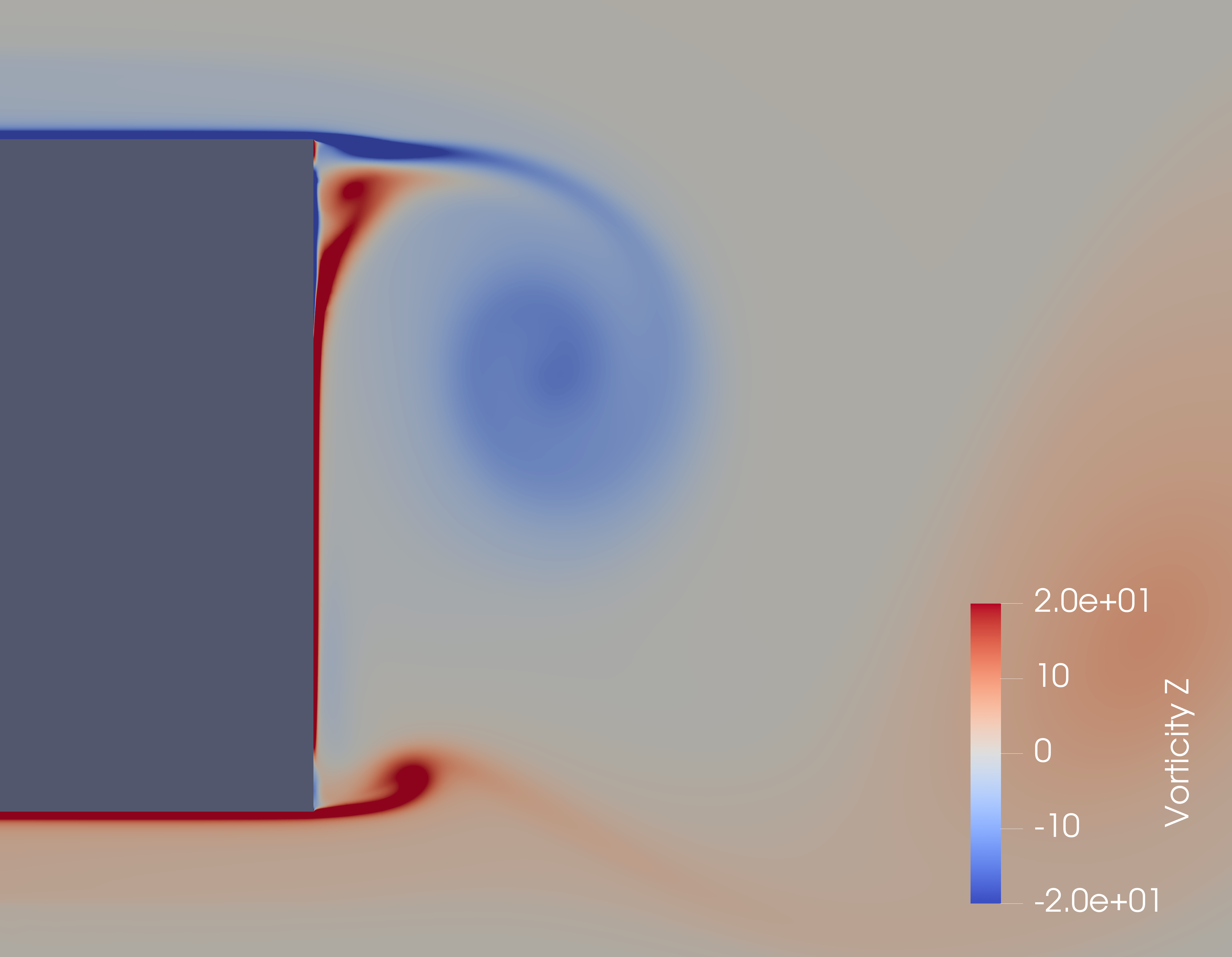}\hfill
	\includegraphics[width=.32\textwidth]{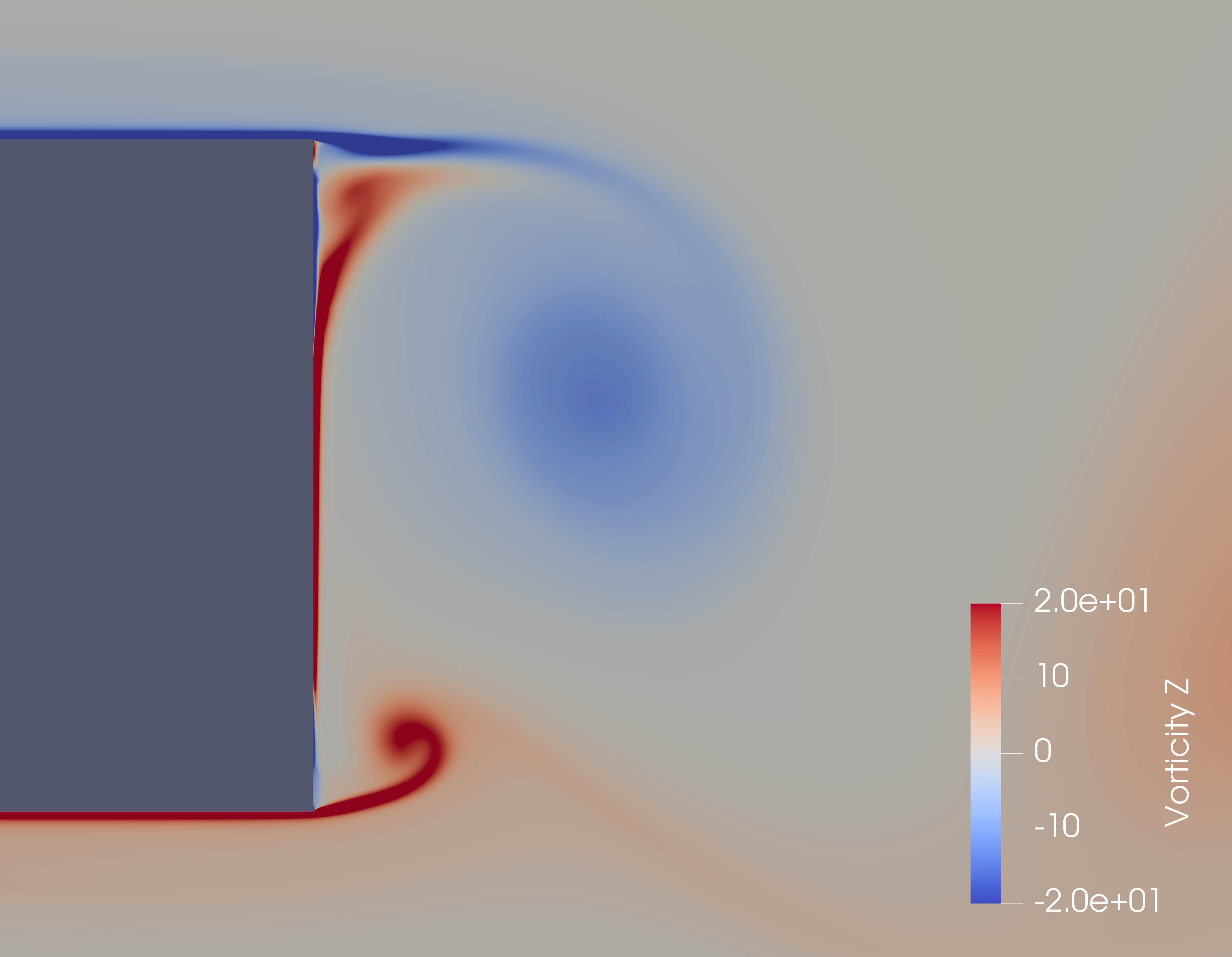}\hfill
	\includegraphics[width=.32\textwidth]{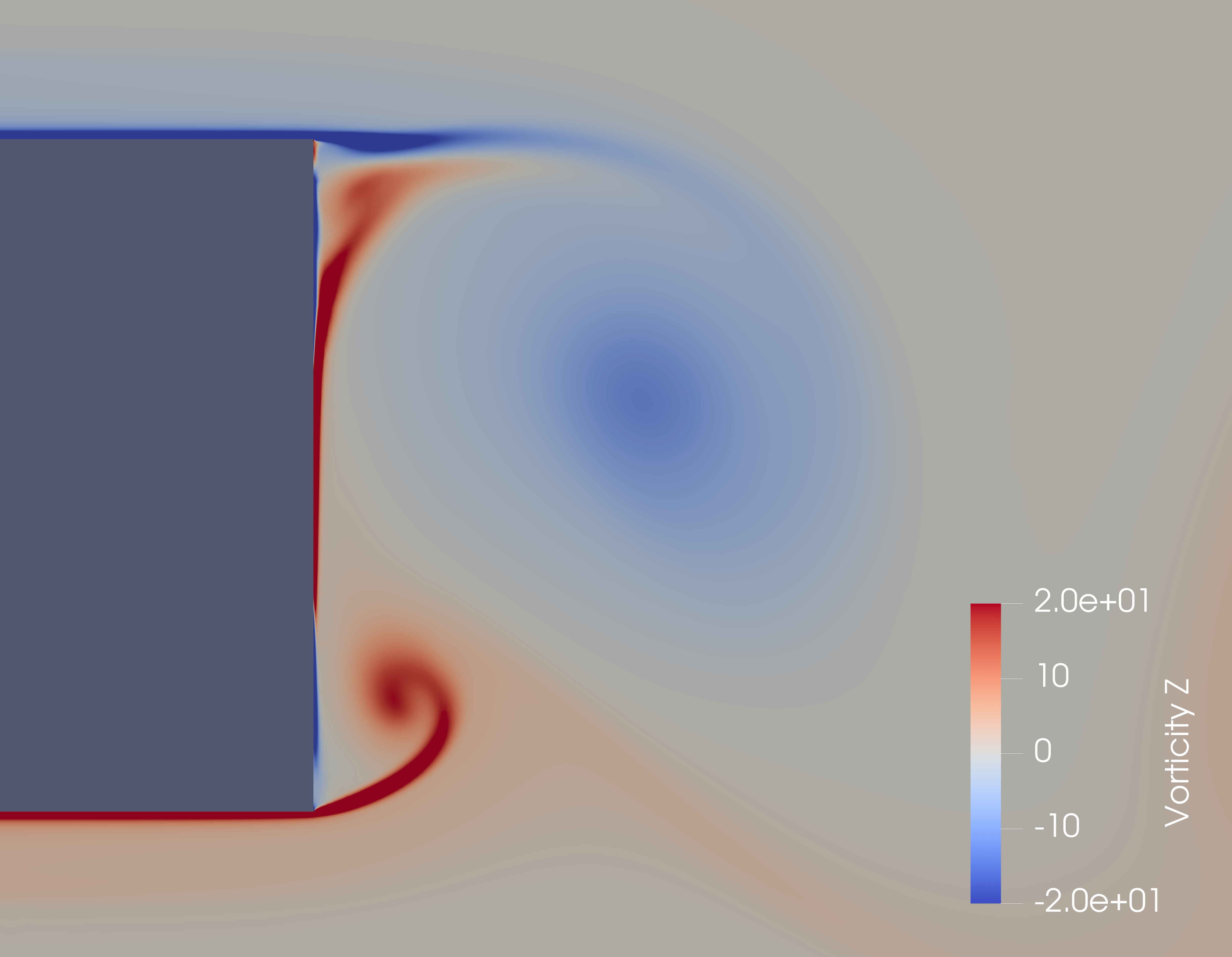}\hfill
	\includegraphics[width=.32\textwidth]{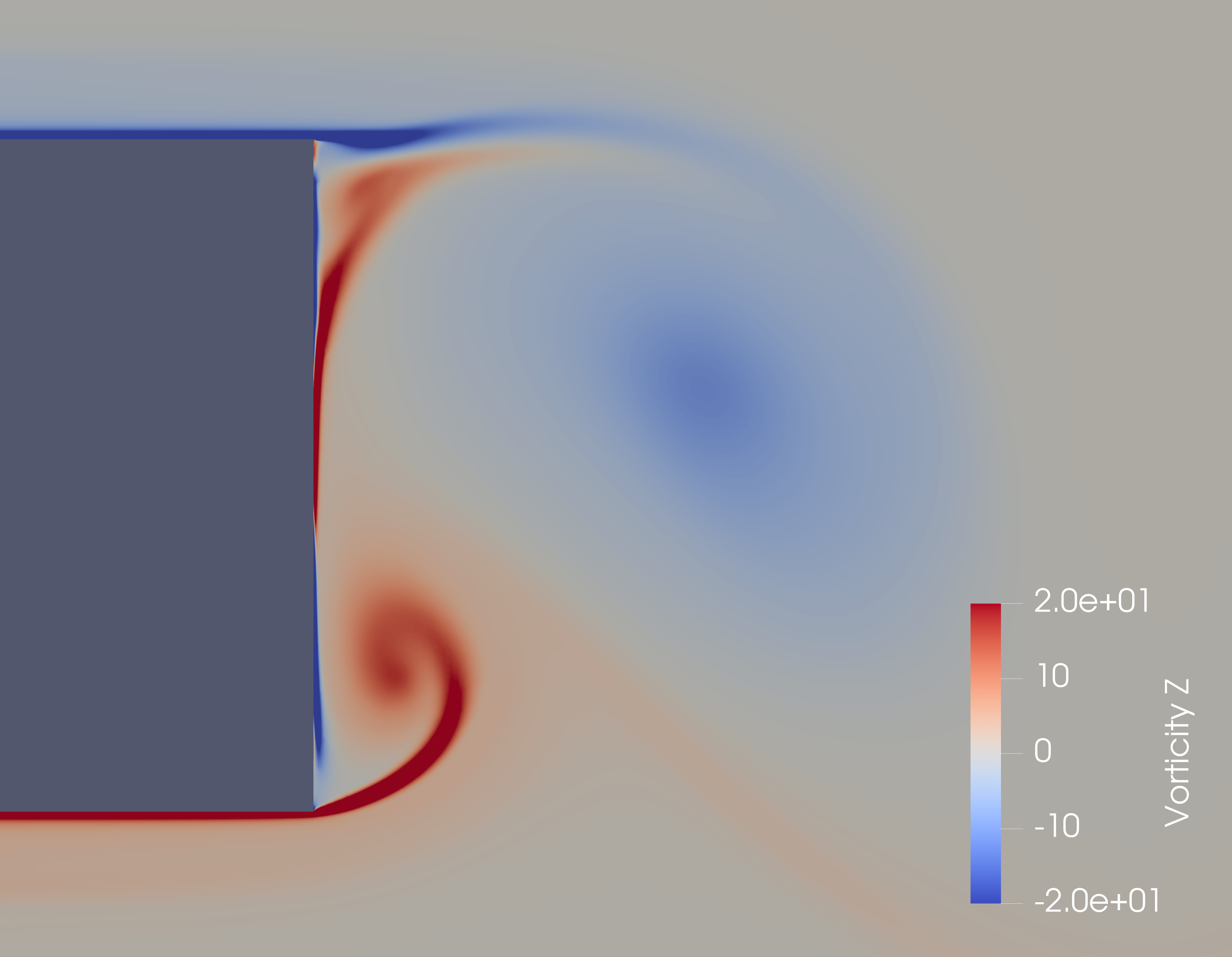}\hfill
	\includegraphics[width=.32\textwidth]{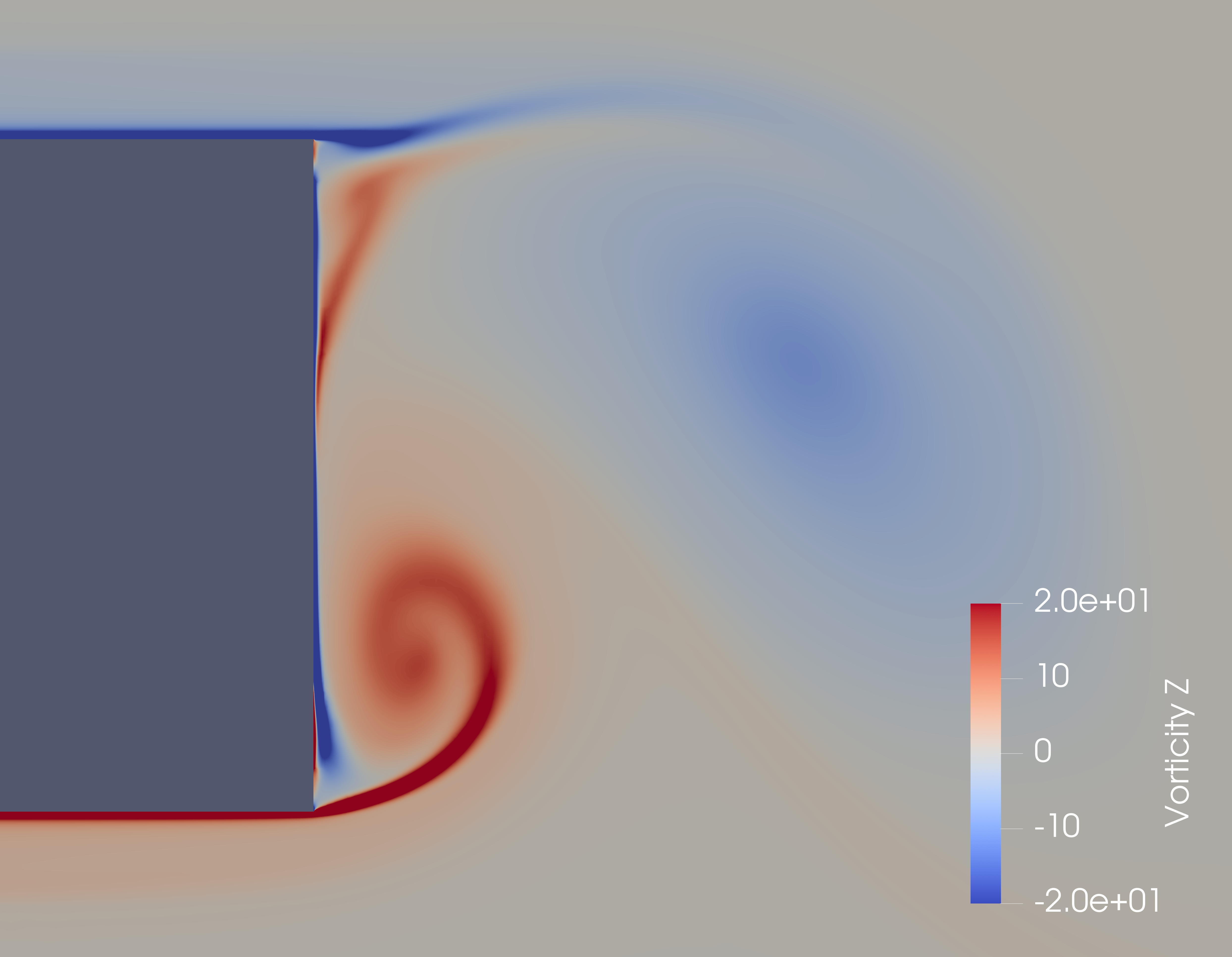}\hfill
	\includegraphics[width=.32\textwidth]{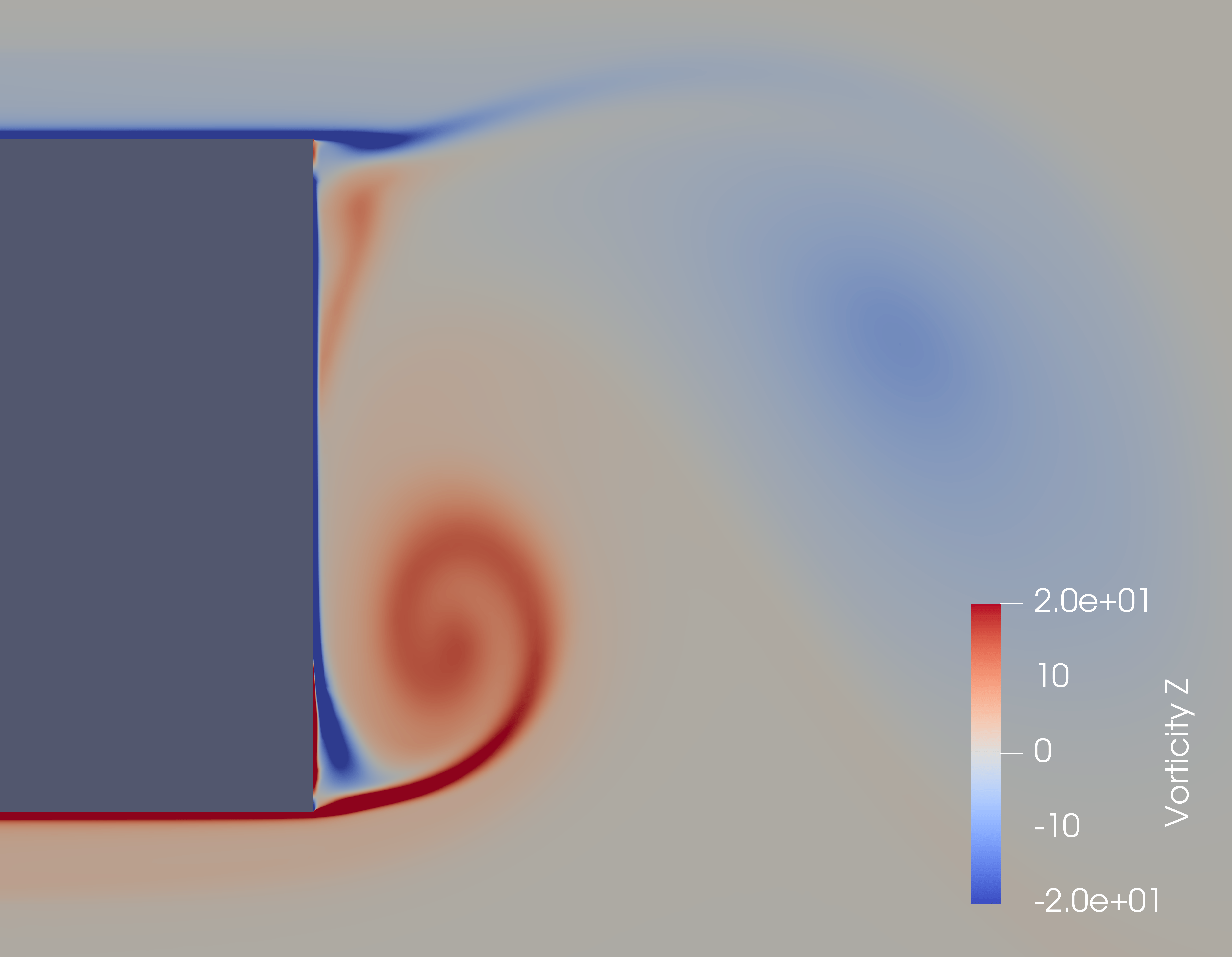}\hfill
	\includegraphics[width=.32\textwidth]{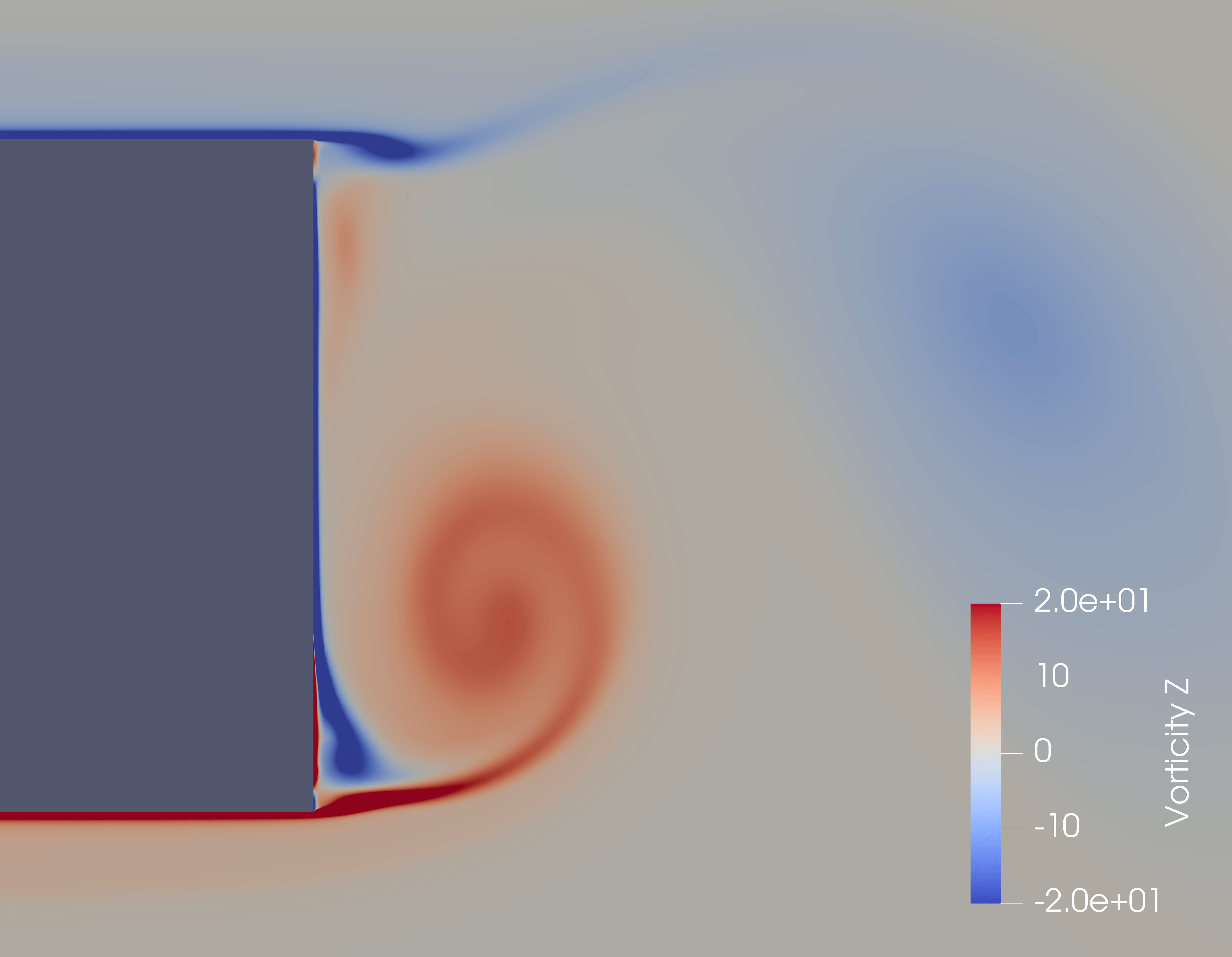}\hfill
	\includegraphics[width=.32\textwidth]{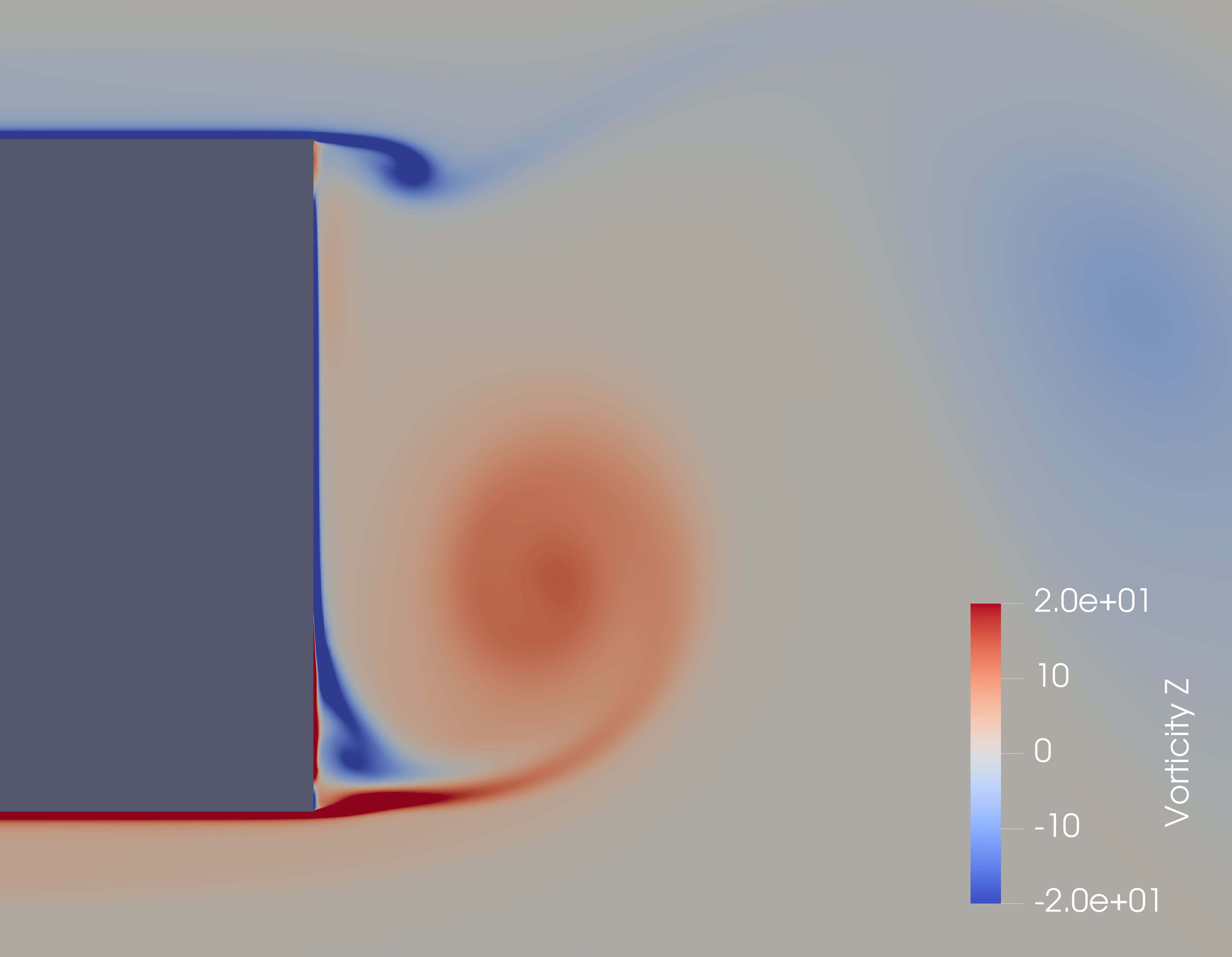}\hfill
	\includegraphics[width=.32\textwidth]{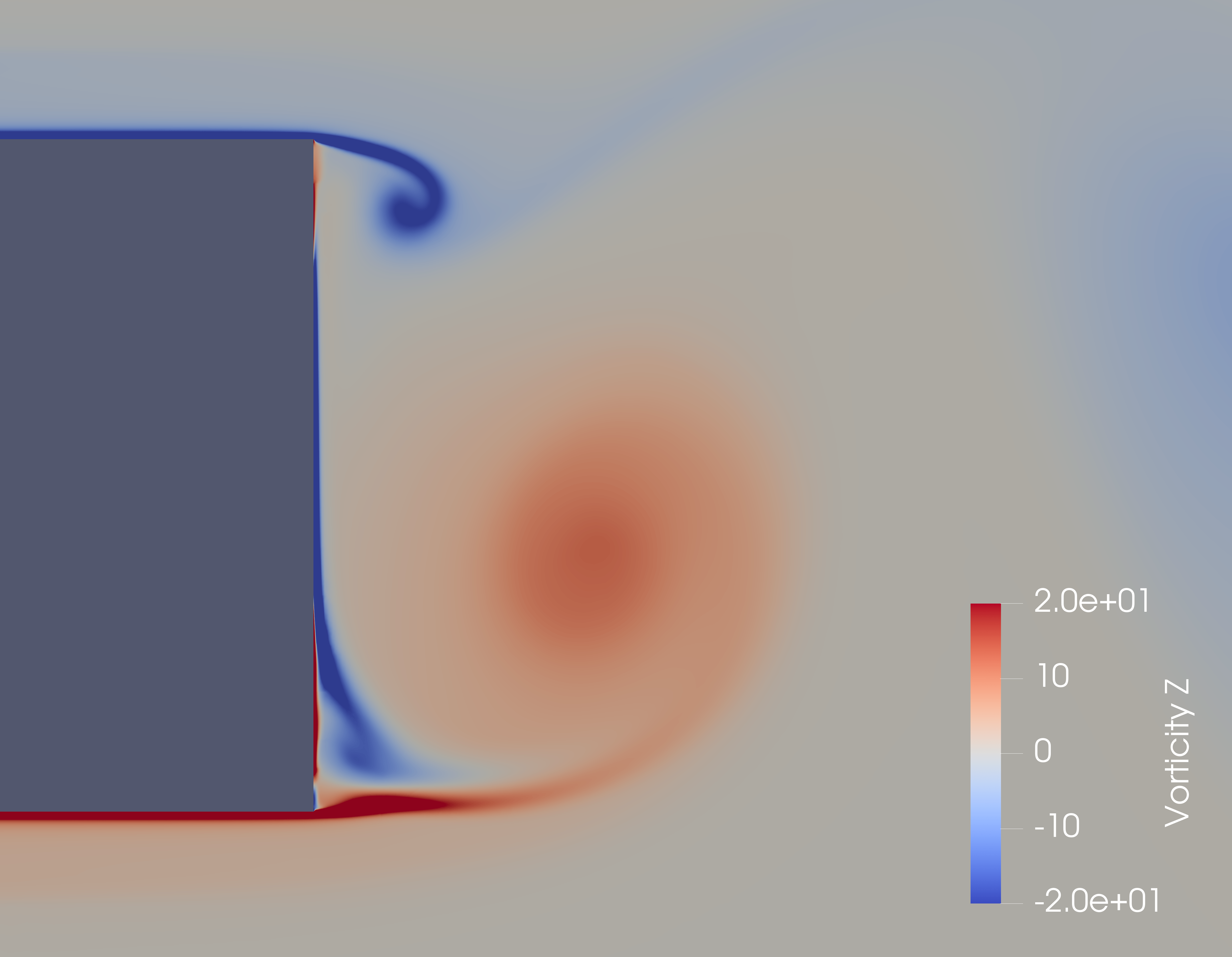}\hfill
	\caption{Time series of vorticity behind the body. The vortex formation and detachment is regular and periodic.}
	\label{fig:vort_clean}
\end{figure*}

\begin{figure*}[h] 
	\includegraphics[width=.32\textwidth]{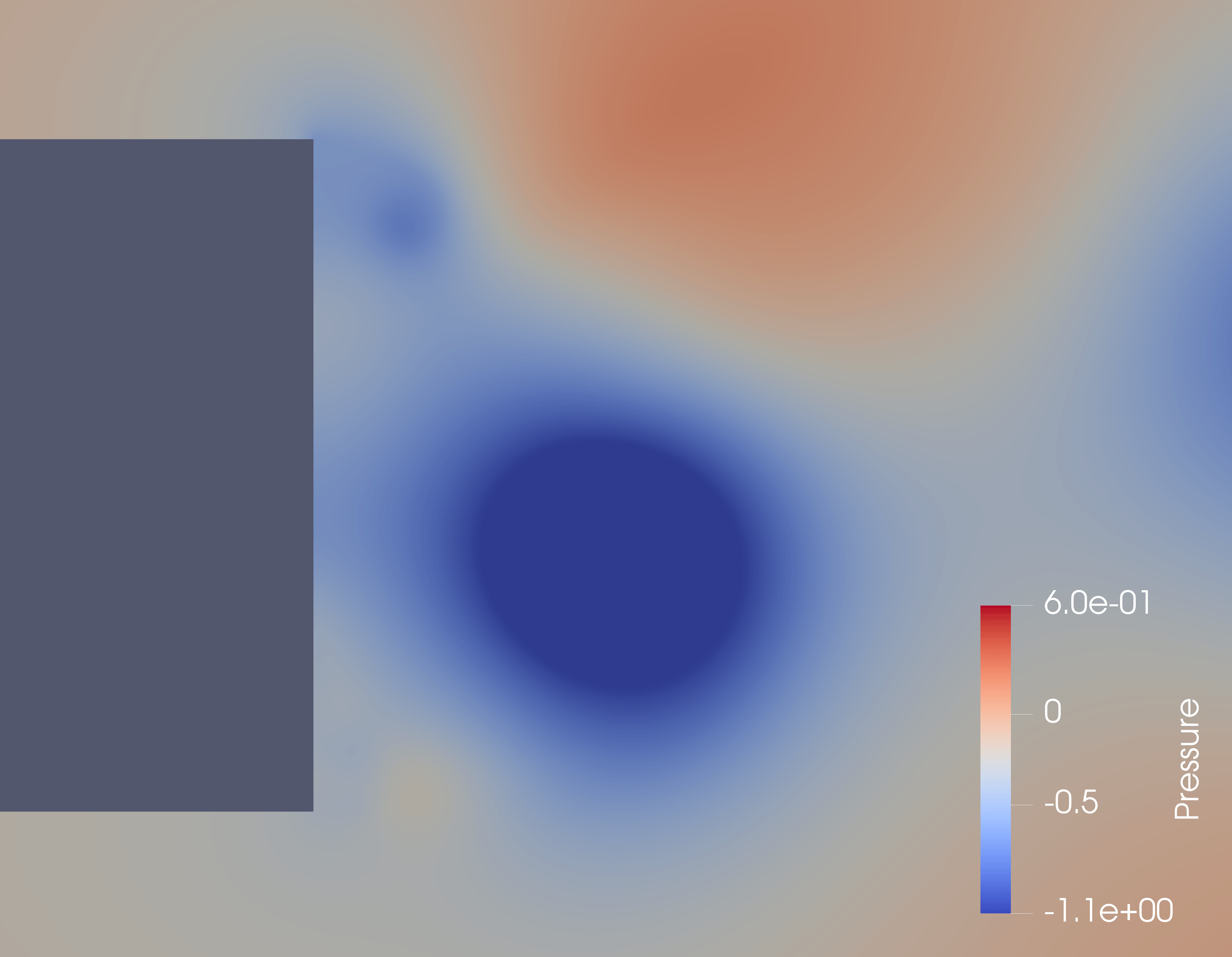}\hfill
	\includegraphics[width=.32\textwidth]{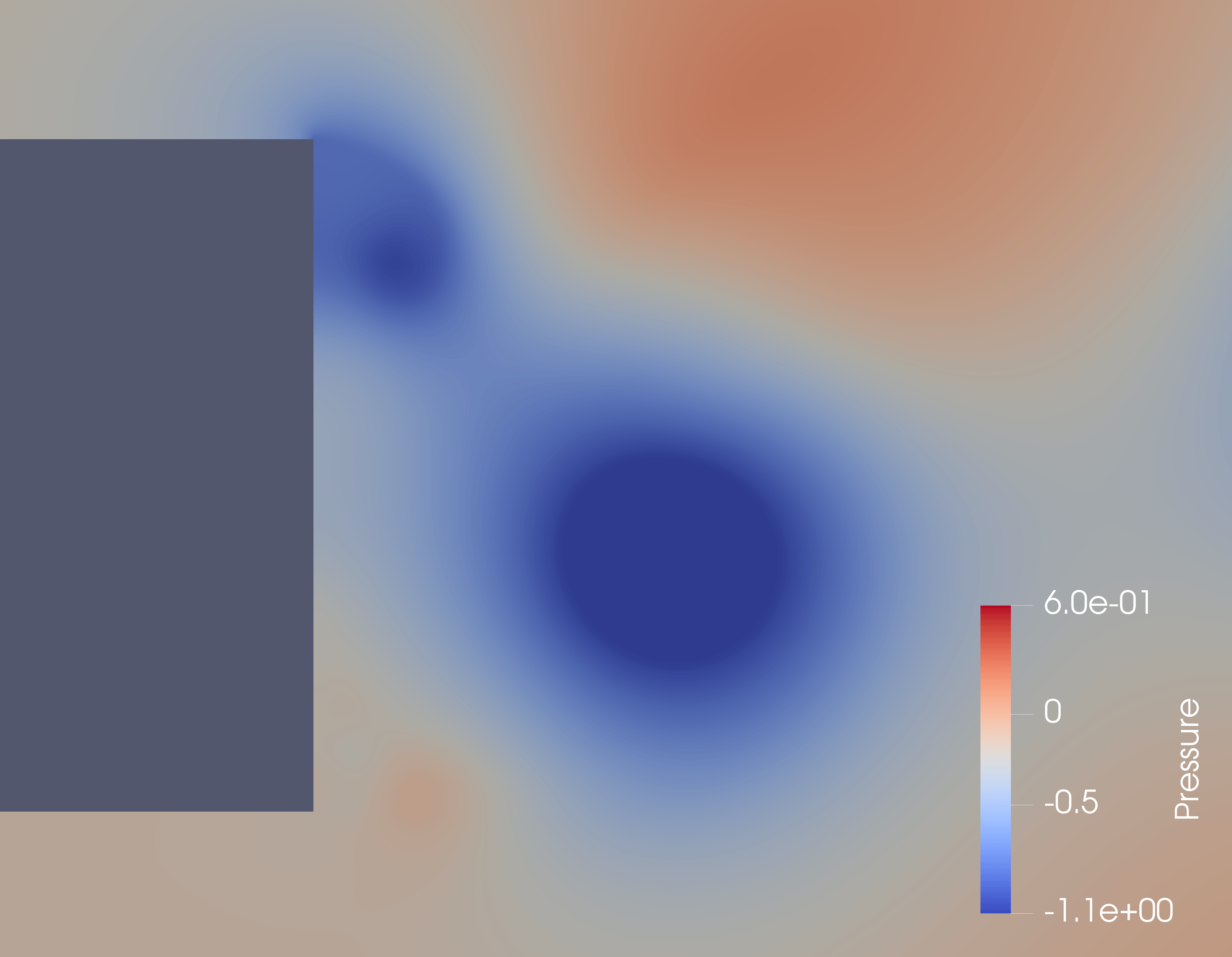}\hfill
	\includegraphics[width=.32\textwidth]{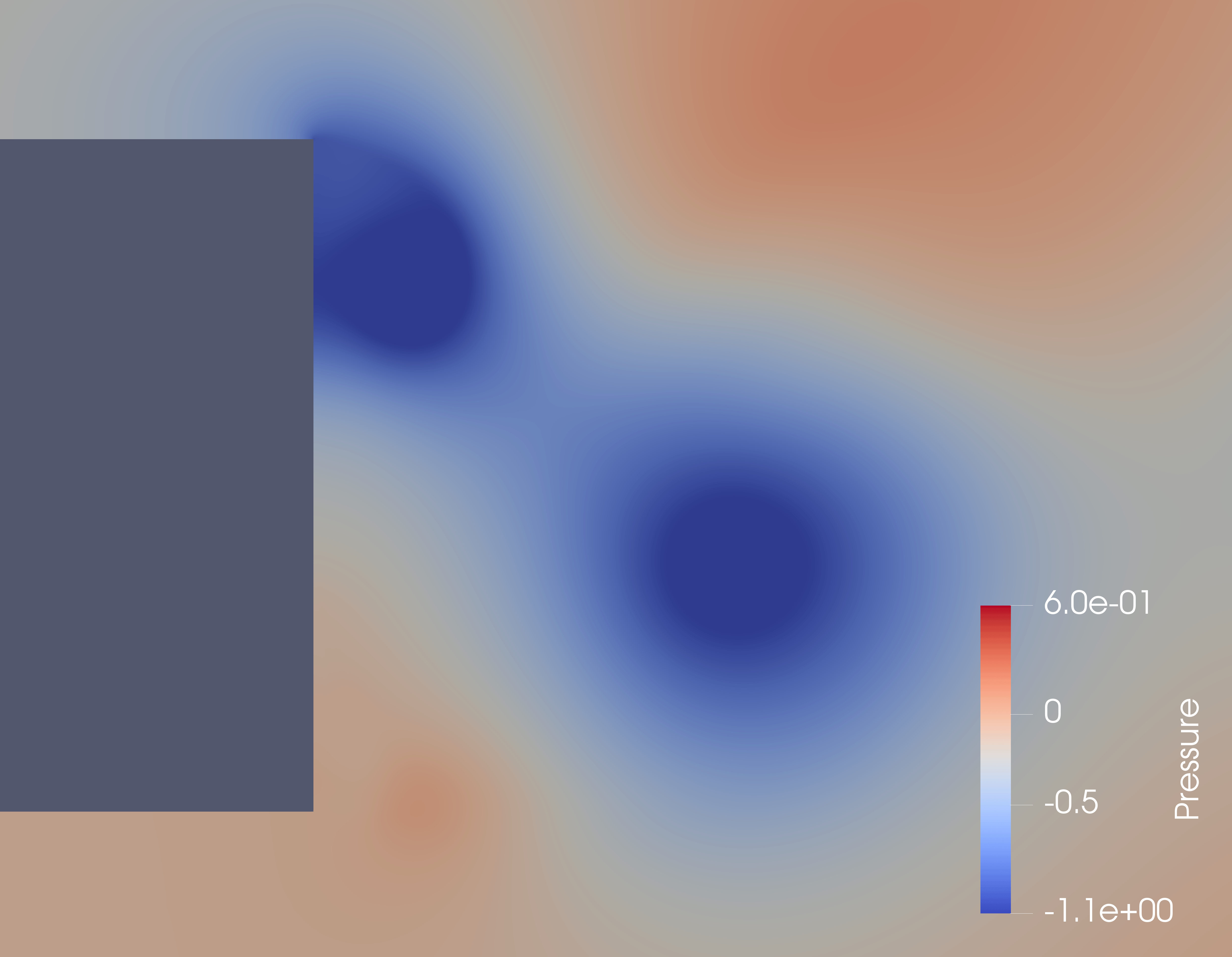}\hfill
	\includegraphics[width=.32\textwidth]{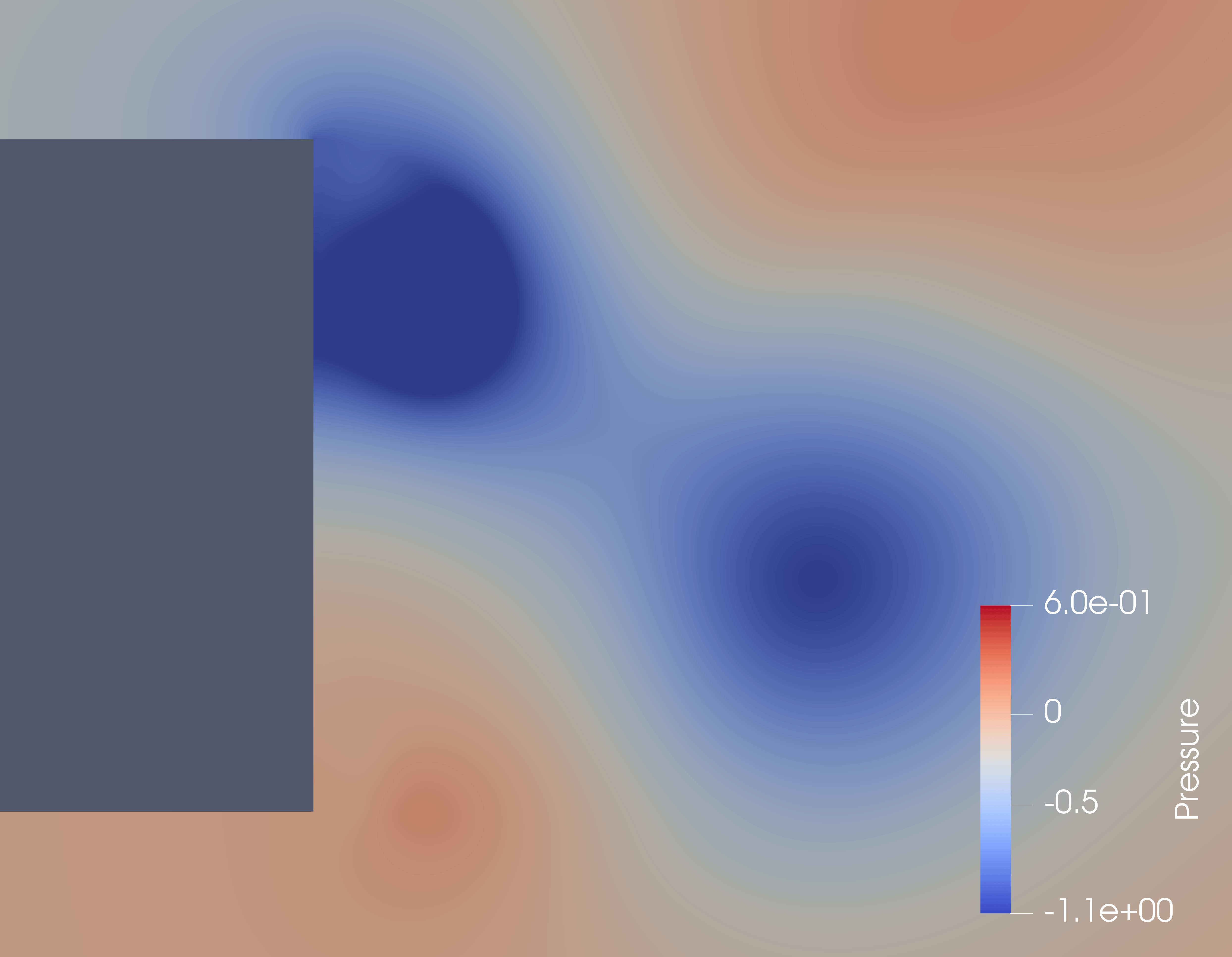}\hfill
	\includegraphics[width=.32\textwidth]{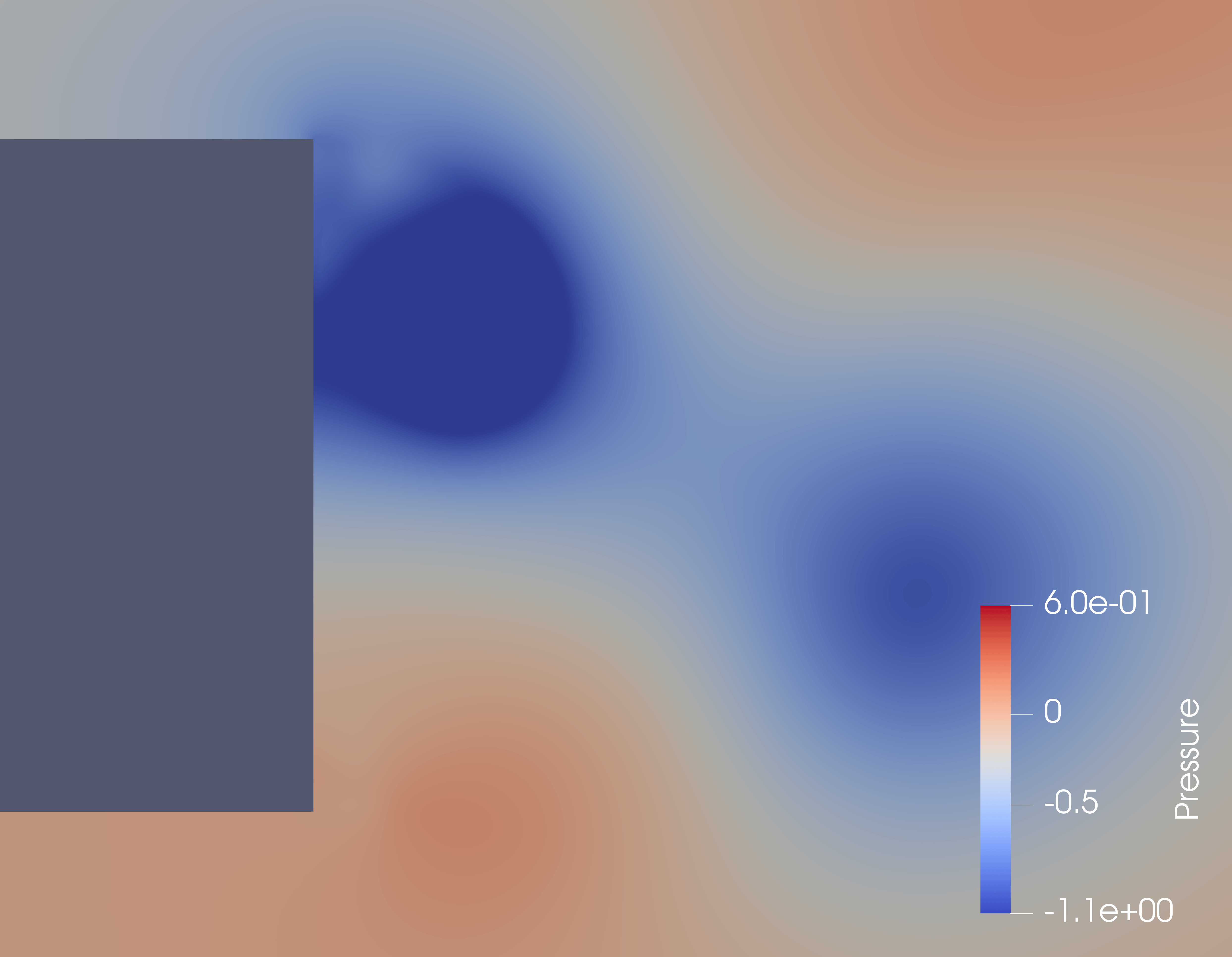}\hfill
	\includegraphics[width=.32\textwidth]{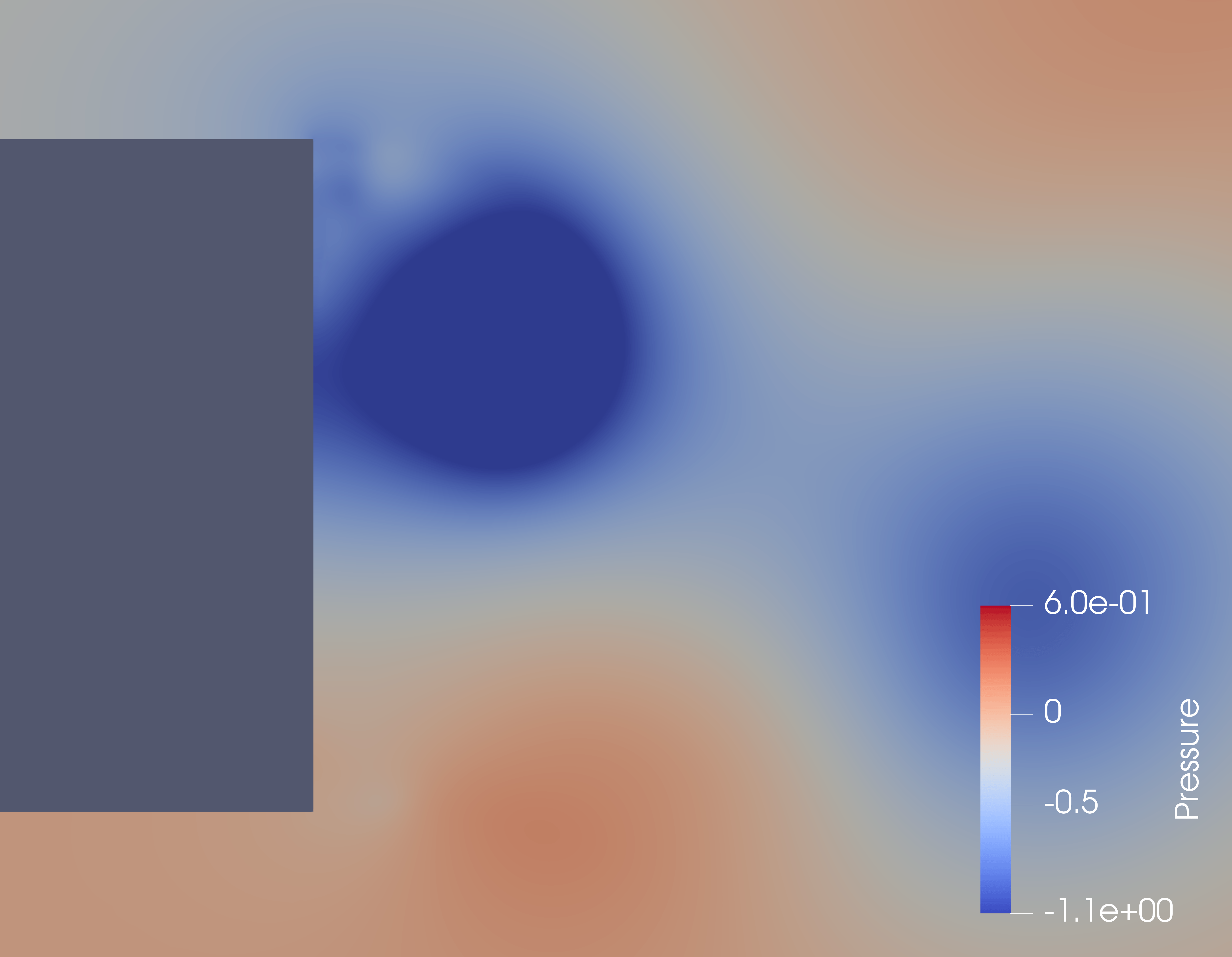}\hfill
	\includegraphics[width=.32\textwidth]{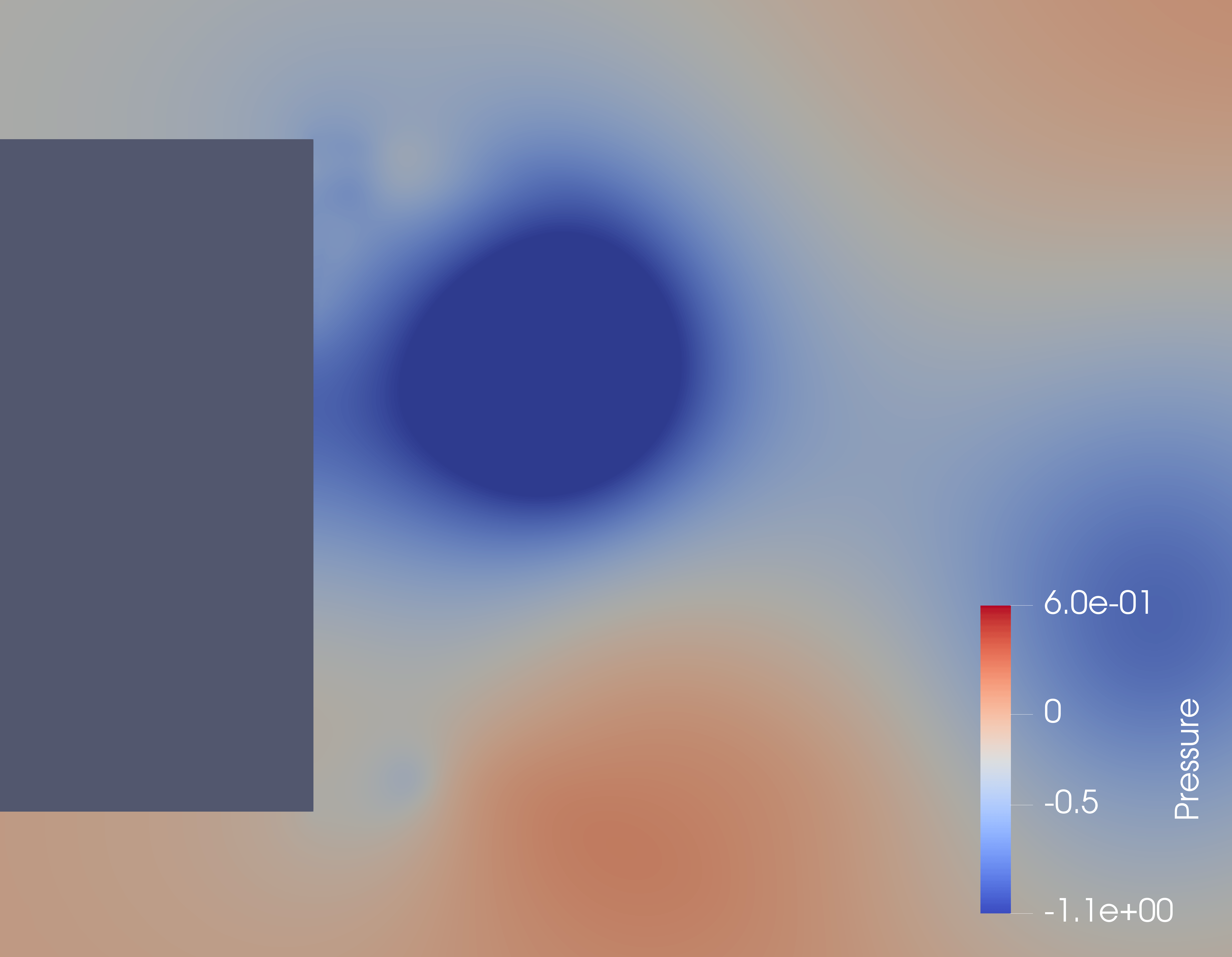}\hfill
	\includegraphics[width=.32\textwidth]{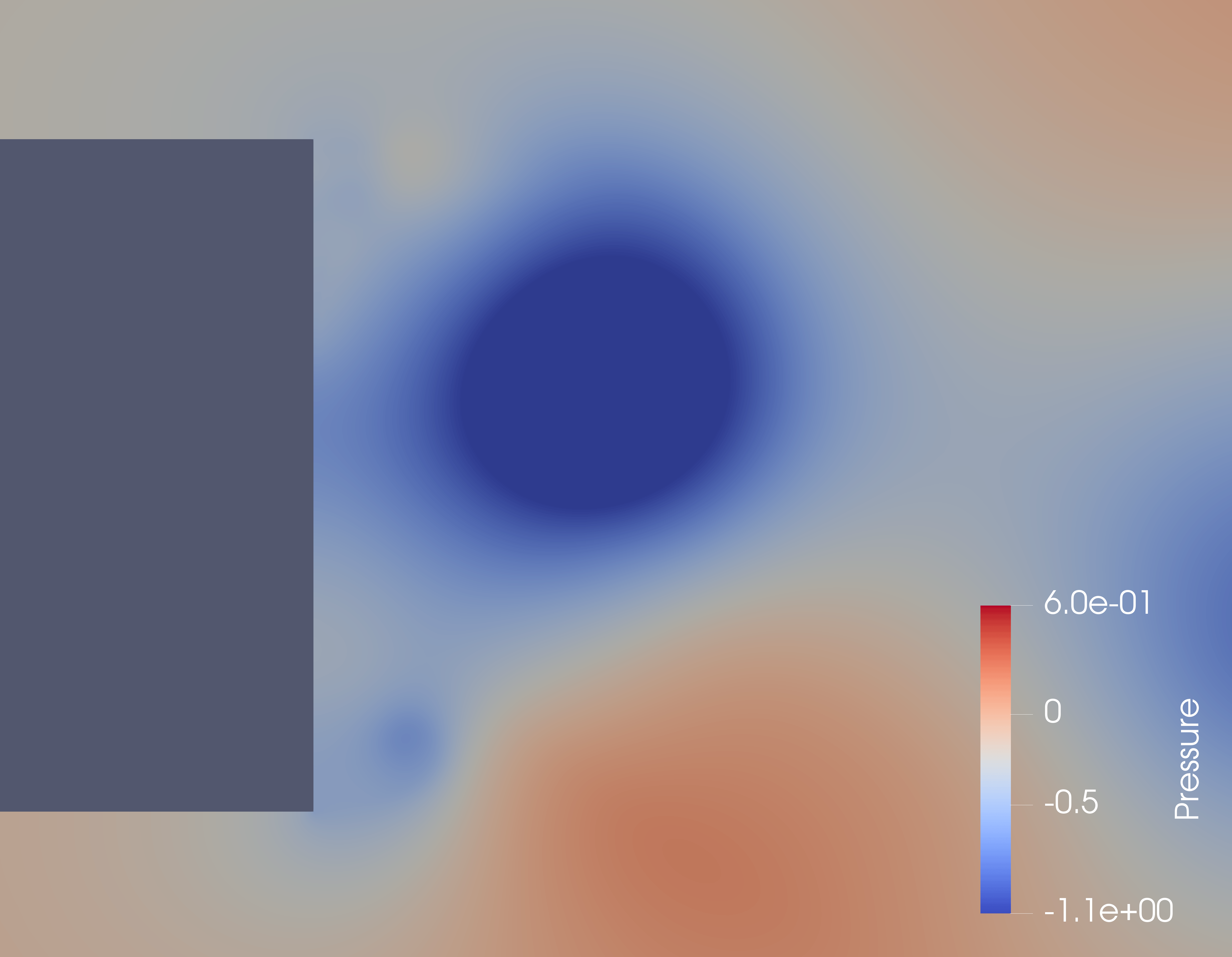}\hfill
	\includegraphics[width=.32\textwidth]{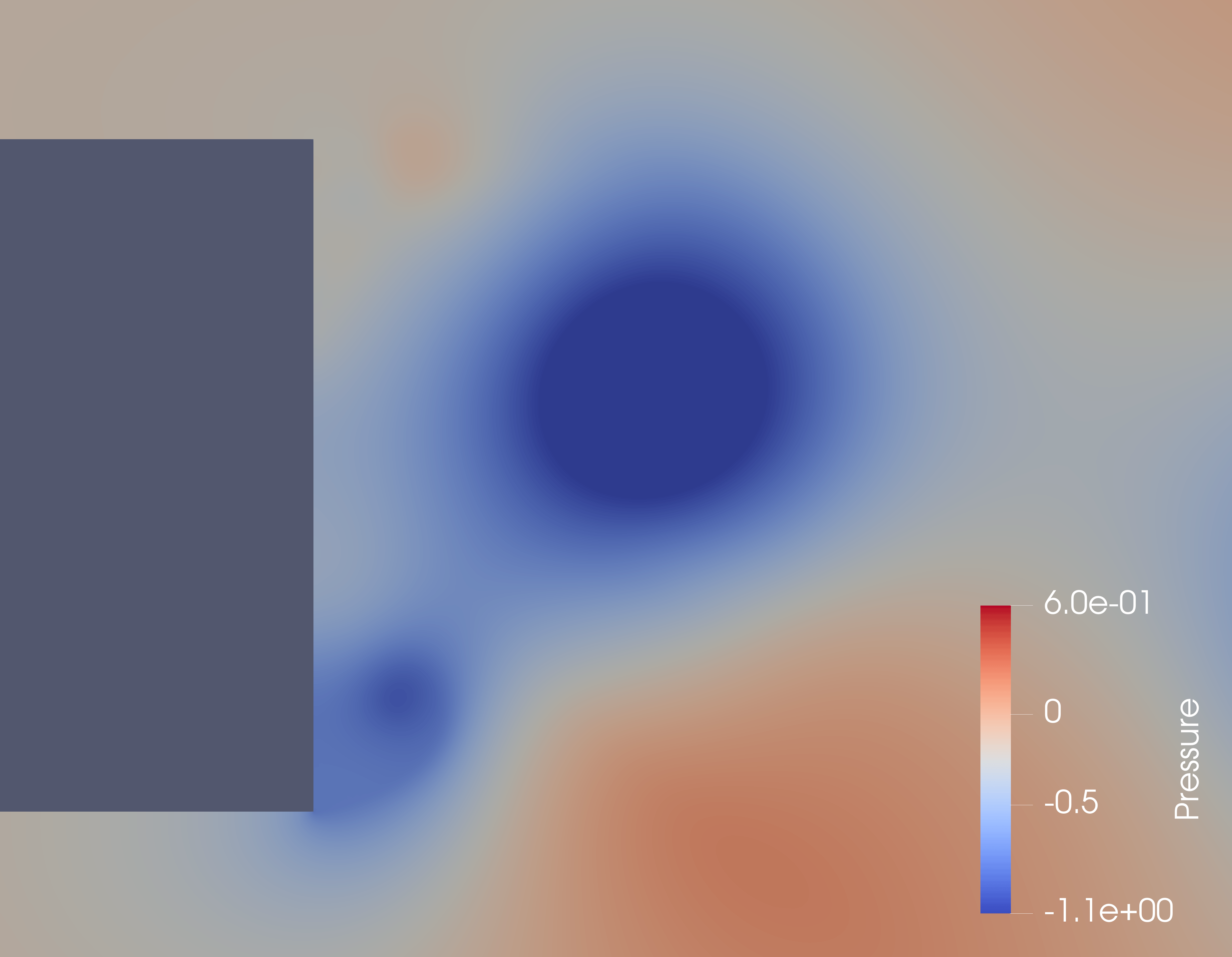}\hfill
	\includegraphics[width=.32\textwidth]{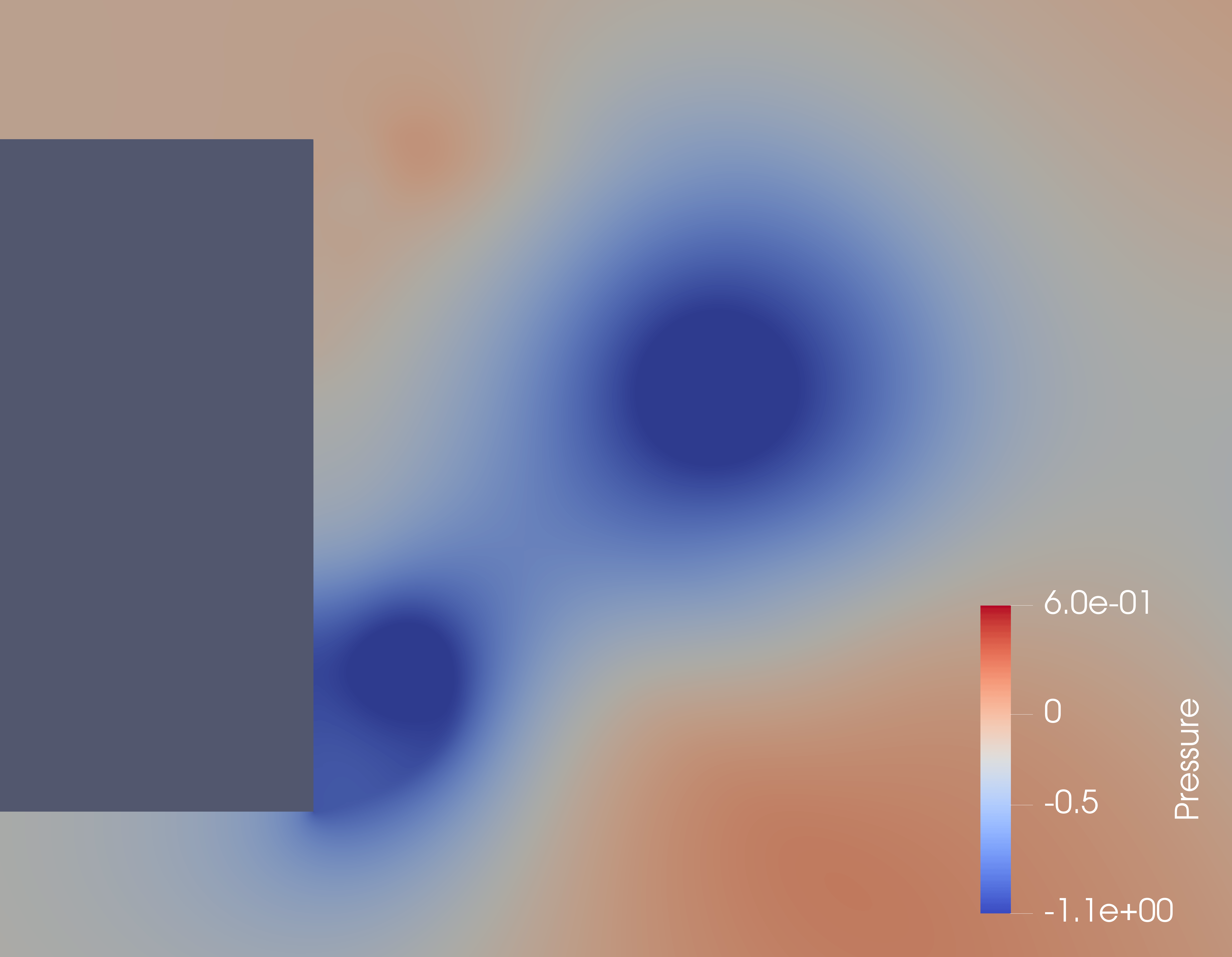}\hfill
	\includegraphics[width=.32\textwidth]{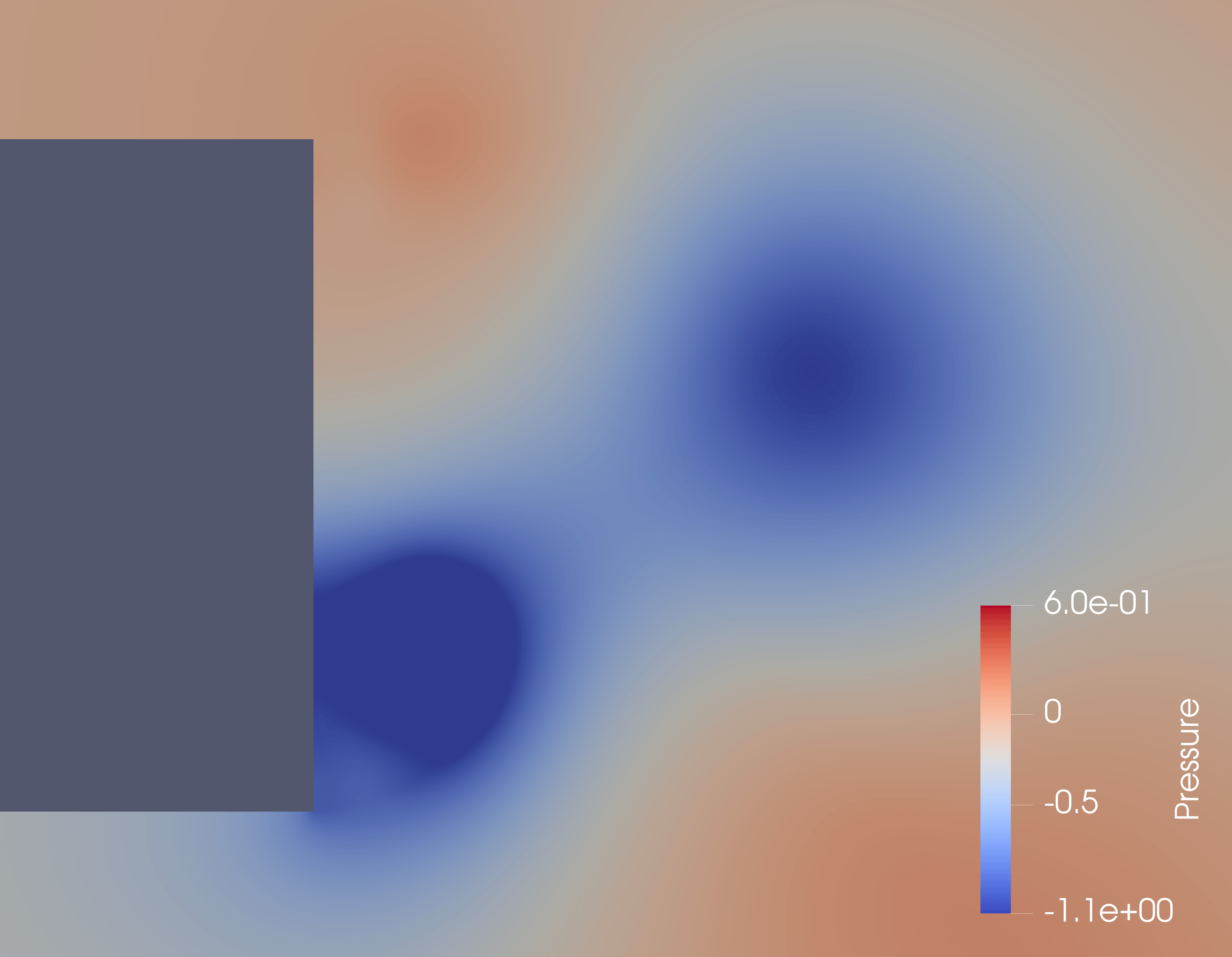}\hfill
	\includegraphics[width=.32\textwidth]{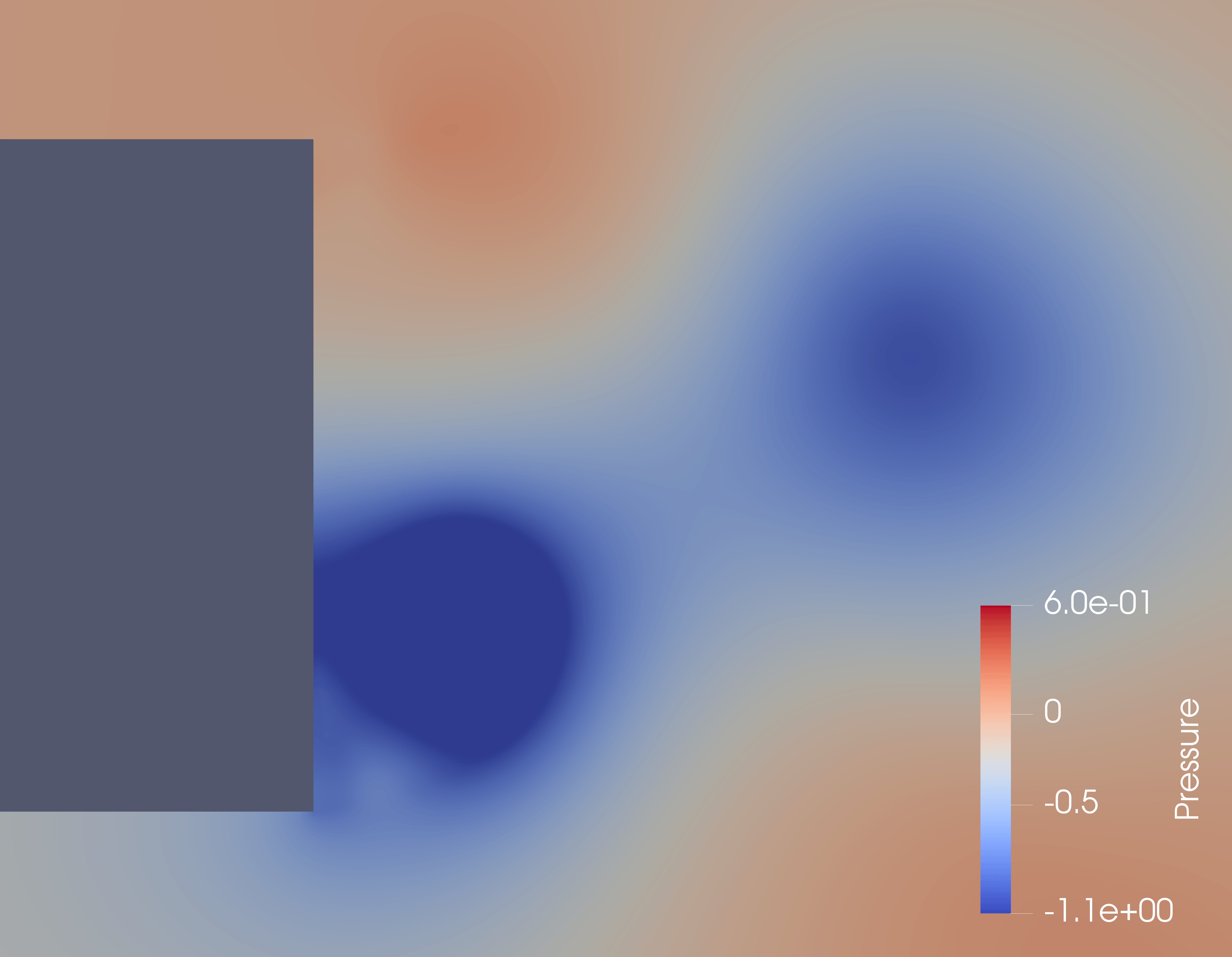}\hfill
	\includegraphics[width=.32\textwidth]{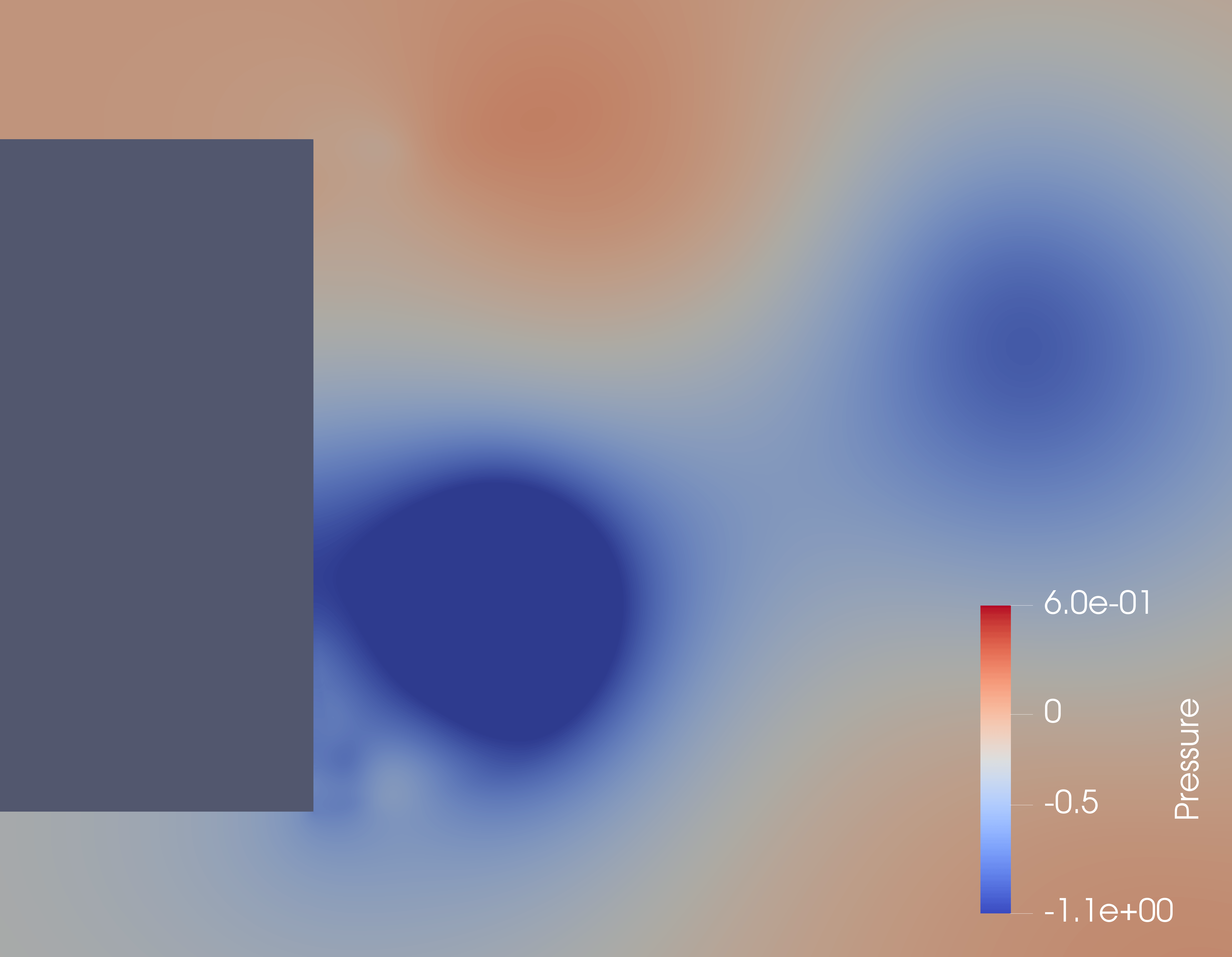}\hfill
	\includegraphics[width=.32\textwidth]{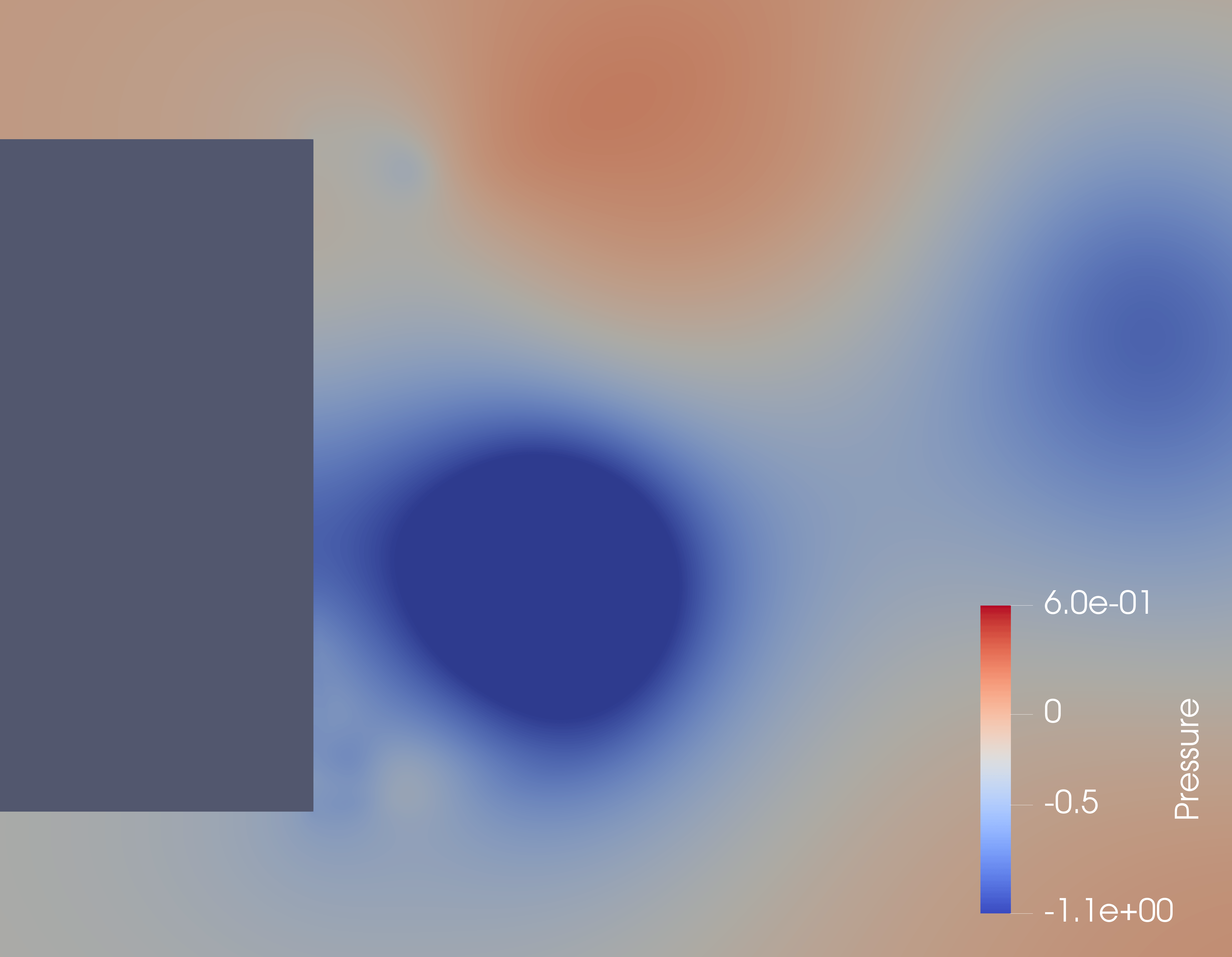}\hfill
	\includegraphics[width=.32\textwidth]{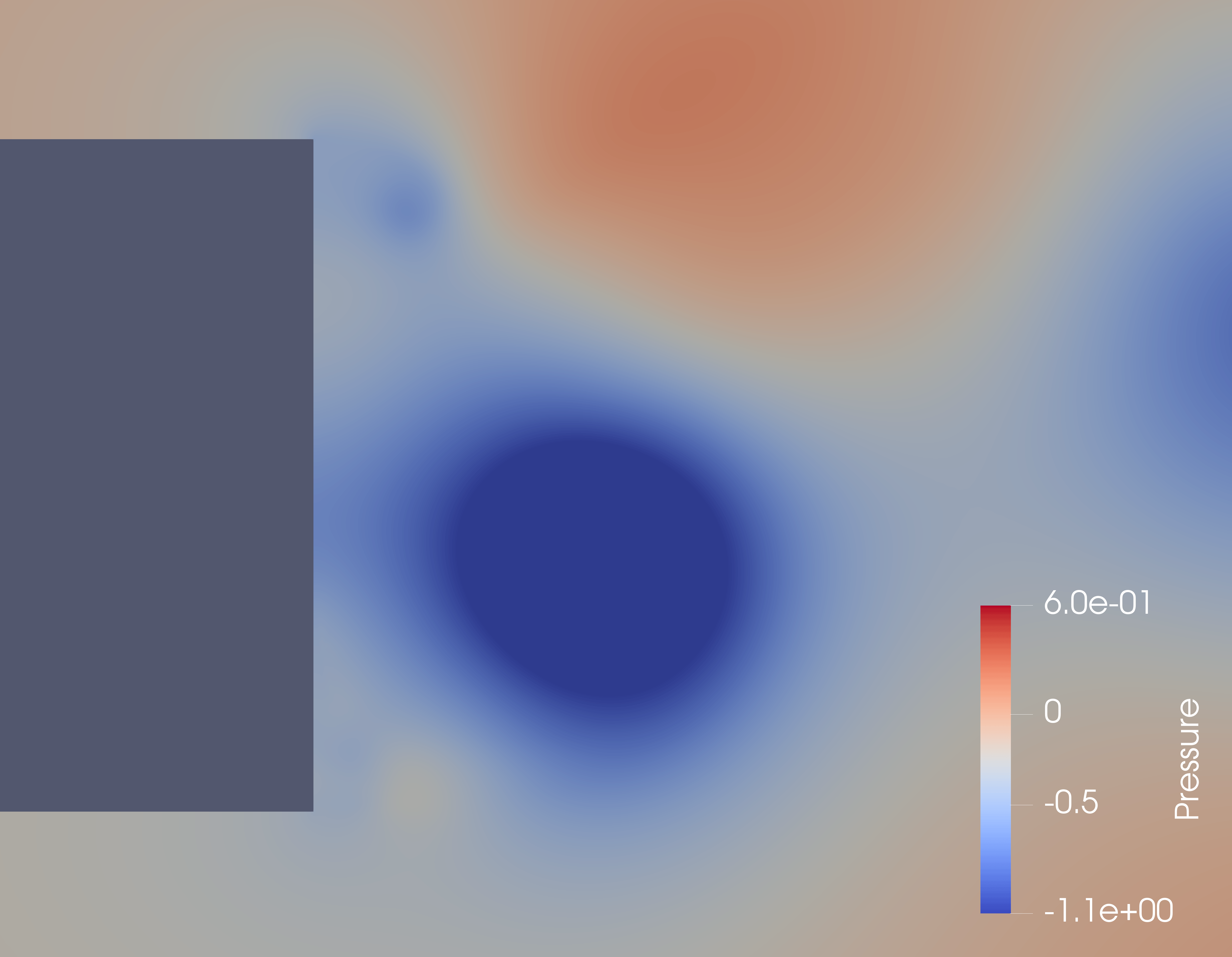}\hfill
	\caption{Time series of pressure behind the body. Low pressure regions caused by the cores of the vortexes touches the rear surface of the body.}
	\label{fig:pres_clean}
\end{figure*}

\begin{figure*}[h]
	\includegraphics[width=.32\textwidth]{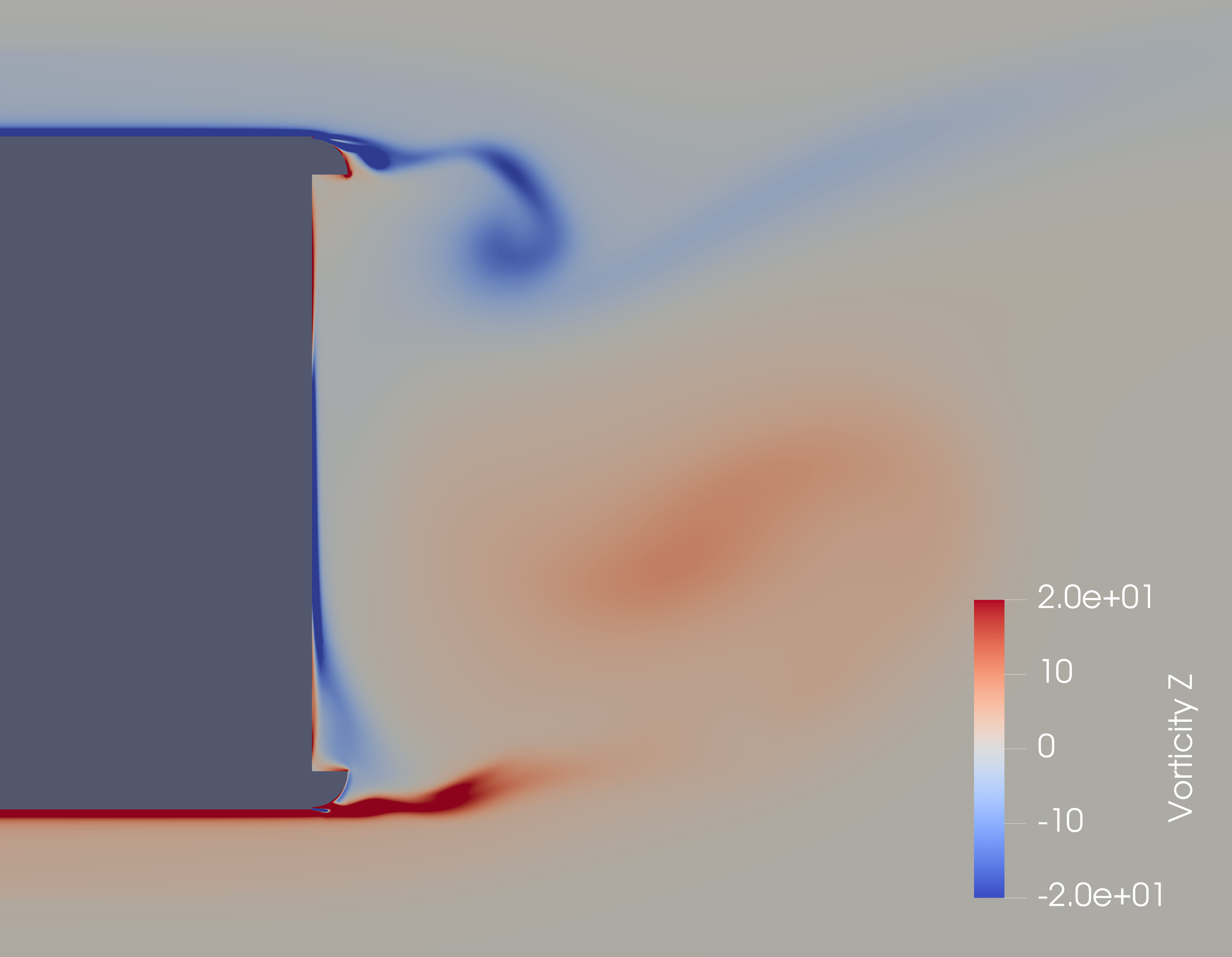}\hfill
	\includegraphics[width=.32\textwidth]{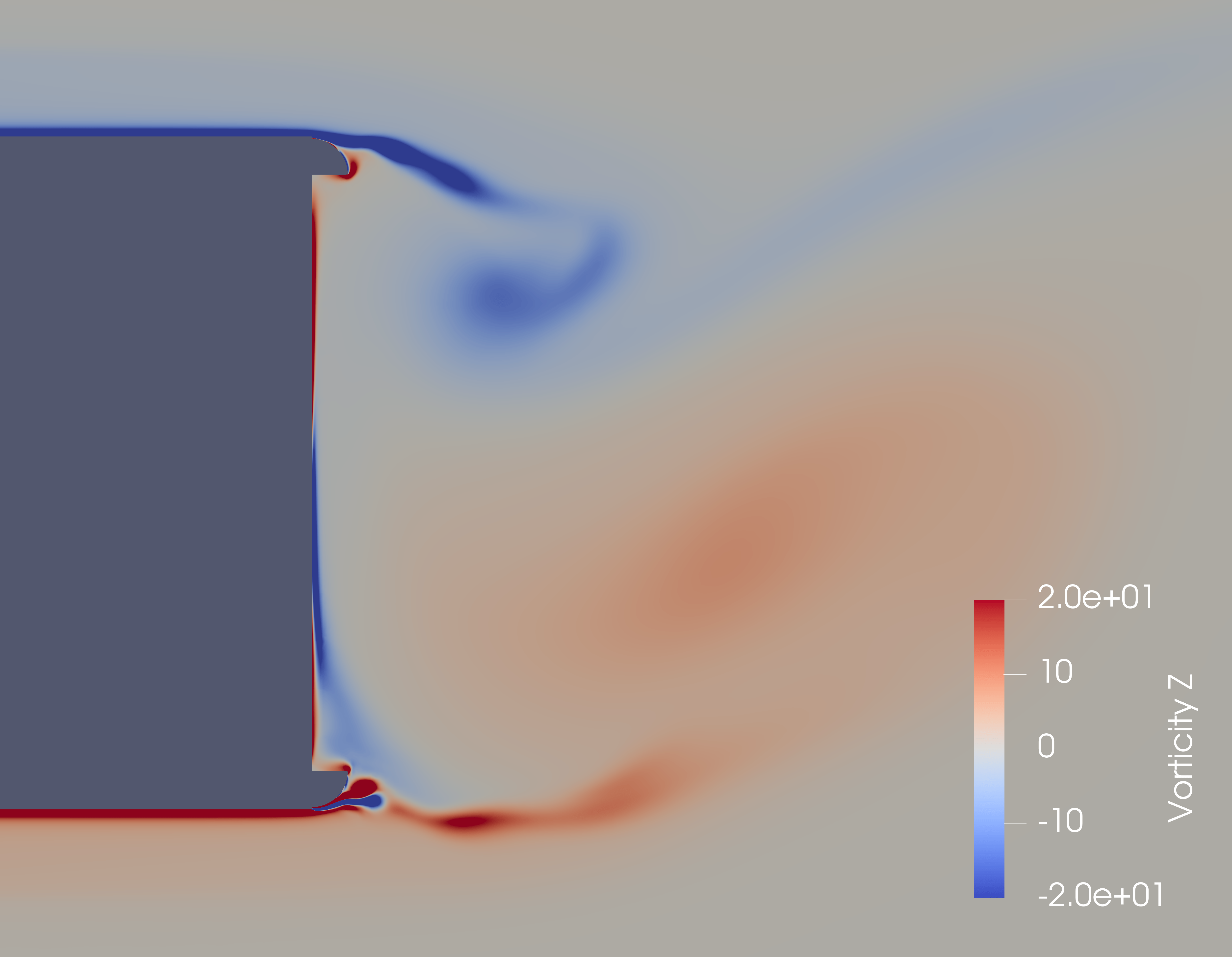}\hfill
	\includegraphics[width=.32\textwidth]{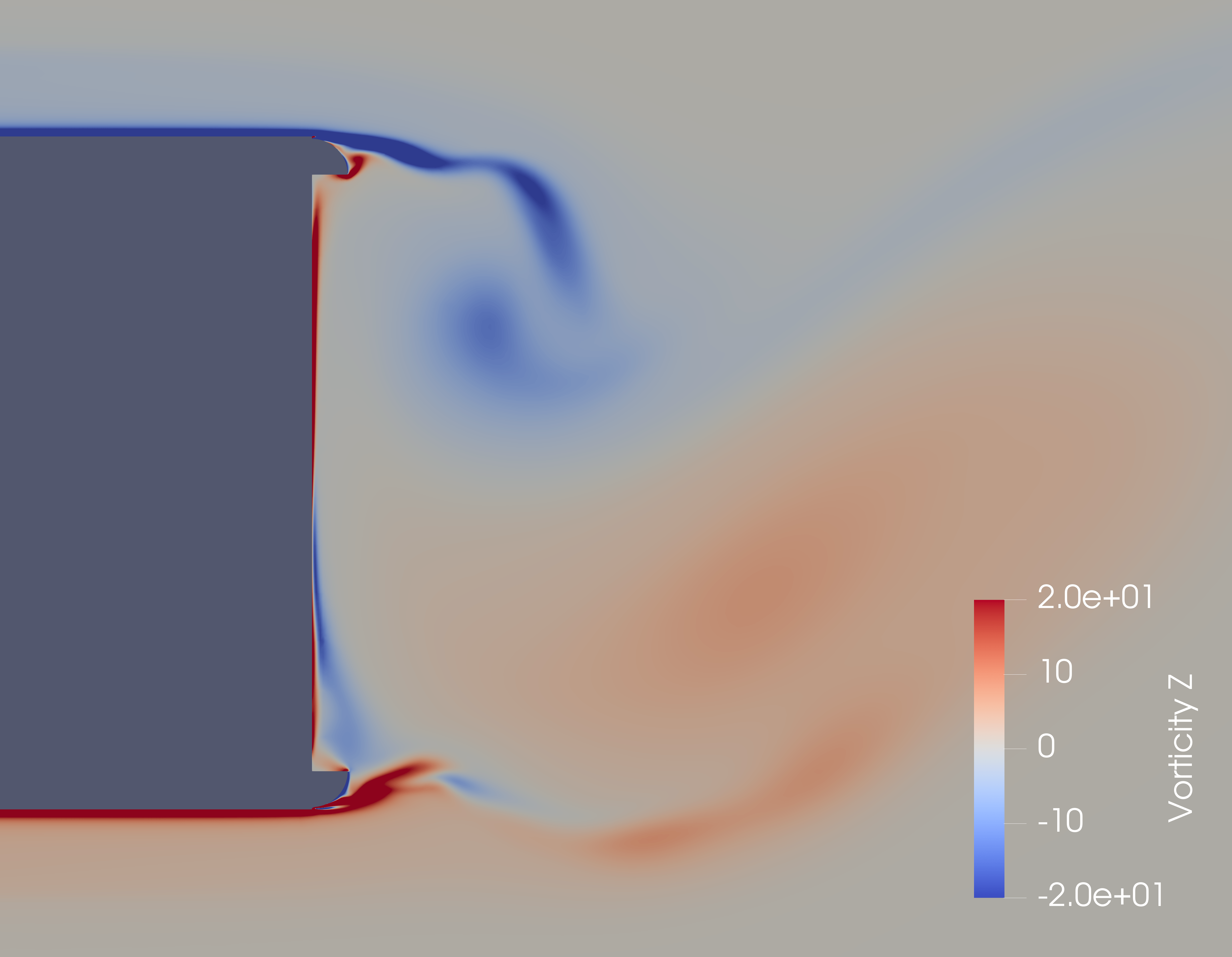}\hfill
	\includegraphics[width=.32\textwidth]{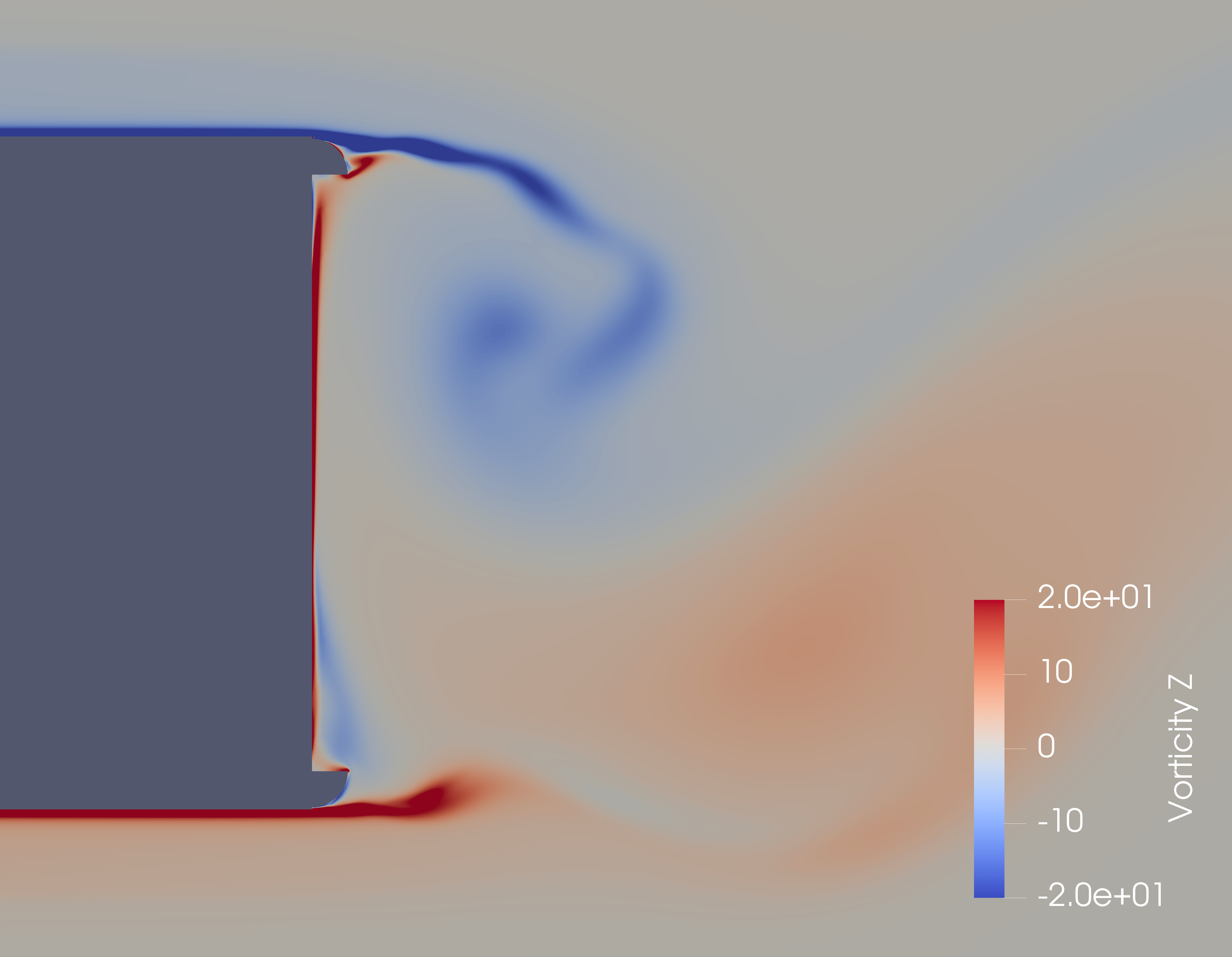}\hfill
	\includegraphics[width=.32\textwidth]{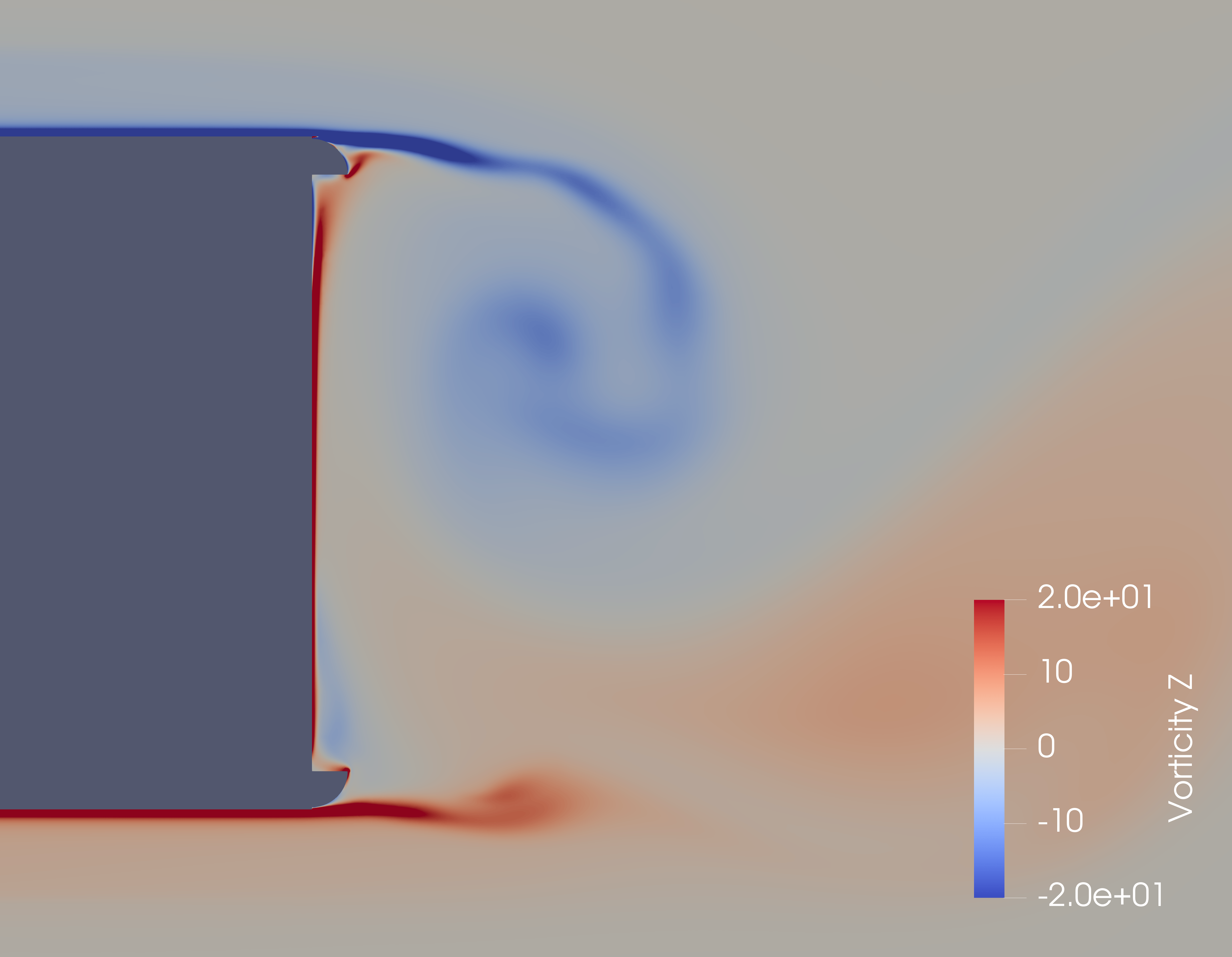}\hfill
	\includegraphics[width=.32\textwidth]{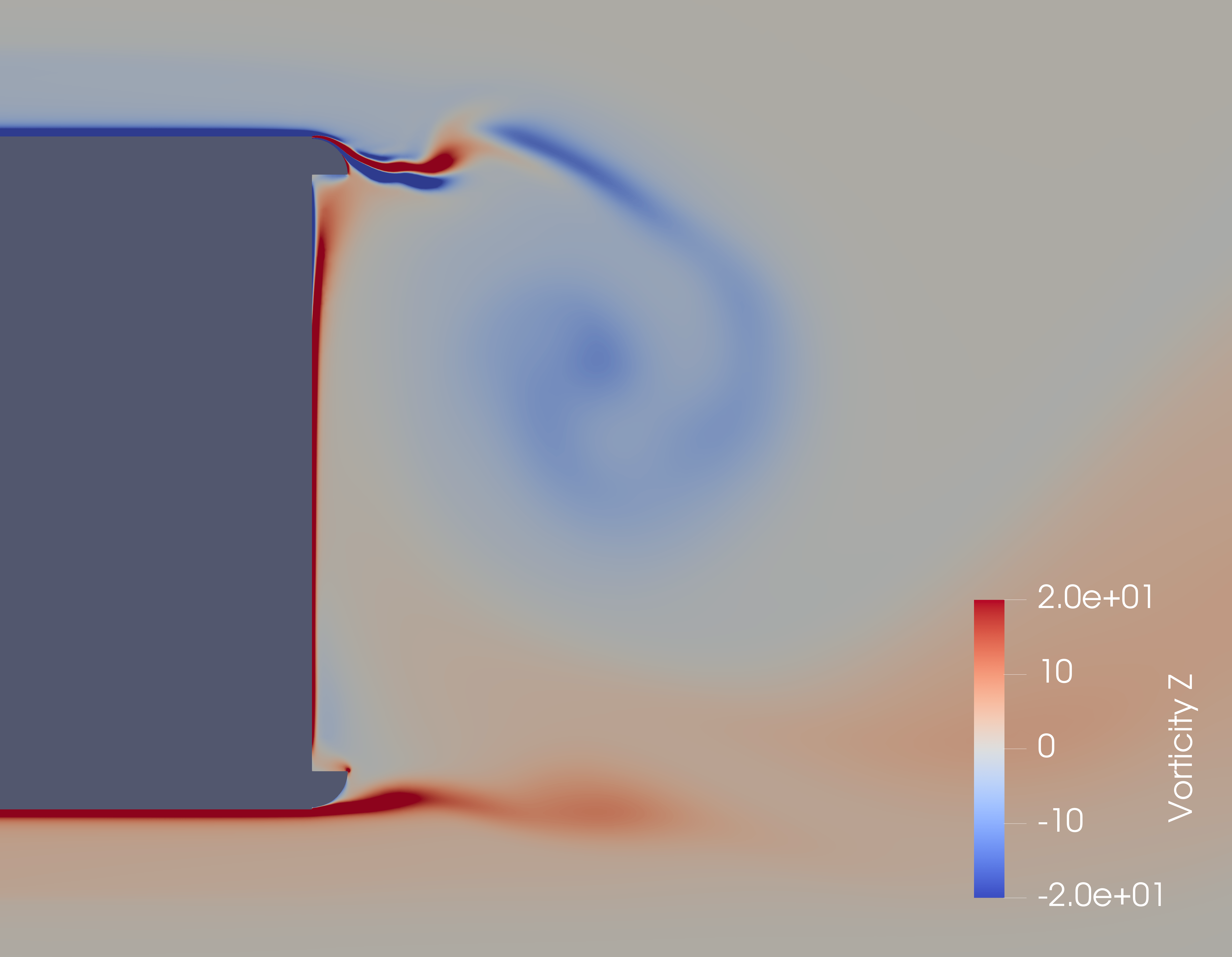}\hfill
	\includegraphics[width=.32\textwidth]{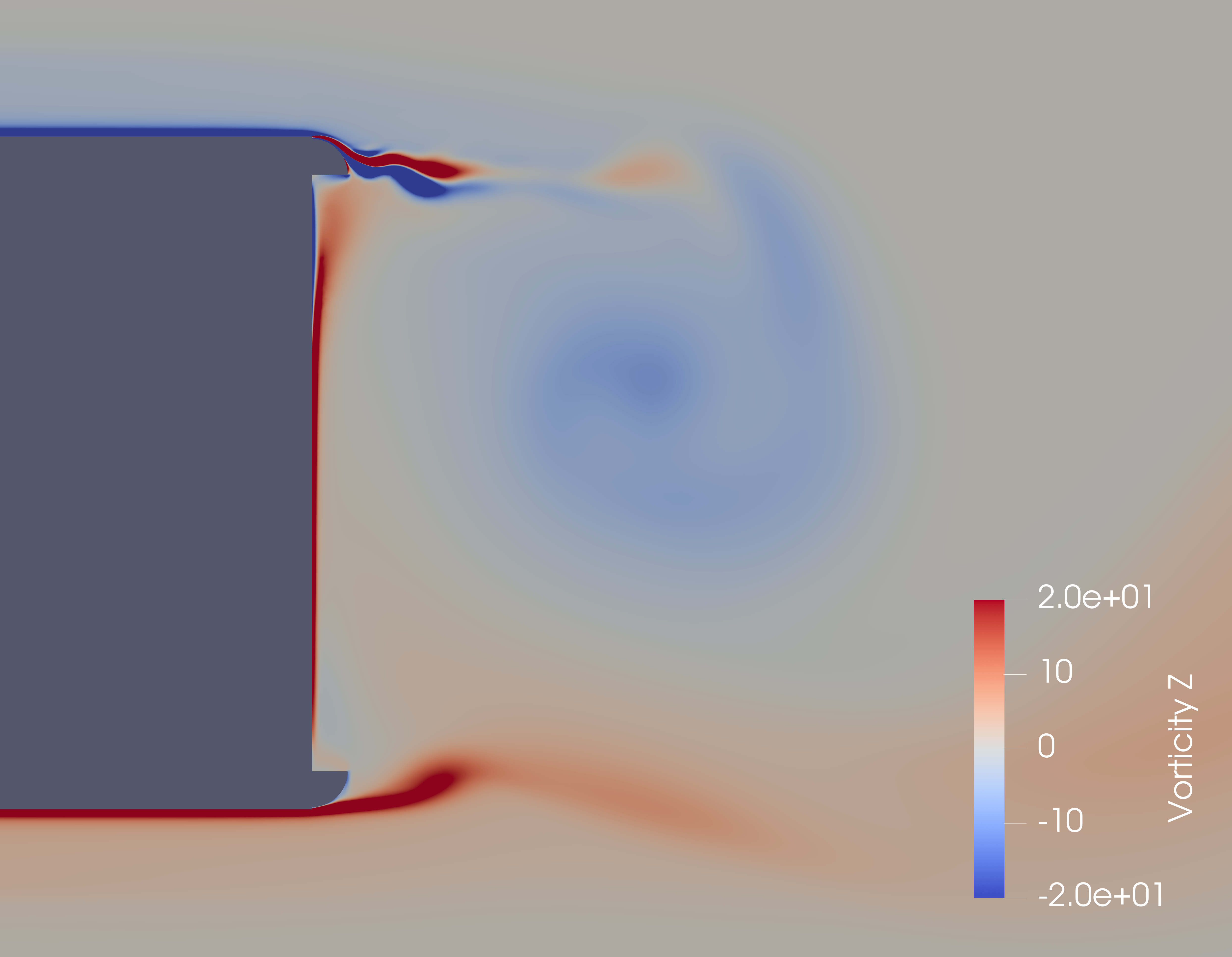}\hfill
	\includegraphics[width=.32\textwidth]{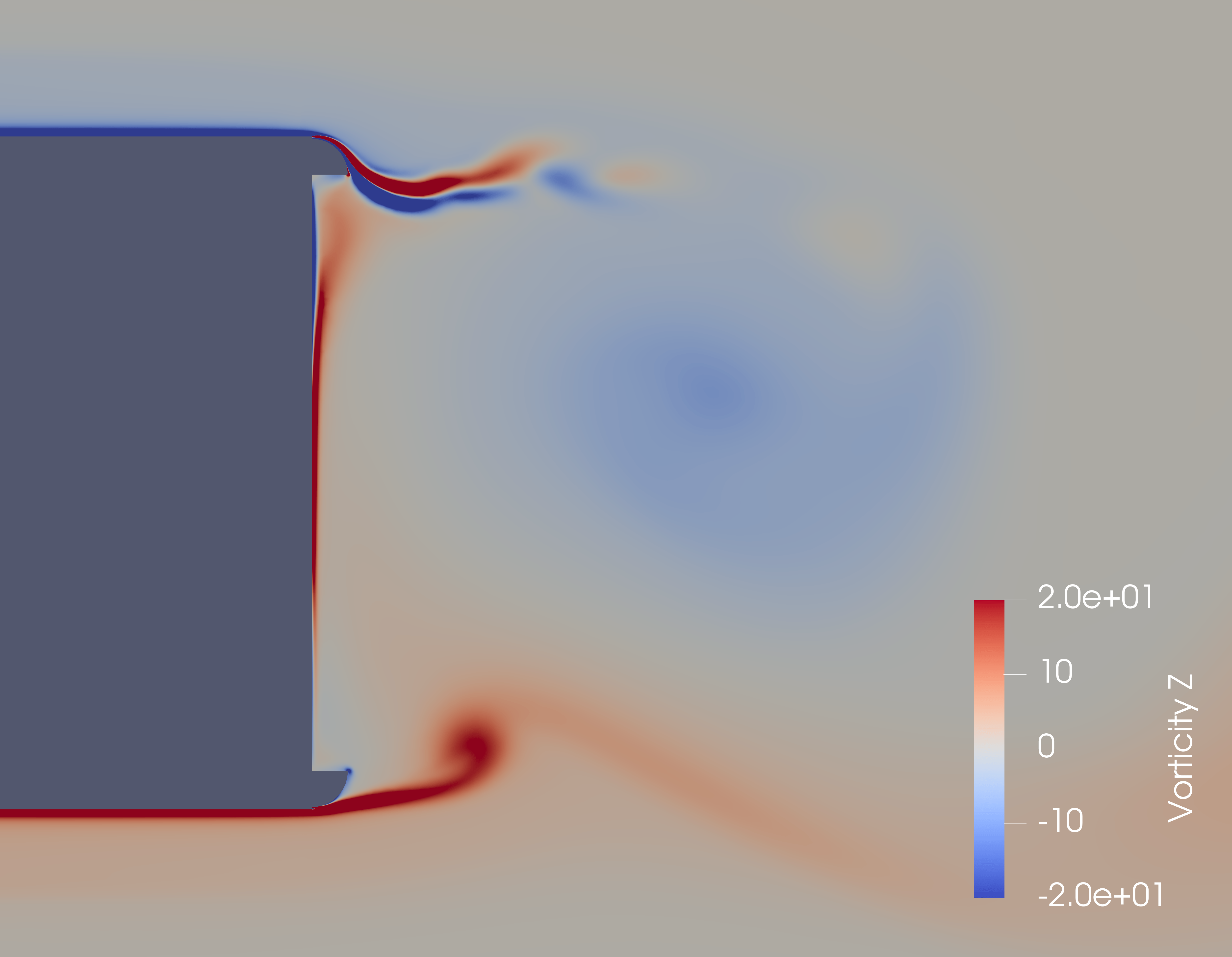}\hfill
	\includegraphics[width=.32\textwidth]{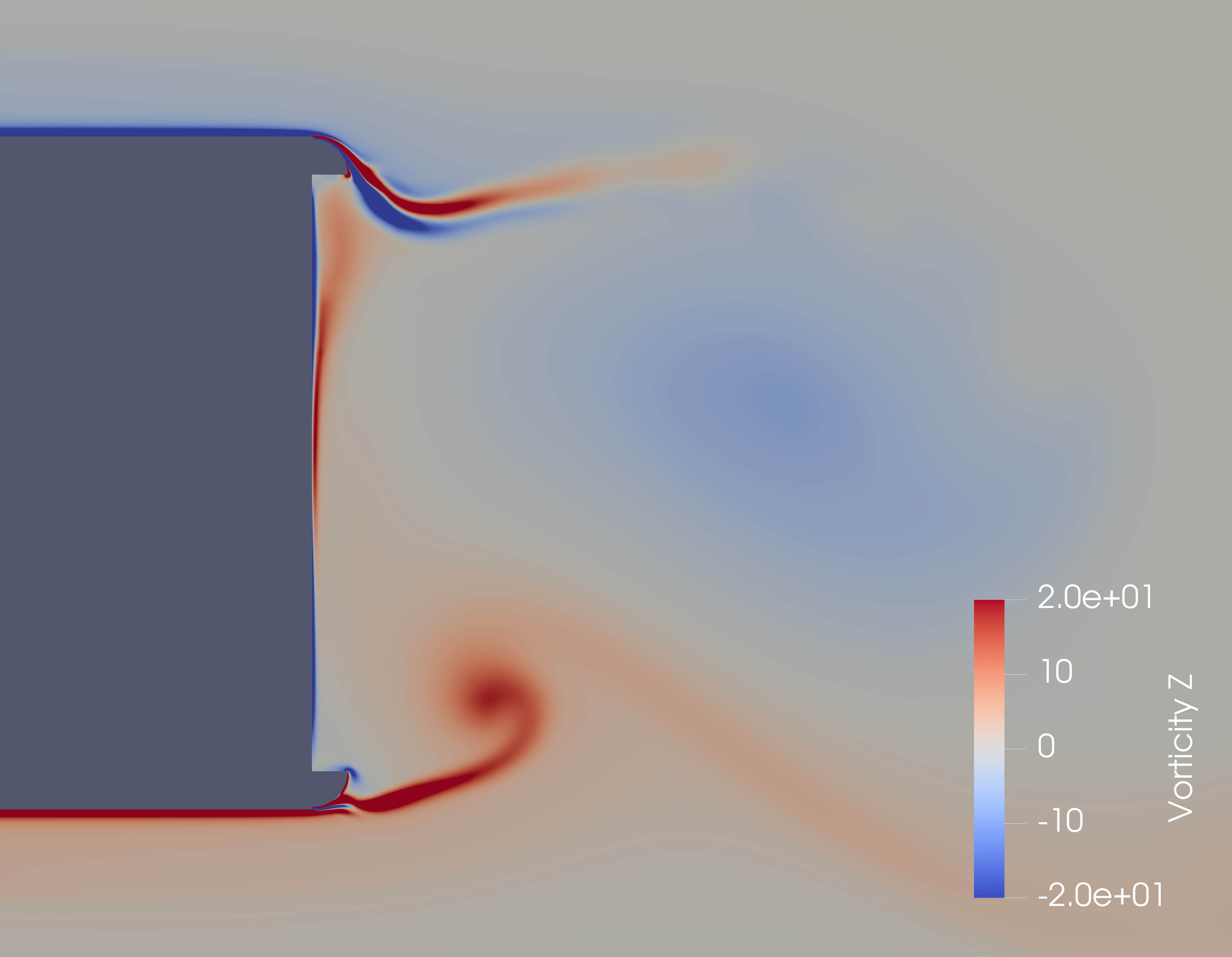}\hfill
	\includegraphics[width=.32\textwidth]{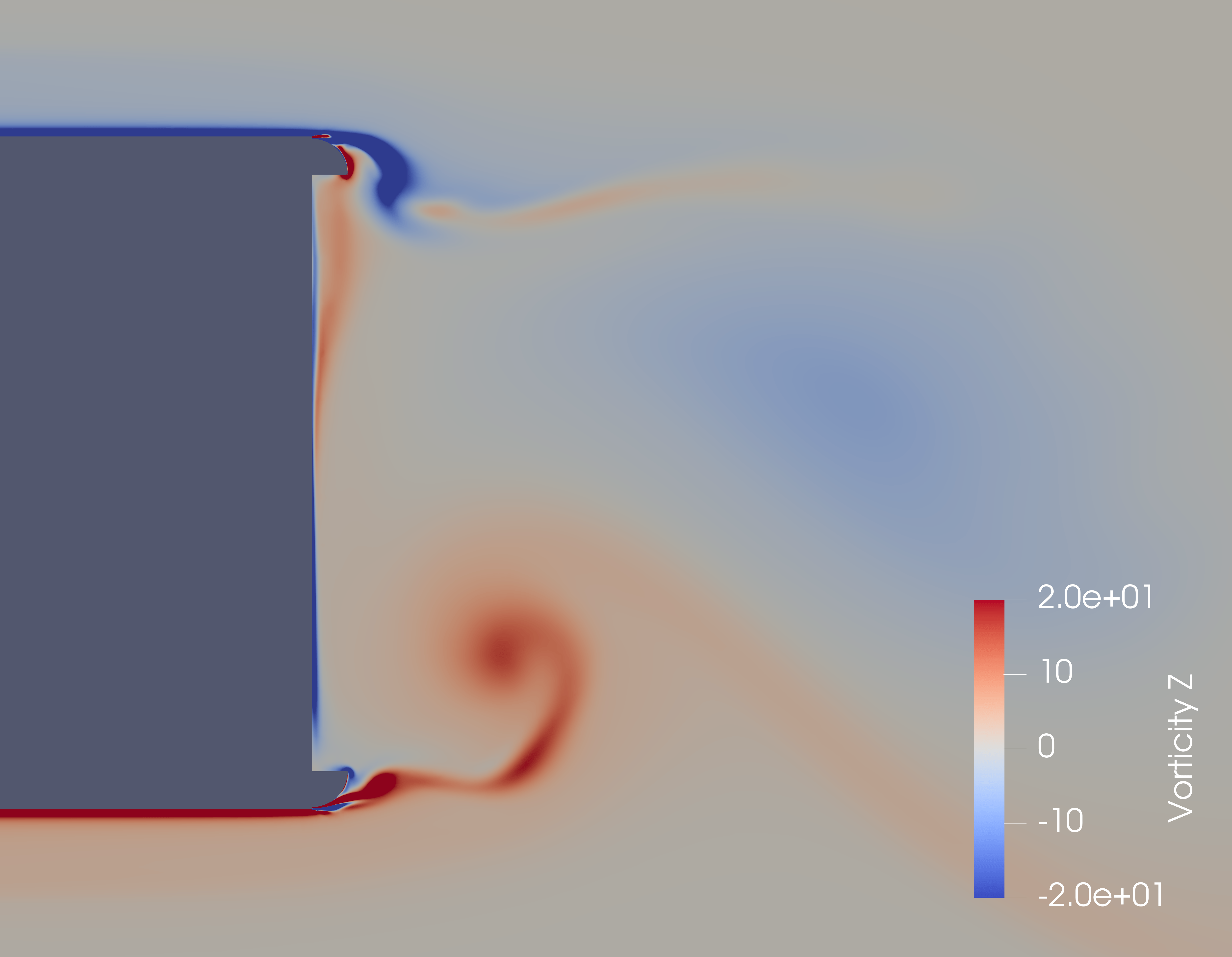}\hfill
	\includegraphics[width=.32\textwidth]{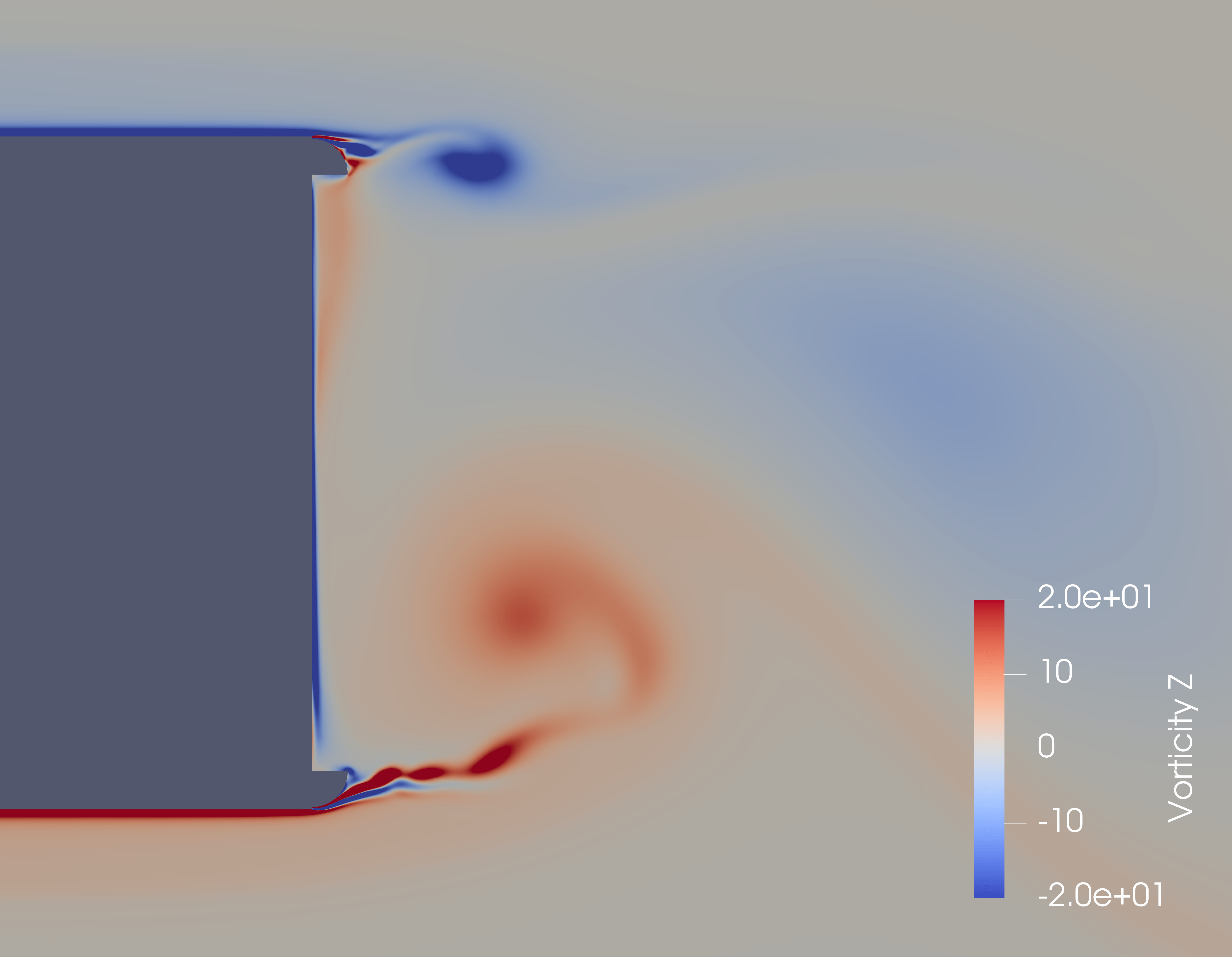}\hfill
	\includegraphics[width=.32\textwidth]{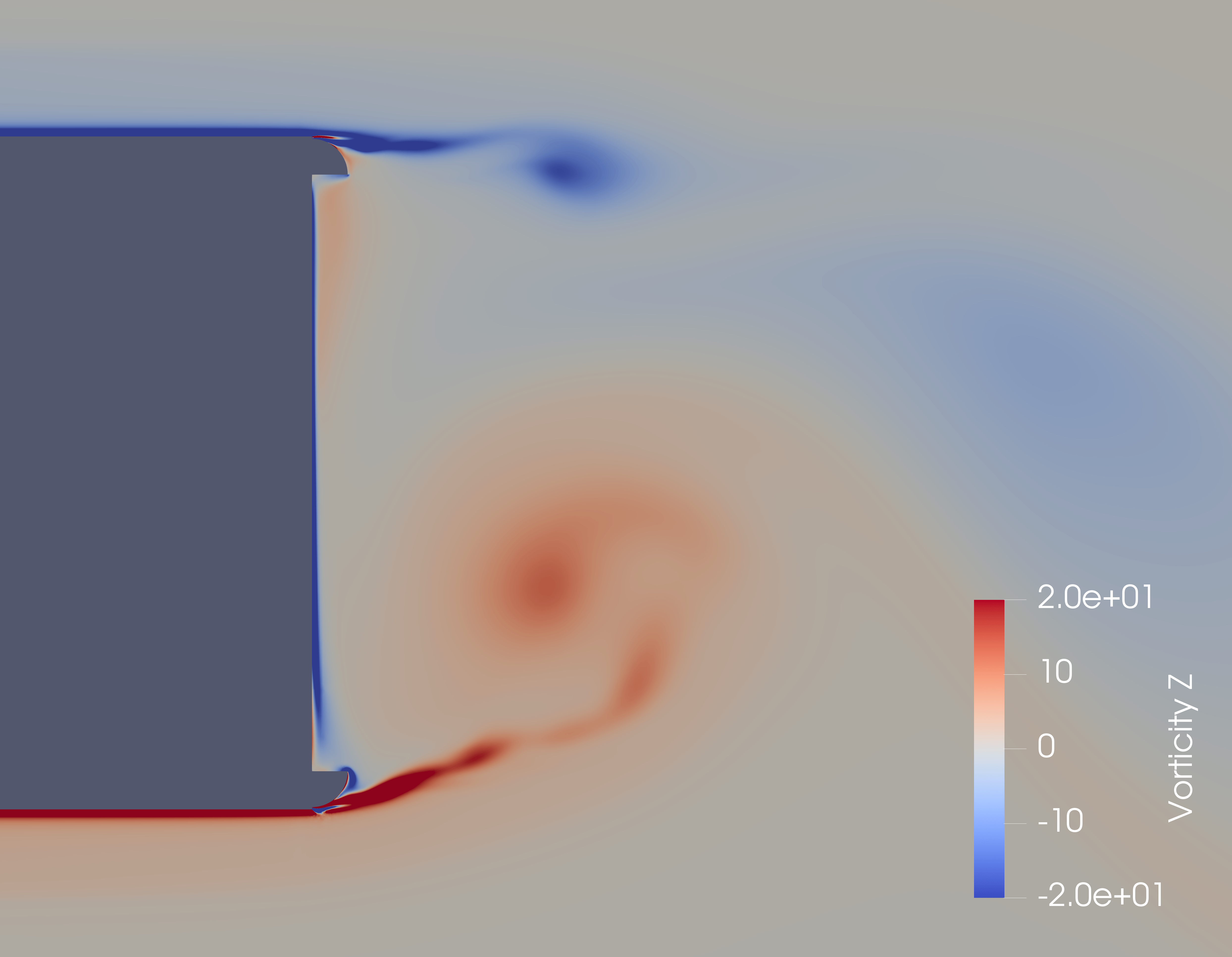}\hfill
	\includegraphics[width=.32\textwidth]{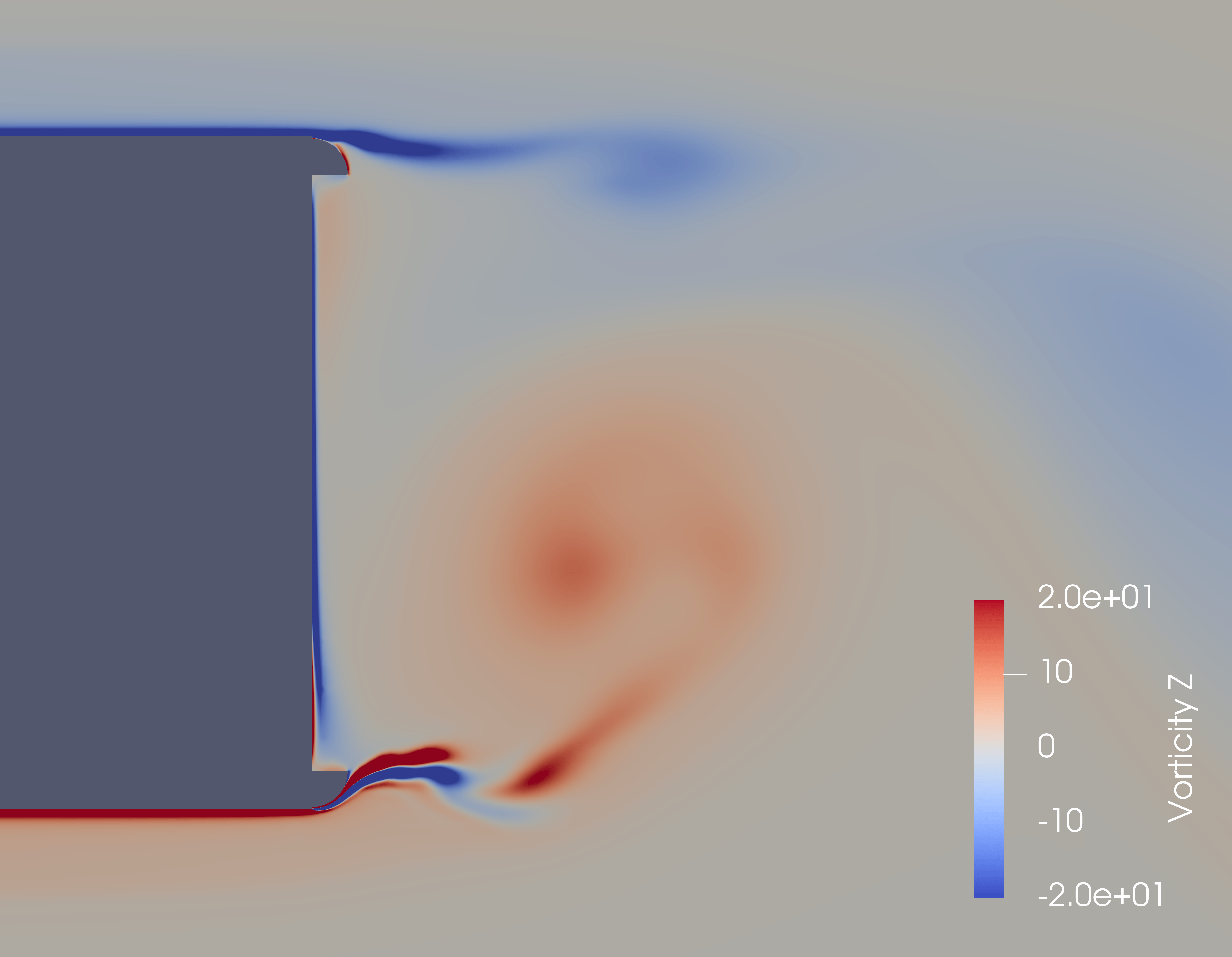}\hfill
	\includegraphics[width=.32\textwidth]{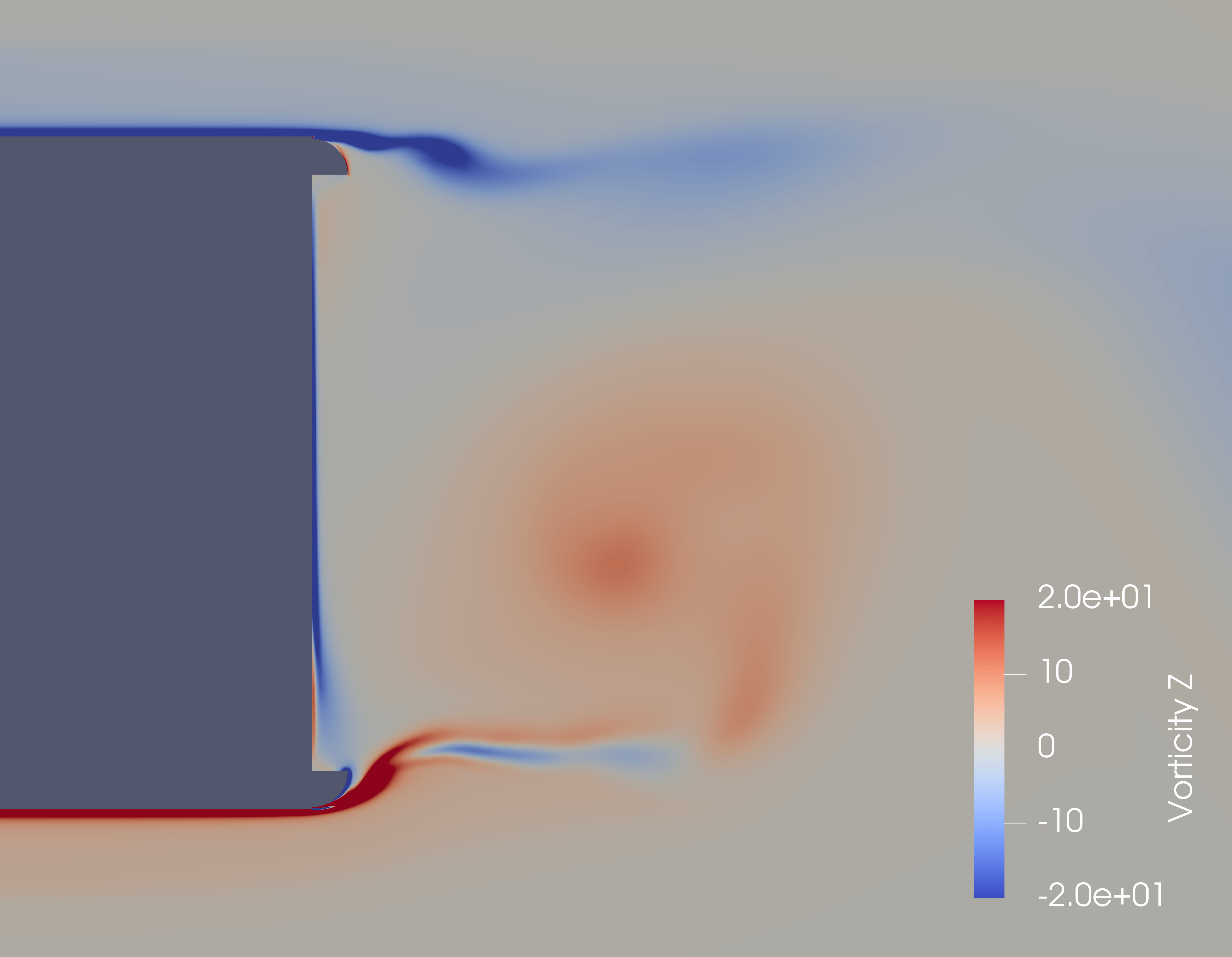}\hfill
	\includegraphics[width=.32\textwidth]{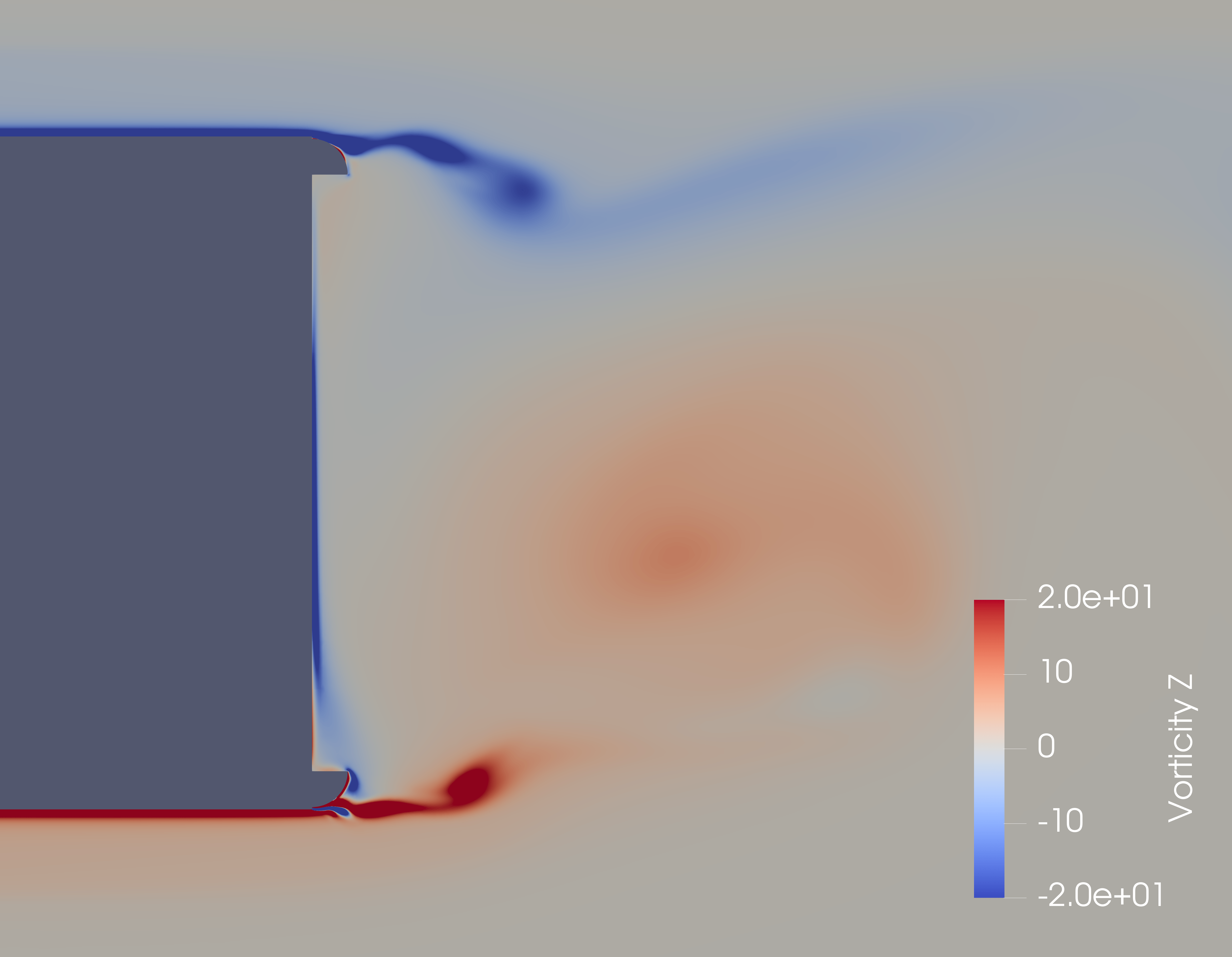}\hfill
	\caption{Time series of vorticity behind the body. The vortexes are strongly influenced by the fluid injection.}
	\label{fig:vort_controlled}
\end{figure*}

\begin{figure*}[h]
	\includegraphics[width=.32\textwidth]{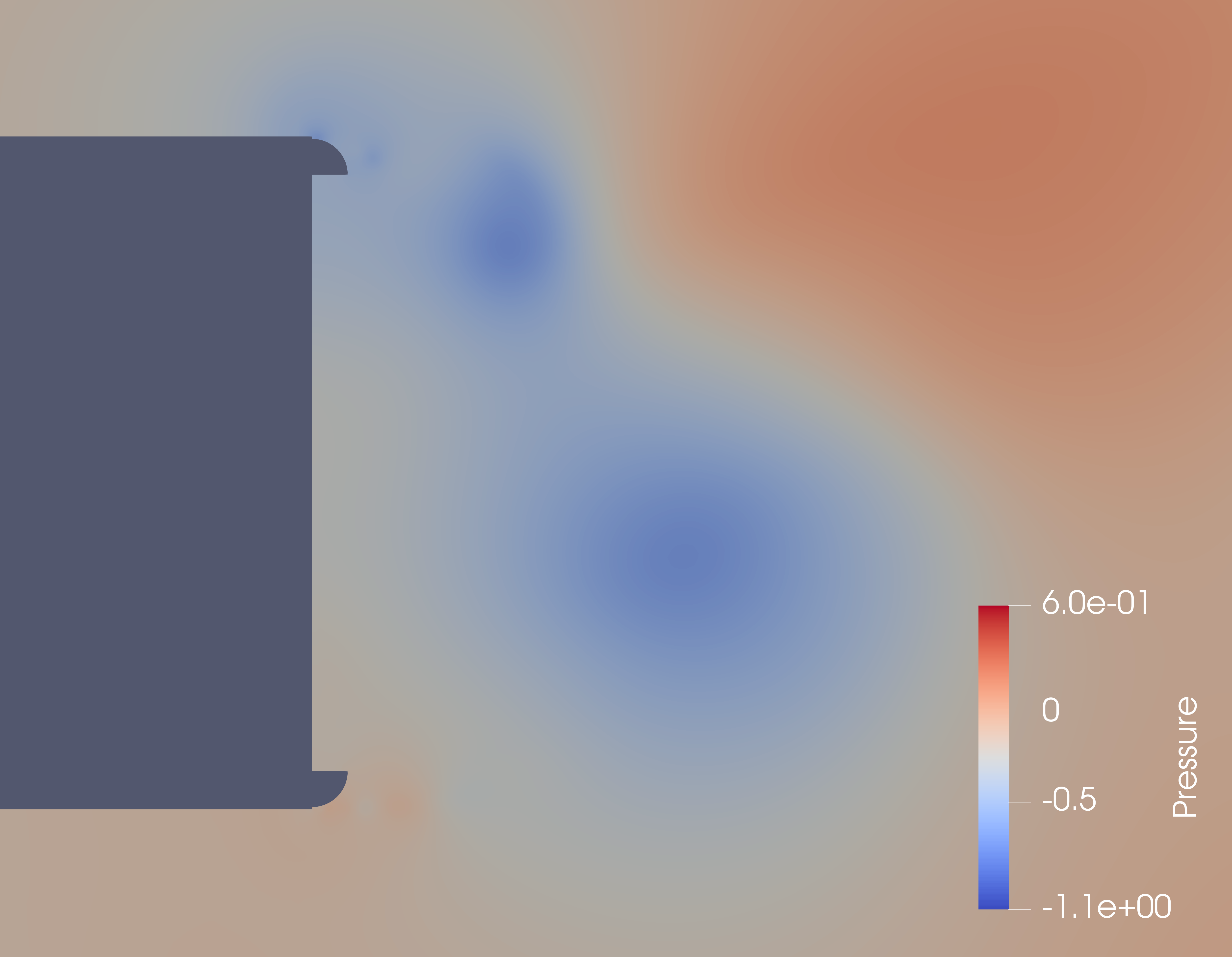}\hfill
	\includegraphics[width=.32\textwidth]{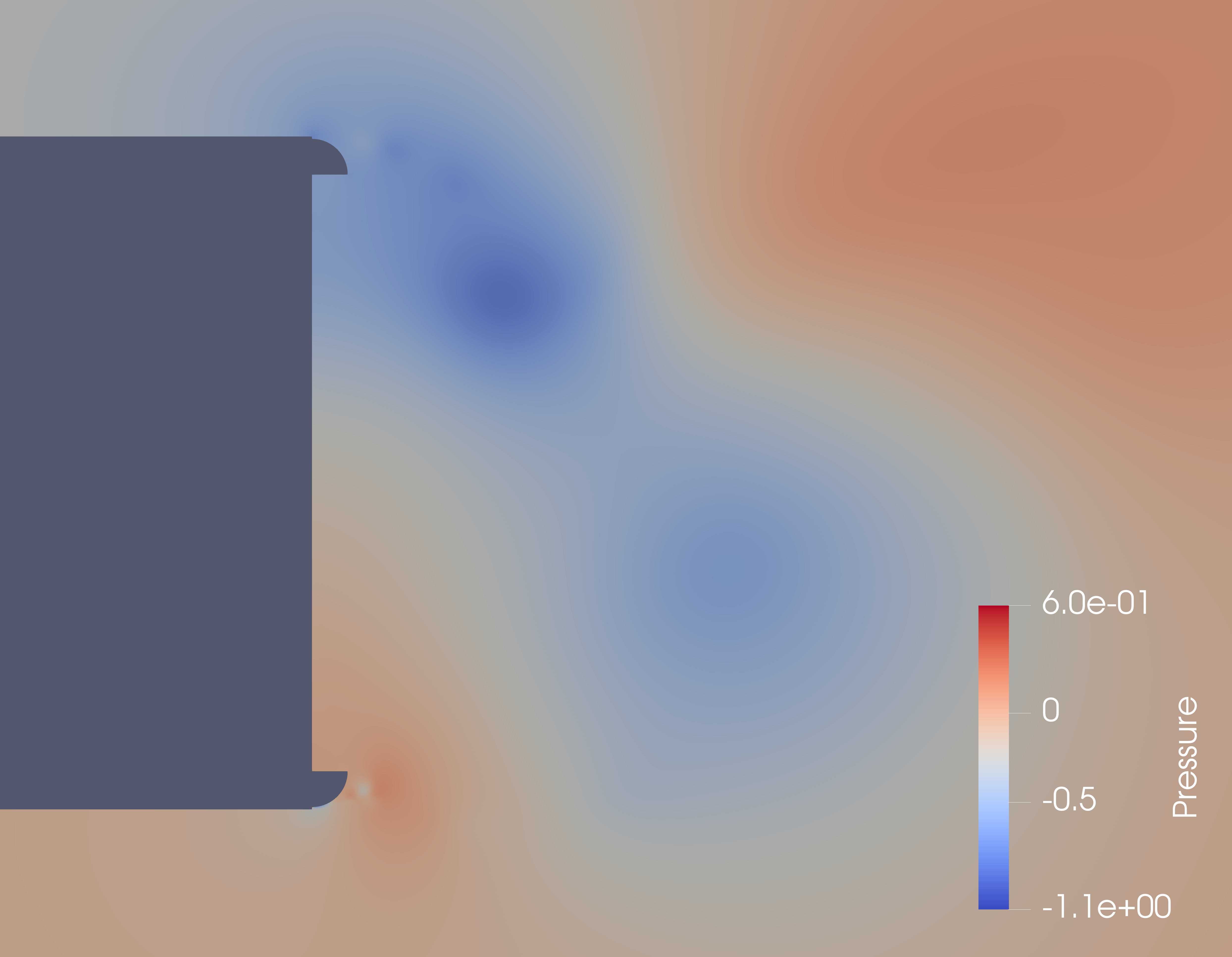}\hfill
	\includegraphics[width=.32\textwidth]{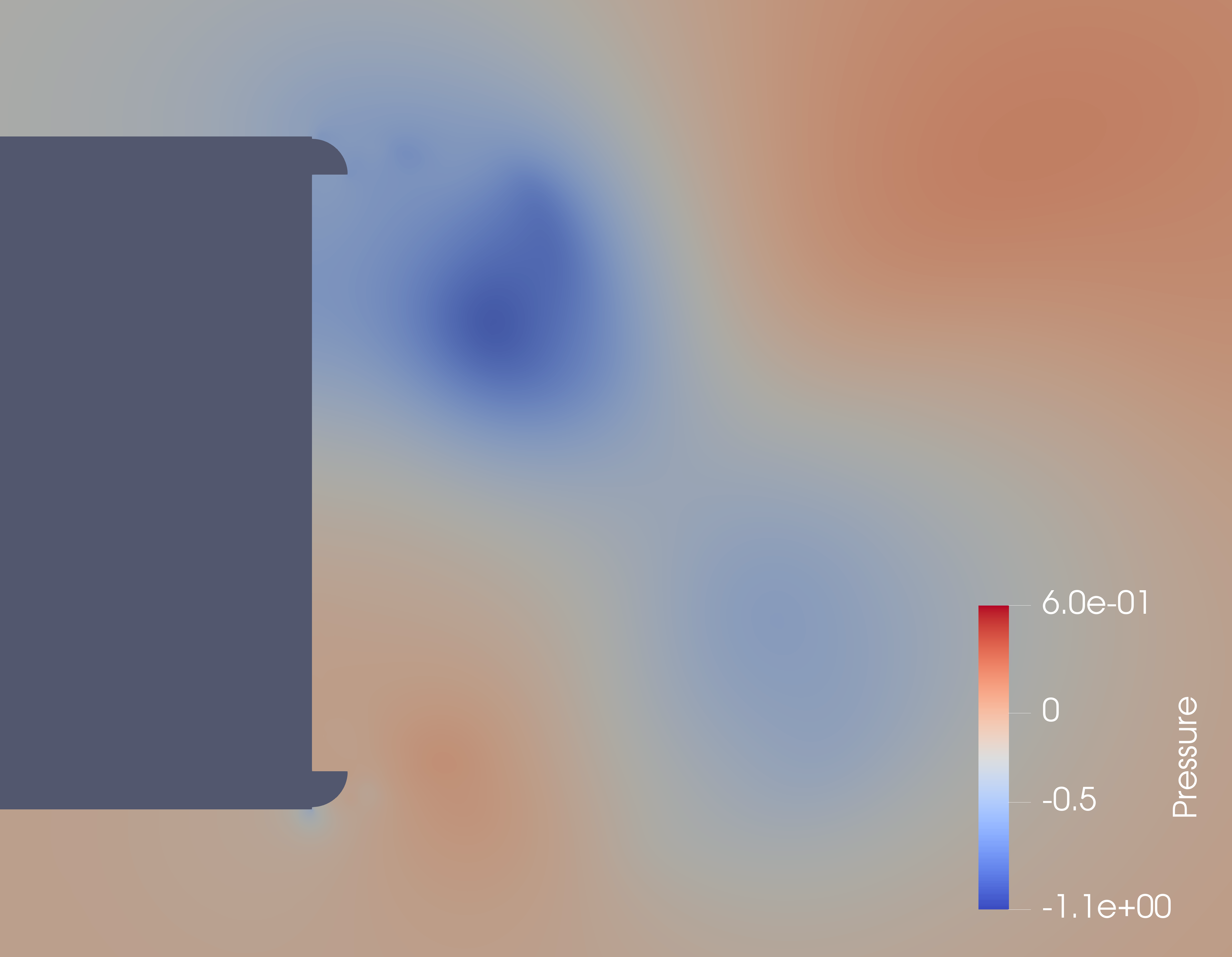}\hfill
	\includegraphics[width=.32\textwidth]{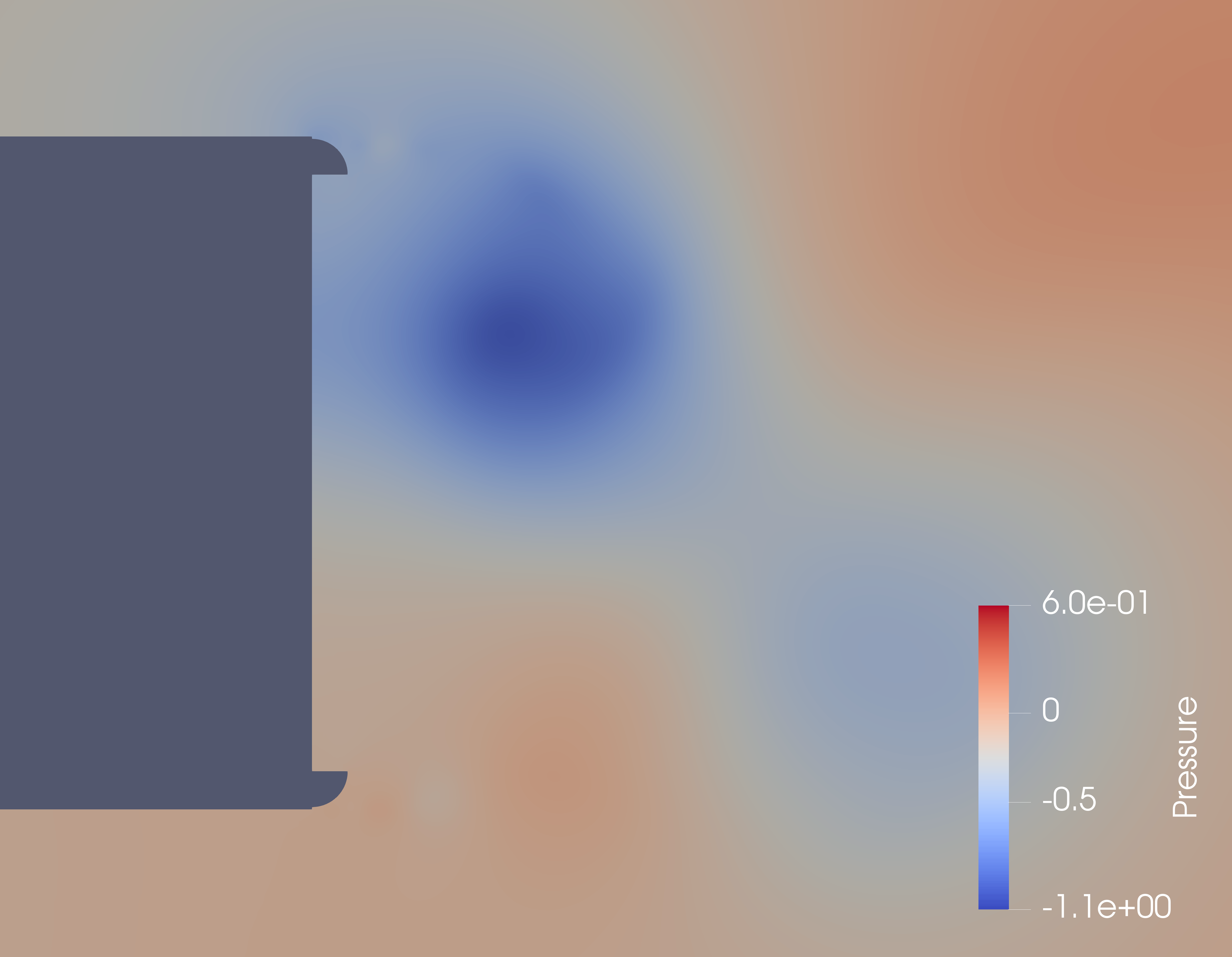}\hfill
	\includegraphics[width=.32\textwidth]{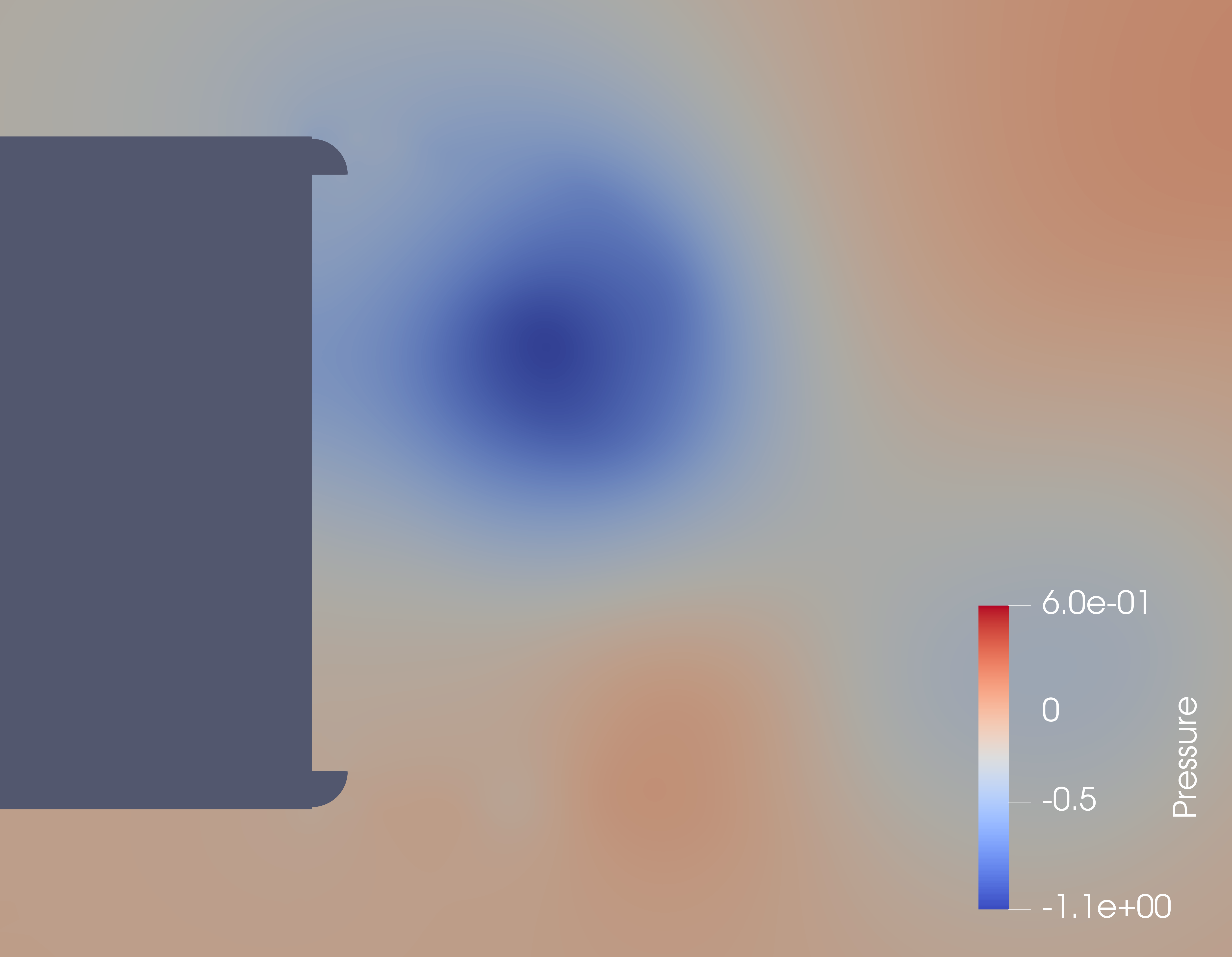}\hfill
	\includegraphics[width=.32\textwidth]{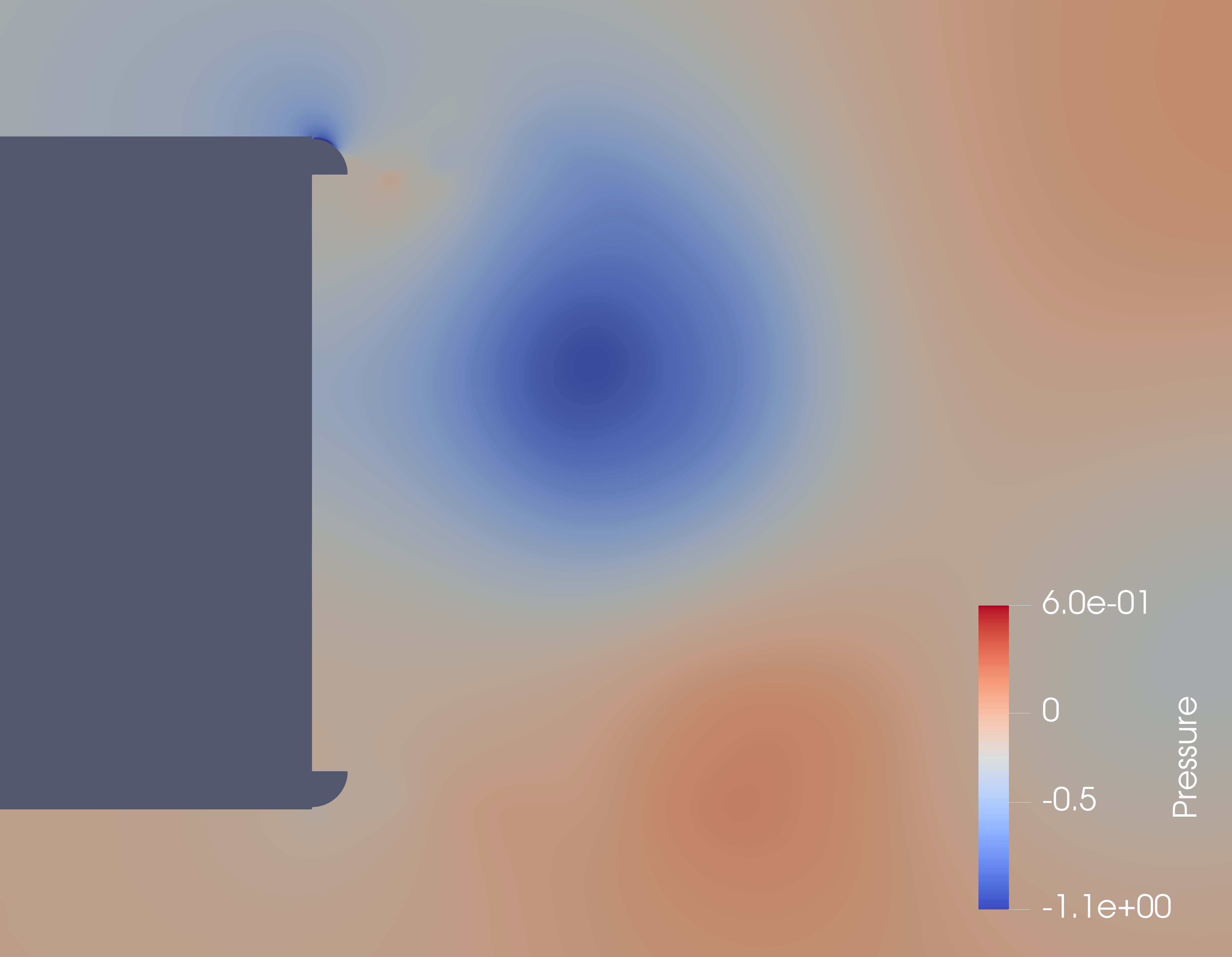}\hfill
	\includegraphics[width=.32\textwidth]{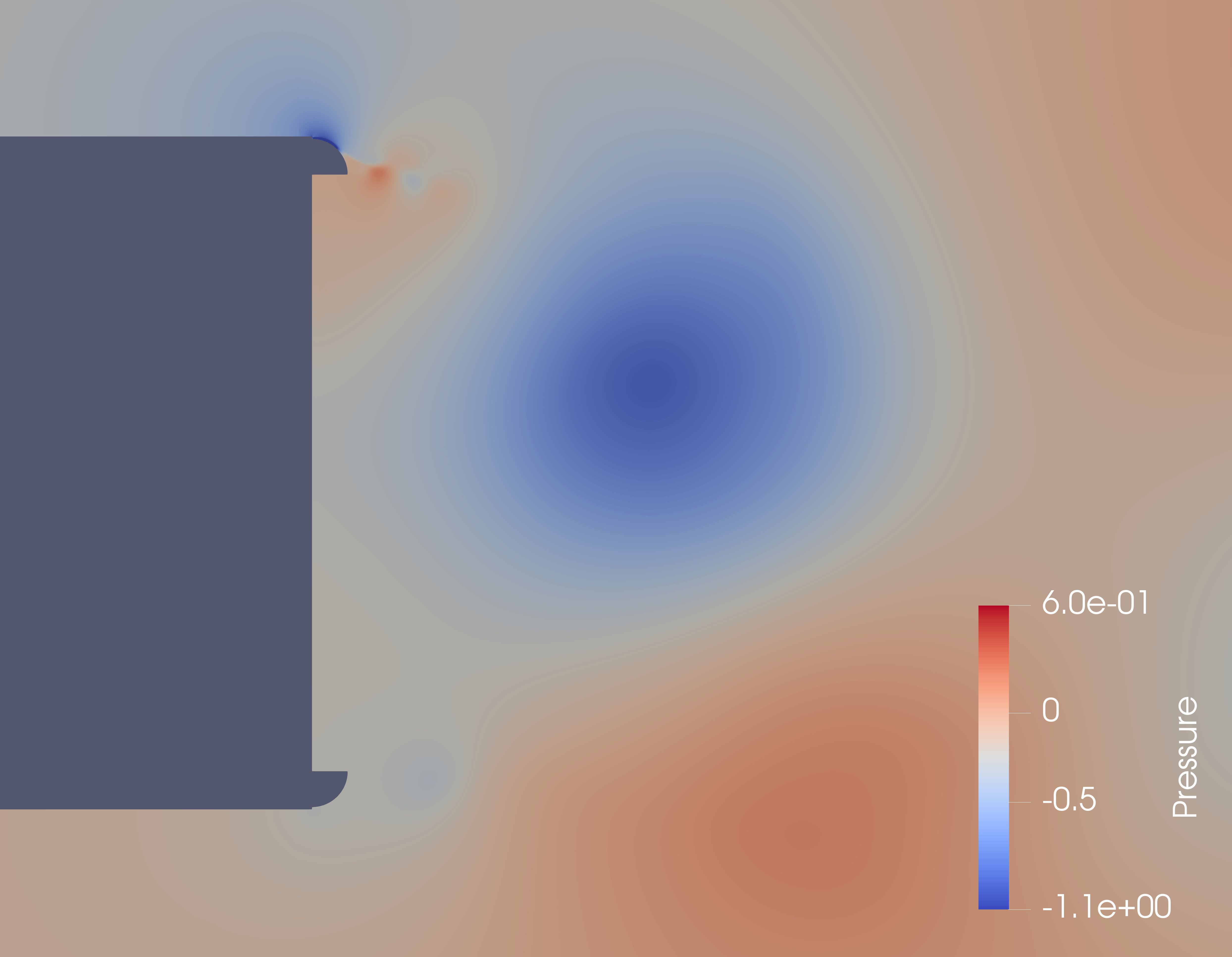}\hfill
	\includegraphics[width=.32\textwidth]{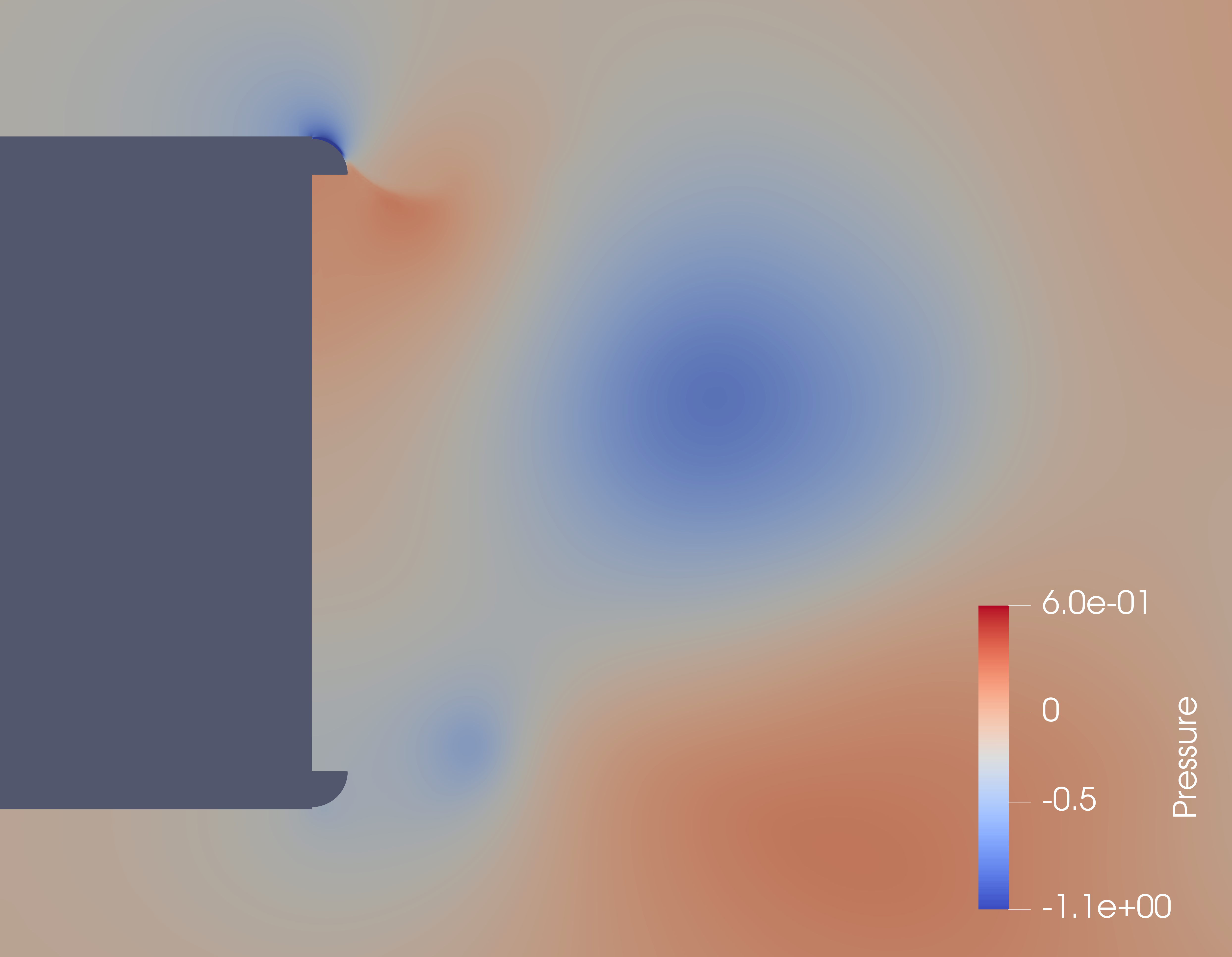}\hfill
	\includegraphics[width=.32\textwidth]{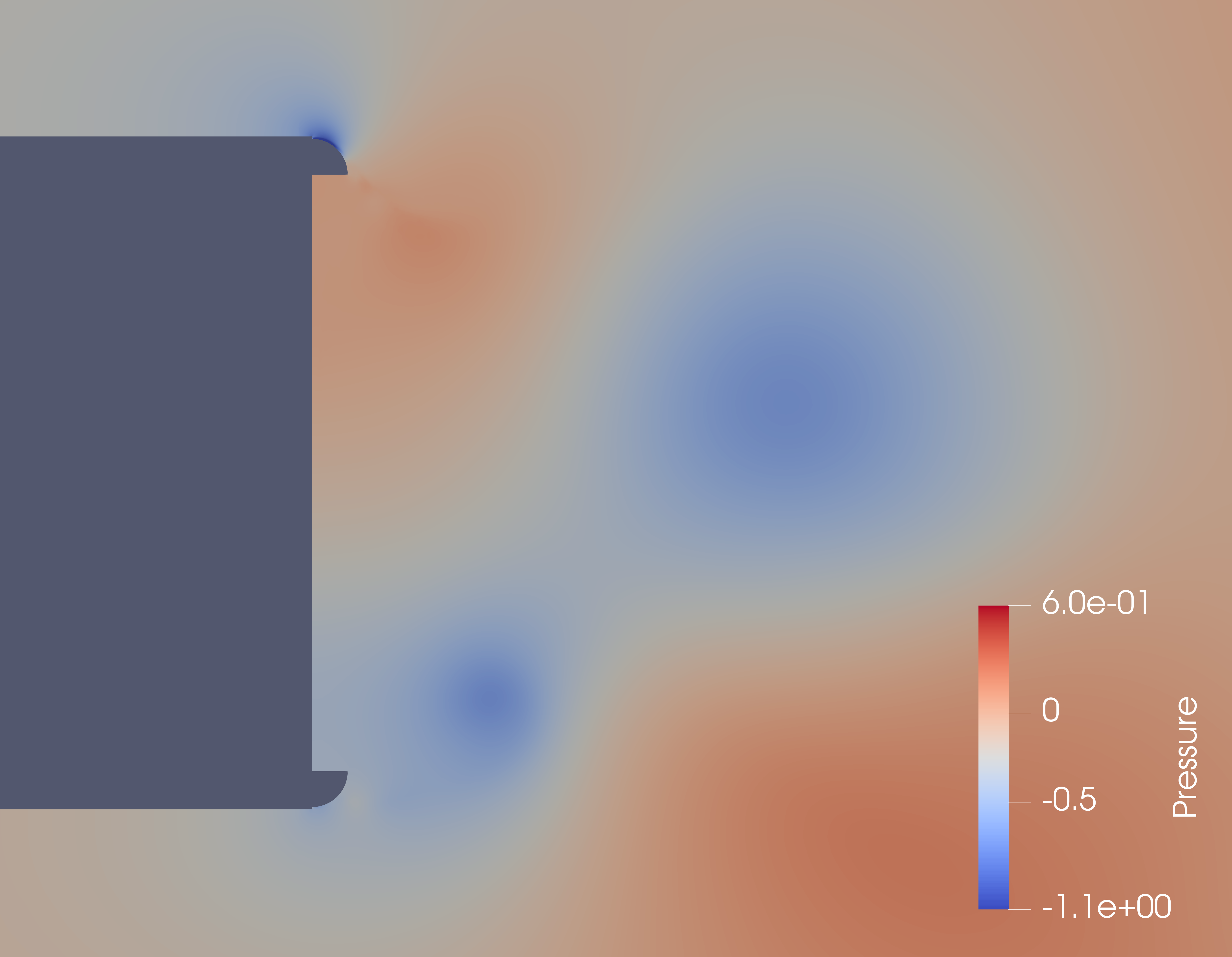}\hfill
	\includegraphics[width=.32\textwidth]{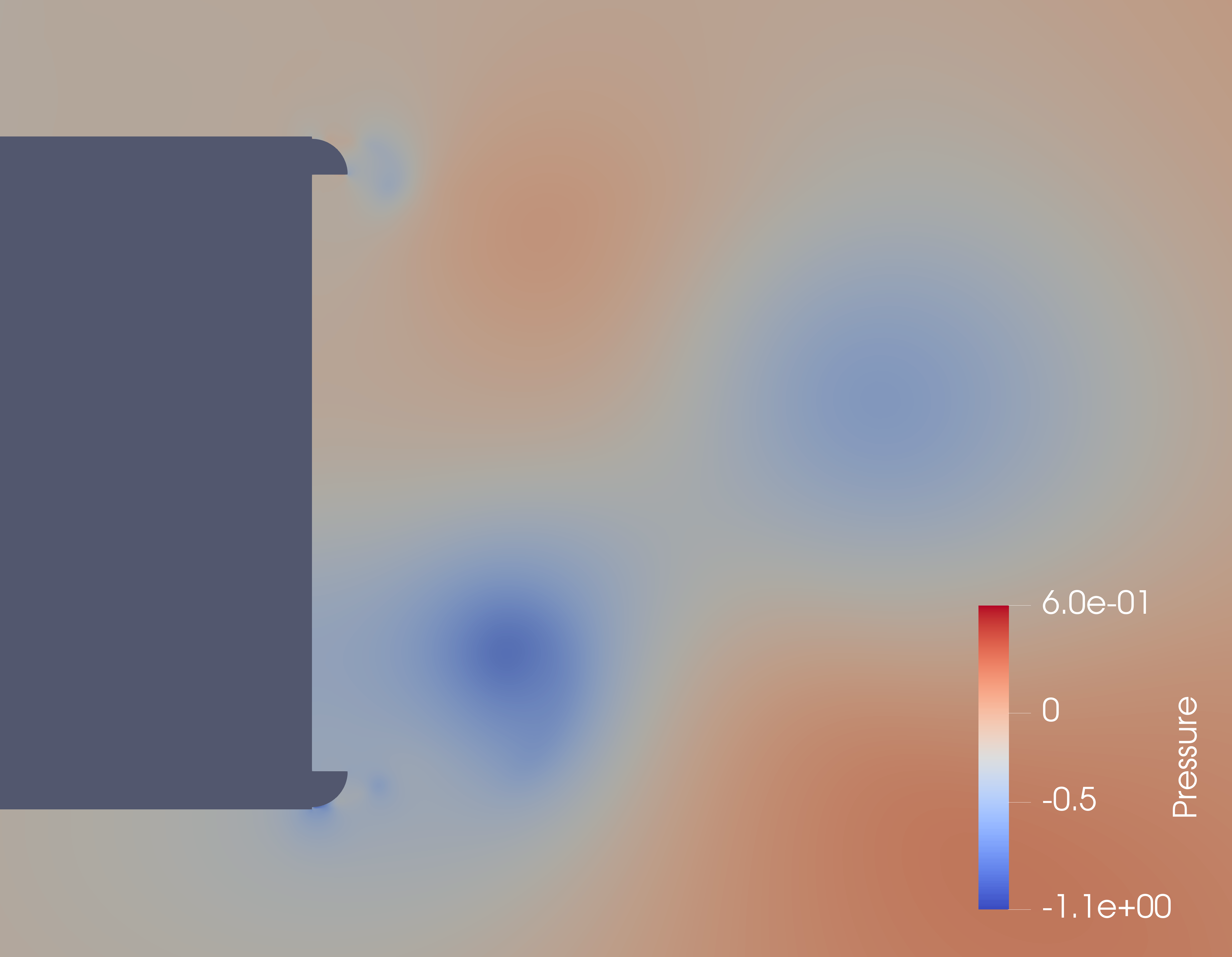}\hfill
	\includegraphics[width=.32\textwidth]{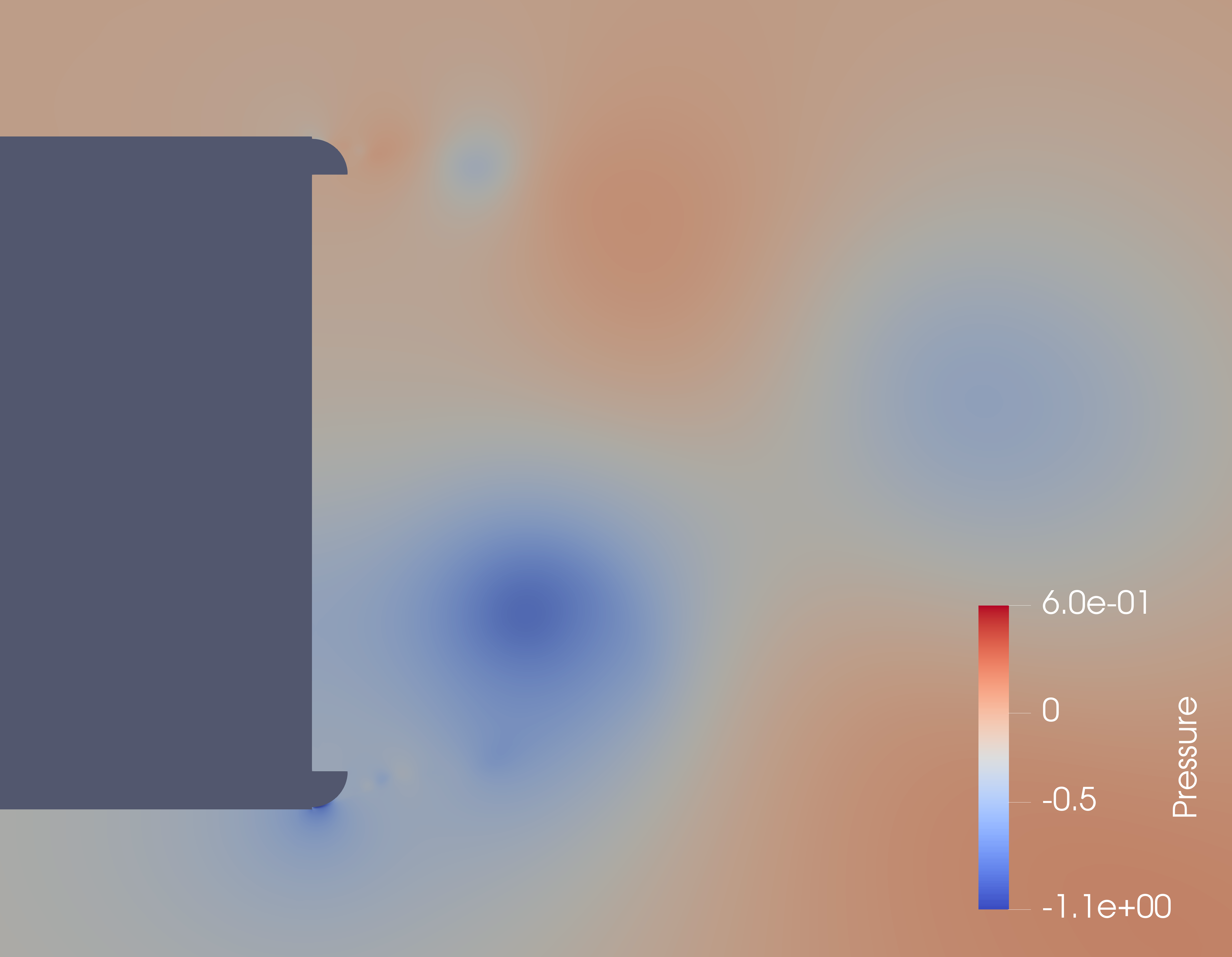}\hfill
	\includegraphics[width=.32\textwidth]{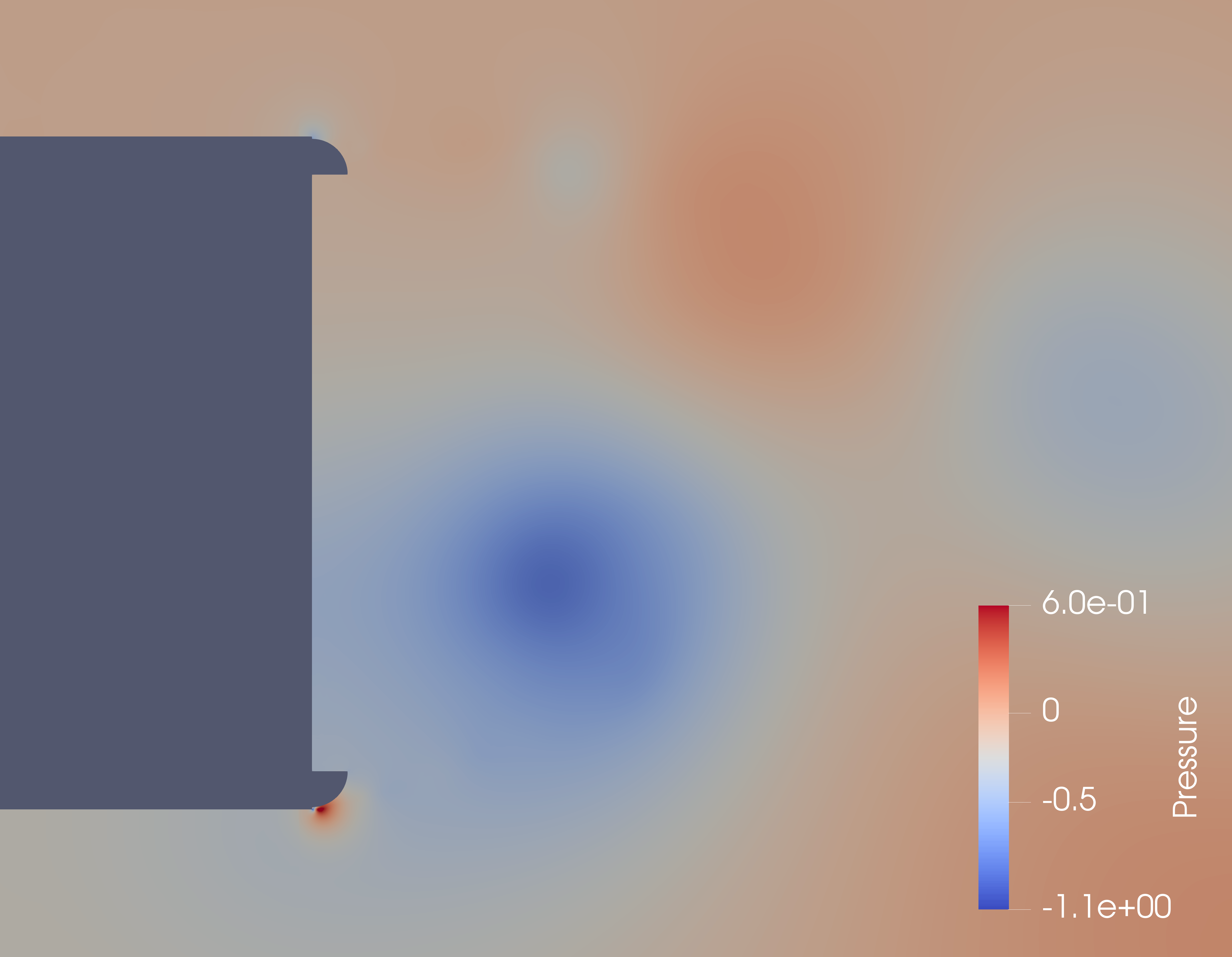}\hfill
	\includegraphics[width=.32\textwidth]{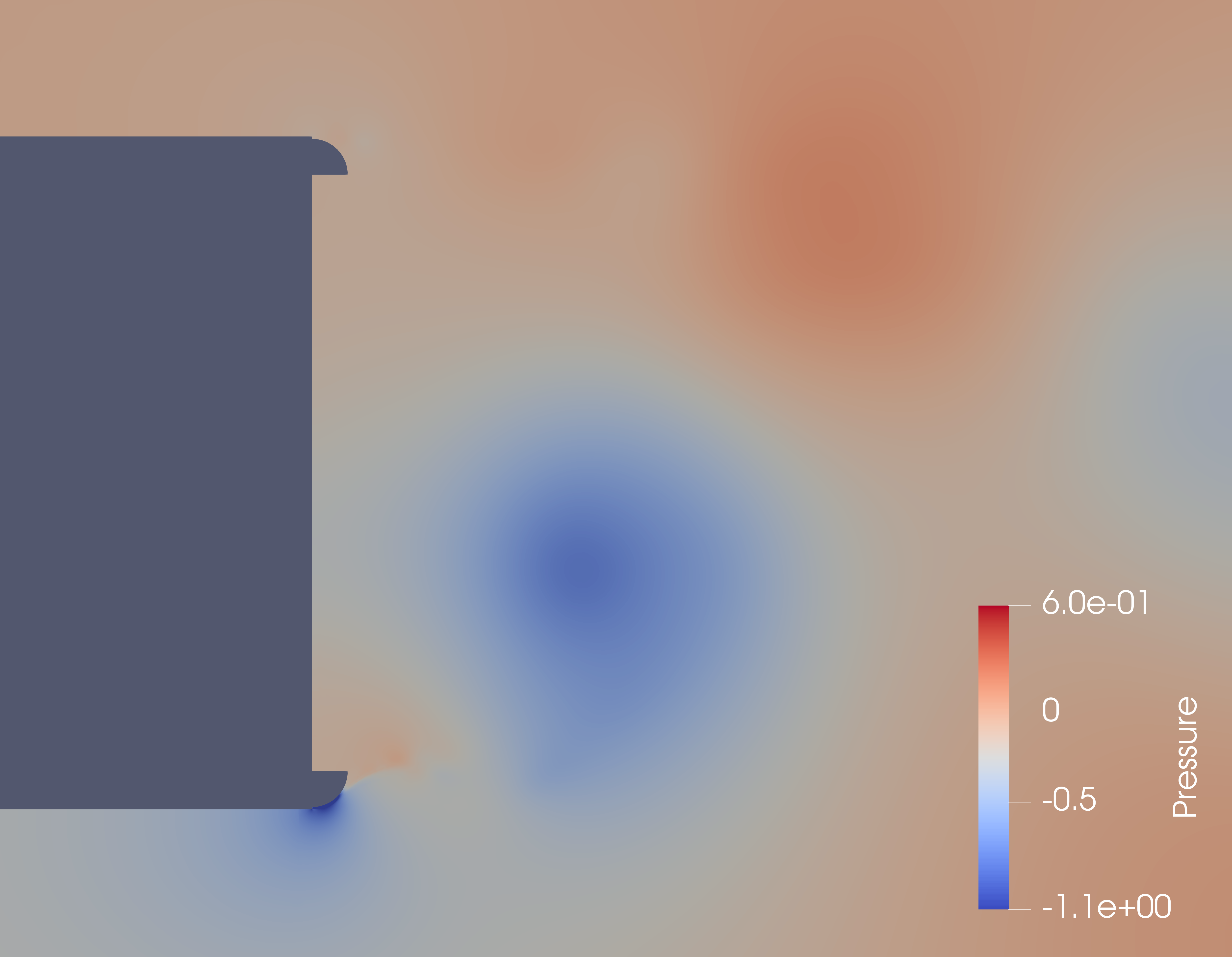}\hfill
	\includegraphics[width=.32\textwidth]{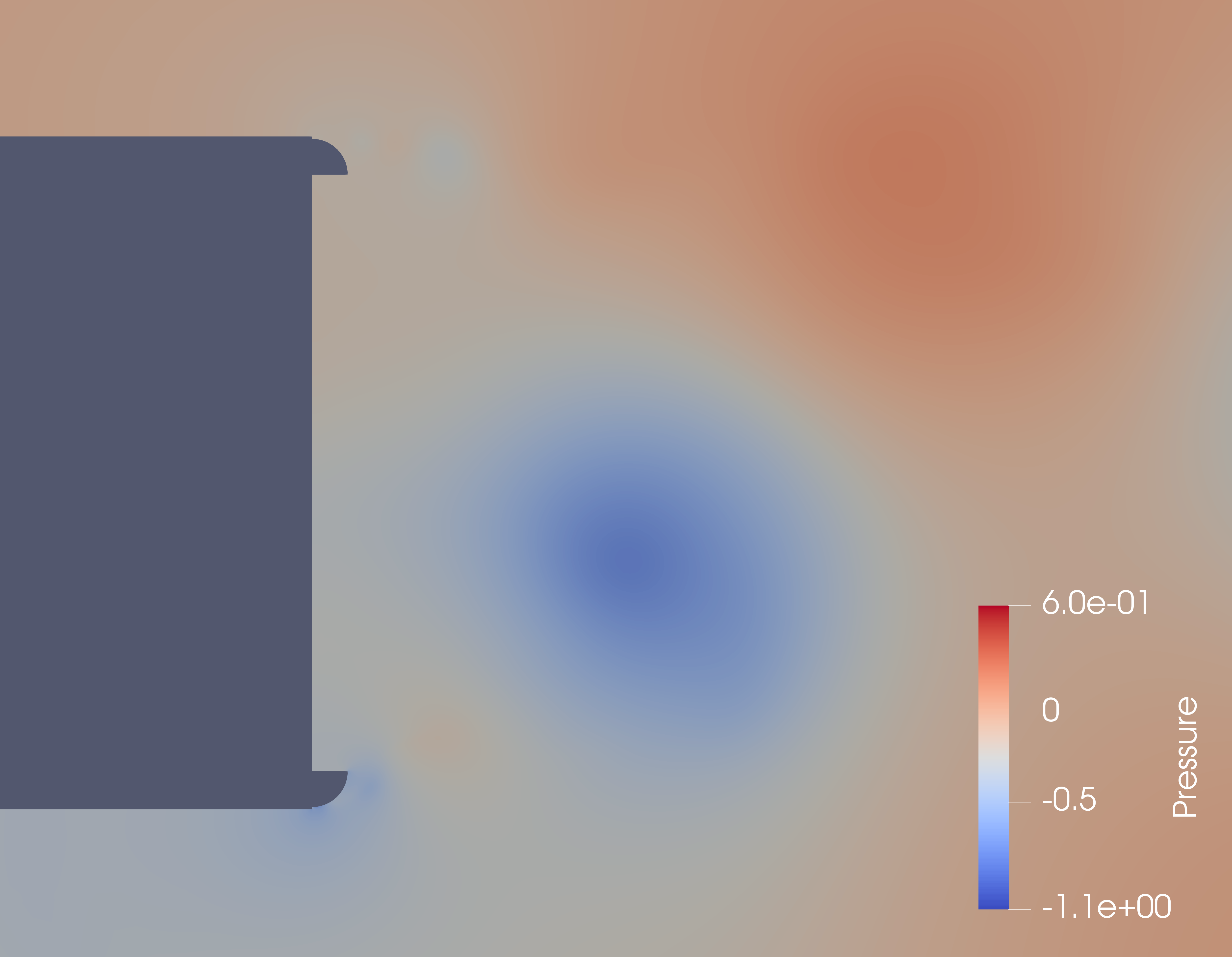}\hfill
	\includegraphics[width=.32\textwidth]{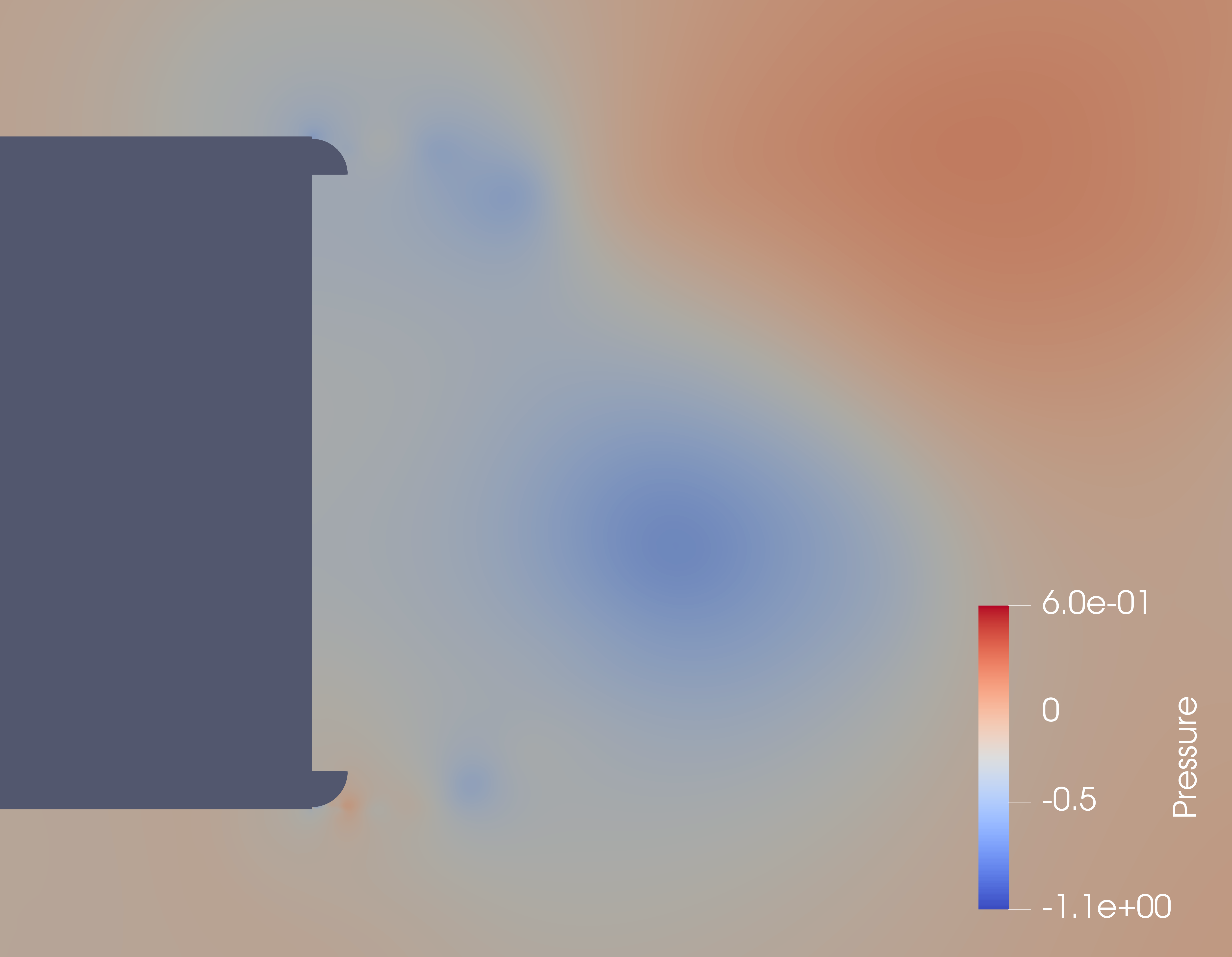}\hfill
	\caption{Time series of pressure behind the body. Thanks to the fluid injection, the vortexes cores are weaker and further than in the clean bluff body.}
	\label{fig:pres_controlled}
\end{figure*}

According to our interpretation, the dynamics governing the jets-vortexes interaction is the following.
The vortexes have low pressure in the core and, in the clean experiment are close to the rear surface as shown in Fig.~\ref{fig:p_vortex} and Fig.~\ref{fig:vort_clean}.
Essentially, the jets have two main effects: firstly, they push downstream the vortexes creation, promoting their rapid motion downstream, and leaving as a consequence a higher pressure near the wall. Indeed, Fig.~\ref{fig:p_vortex}(b) shows the distance from the wall at which the vortex is created and the high-pressure inter-space between them. Secondly, they reduce the vortexes intensity, as can be inferred from Figs~\ref{fig:p_vortex} - ~\ref{fig:pres_controlled}.

We shall consider the Figs~\ref{fig:vort_clean}, which are few frames taken inside a period, looking at the vorticity to better appreciate the phenomena. Before the lower vortex is fully developed, the upper jet turns on, thus lowering the potential flow and pushing away the vortex in incipient creation. After its activation, an upper vortex arises and the lower jet turns on. Towards the end of the cycle, there are two vortexes not distinctly separated along the $x$ coordinate, but after few instants, the positive vortex moves away vanishing and the negative one takes place recreating the initial situation.
Moreover, this mechanism reduces the vertical oscillations of the wake (similar results were obtained in \cite{longa_2017}) and makes the vortexes less intense and more irregular. Even the Strouhal number of vortex shedding is affected, decreasing from $0.24$ to $0.23$.

%

% \clearpage %

% vorticity clean:

% \clearpage %

% pressure clean:

% \clearpage %

% vorticity controlled:

% pressure controlled:

\clearpage %

% ================================================
% CONCLUSIONS =====================================
% ================================================
\section{\label{sec:conclusion}Conclusions} 

% really short summary of the work done:
In this work, we have applied reinforcement learning to the fluid dynamics problem of drag reduction of a bluff body. To achieve this, we focused on controlling two rear jets.

% results obtained:
The reinforcement learning proved to be effective also in this fluid dynamics context. We have managed to train a neural network so that it interacts correctly with the environment and in particular with the wake vortexes, despite the absence of a model of the environment but having only a poor vision of its response to the actions taken (data-driven approach). This interaction involves pushing downstream the vortex formation point and consequently removing the low-pressure region caused by the vortex core. In fact, the base pressure is mainly responsible for the drag reduction. Considering the overall system, i.e., the sum of the power of the drag force and the power expended by the controller, we have obtained a considerable saving of 40\% compared to the uncontrolled case. 
% It was not possible to demonstrate the optimality of this control. However, the non-symmetrical interaction with vortexes suggests the possibility of further improvement.

% future developments:
The results obtained amply justify the continuation of the research in this direction. Some research fields may be:\\
- Further improvement of the performances, with simultaneous optimization of geometry and control. In fact, during the jet-vortex interaction, the shape of the jet is also important, which is associated with the shape of the curved edge. Its optimization would improve the effectiveness of the jet. This simultaneous optimization should be performed within the training phase of the reinforcement learning, and the process could be enriched with the information deriving from the adjoint problem.\\
- In-depth study of the optimal control obtained and comparison of the results with other control methods to better define the potential of the reinforcement learning.\\
- Study on the memory of the neural network deepening the usefulness of recurrent neural network or multiple memory levels, i.e., saving also the antecedent states to the previous one.\\
- Test of control robustness and adaptation to other boundary conditions.\\
- Increased realism and accuracy of simulation and/or wind tunnel experiments, analyzing a more realistic geometry and possibly conforming to a practical case.

%- Indagine sulla specificazione dei concetti esenziali del RL, ossia scelta dello stato, delle azione e reward shaping. Questo studio andrebbe ad indagare l'osservabilita e la contro di ... per ottenere un migliramento della scelta delle osservazioni.\\

\section{Acknowledgments}
S. Micheletti thanks the PRIN research grant n. 20204LN5N5, {\it Advanced
	Polyhedral Discretisations of Heterogeneous PDEs for Multiphysics Problems} and the
INdAM-GNCS 2022 Project, {\it Metodi di riduzione computazionale per le scienze applicate: focus su sistemi complessi}.

\section{Conflict of interest statement}
The authors have no conflicts to disclose.

% ================================================
% APPENDICES =====================================
% ================================================
\appendix
\section{\label{app:cfd_settings}CFD settings}

The constants used in the $k-\omega$ SST turbulence model, listed inf Table~\ref{tab:coeffs}, are found in \cite{nasa_sst}.
\begin{table}[htb]
	\caption{\label{tab:coeffs}Constants used for the SST turbulence model.}
	\centering
	%\begin{ruledtabular}
		\begin{tabular}{lr}
			coefficient & value \\
			\hline \\
			$\sigma_{k1}$         & 0.85 \\
			$\sigma_{k2}$         & 1 \\
			$\sigma_{\omega1}$  & 0.5 \\
			$\sigma_{\omega2}$   & 0.856 \\
			$\beta_1$           & 0.075 \\
			$\beta_2$           & 0.0828 \\
			$\gamma_1$          & $\frac{\beta_1}{C_\mu} - \frac{\sigma_{\omega1}\kappa^2}{\sqrt{C_\mu}}$ \\
			$\gamma_2$          & $\frac{\beta_2}{C_\mu} - \frac{\sigma_{\omega2}\kappa^2}{\sqrt{C_\mu}}$ \\
			$C_\mu$         & 0.09 \\
			$\kappa$        & 0.41 \\
			$a_1$          & 0.31 \\
		\end{tabular}
	%\end{ruledtabular}
\end{table}

The fluid dynamics equations are solved with a Finite Volume Method (FVM) set with the following main characteristics:

- The convective flux \cite{SU2_2016} is evaluated using the Flux Difference Splitting method with a MUSCL \cite{leerV_1979} second order reconstruction.

- The turbulent convective flux \cite{SU2_2016} is solved without the second-order reconstruction because of strong instabilities.

- Venkatakrishnan slope limiter \cite{venka_1993} is applied.

- The derivatives of the velocity field in the viscous flux \cite{blazek_2015, SU2_2013, SU2_2014, SU2_2016} are retrieved from the Green-Gauss theorem.

- The time integration is performed using the first-order implicit Euler scheme.

- The linear system that arises after spatial discretization \cite{SU2_2016} is solved through BiCGSTAB with ILU preconditioner. The accuracy is set such that the error on the meaningful quantities (such as the drag coefficient) are accurate to within the fourth digit. Decreasing this value would imply an increase of the computation time without an appreciable variation in the results.

\section{\label{app:hardware}Hardware}
The following is a list of the hardware used:
\begin{itemize}
	% 	\item Personal cluster in beowulf style made of two heterogeneous computers connected by a Cat5e crossover cable.\\ % The process manager is Hydra \ref{hydra}
	% 	- PC 1: i7-4770 @3.40GHz, RAM 16GB 1600MHz (overclocked to 1866MHz) latencies 9-9-9-24\\
	% 	- PC 2: i7-3630QM @2.40GHz, RAM 16GB 1600MHz latencies 11-11-11\\
	% 	The performance of this hardware was not sufficient and therefore, the HPC resources of MOX - Dipartimento di Matematica of Politecnico di Milano were subsequently used. There is no advantage in distributing the work on GPUs, as usually done for training neural networks, because the CFD is the heaviest computation.
	\item Gigat (MOX - Dipartimento di Matematica of Politecnico di Milano): Xeon E5-2640 v4 @2.4GHz, RAM 64GB per node, 20 cores per node, max 2 nodes
	
	\item Gigatlong (MOX - Dipartimento di Matematica of Politecnico di Milano): Xeon E5-4610 v2 @2.3GHz, RAM 256GB per node, 32 cores per node, max 1 node at the time of this project.
	
\end{itemize}

\section{\label{app:rl_settings}Reinforcement Learning settings}
The main reinforcement learning settings are listed below:
- Number of rollout workers, that is the number of environment run in parallel, equal to $4$. From each rollout worker, a trajectory fragment of $160$ samples is selected to fill the train batch of size $4 \times 160 = 640$. They are not ordered in a sequence but they are joint shuffled to avoid local overfitting.\\
- The number of epochs to execute per train batch is $30$. \\
- The specific expression of objective function can be found in \cite{schulman_2017}. We use the Generalized Advantage Estimator (GAE) \cite{schulman_2018} as advantage. \\   
- The learning rate is $0.0001$ and the discount factor, $\gamma$, is 0.99.\\
- The initial coefficient for KL divergence is $0.2$, while the target value for KL divergence is $0.01$.\\
% - vf_loss_coeff  (epsilon del paper https://arxiv.org/pdf/1707.06347.pdf ?)
% - vf clip param (inutile)s
The size of the training dataset and the fragment length were object of study because they highly affect the training time. A small training dataset takes short time to be filled, while a big one requires computing more samples. From this aspect, smaller dataset should be preferred but, as confirmed by practice, they could not work well due to the physics of the problem: it becomes harder and harder to distinguish good actions from bad ones if the environment does not show their effects. This would lead to a poor estimation of the objective function gradient, and, consequently, to a bad weight update. The correct compromise between computation time and quality of the result is attained with a rollout worker fragment length of $160$.
The length of the trajectory fragment corresponds to, approximately, $8$ $C_L$-periods. A shorter fragment length of, for example, $10$ samples ($0.5$ $C_L$-periods) makes the training impossible, whereas fragment length of, for example, $80$ samples ($4$ $C_L$-periods) is fine, but high rewards are not reached. 
Formally, our setup is episodic, with an episode maximum length of $160$ samples. Despite the end of an episode, the environment is not reinitialized, so an episode starts from the last CFD-time steps of the previous episode. The episodes define only a cut in the continuous trajectory. This does not hold in case of numerical instability: if it happens, the CFD simulation is interrupted and relaunched from a given fixed initial condition. This is quite a rare event, so there is no manifestation of overfitting of the initial condition.

% ================================================
% REFS ===========================================
% ================================================
% \bibliography{rl_bb_aip}
% \bibliographystyle{rl_bb_aip}

% \clearpage %
\bibliographystyle{siam}
\bibliography{rl_bb_aip}% Produces the bibliography via BibTeX.

\end{document}